\documentclass[a4paper,11pt]{article}
\usepackage[english]{babel}
\usepackage{jheppub}
\pdfoutput=1
\usepackage[T1]{fontenc}
\usepackage{bbold}
\usepackage{float}
\usepackage{amsmath}
\usepackage{empheq}
\usepackage{booktabs}
\usepackage{color}
\usepackage[utf8]{inputenc}
\usepackage{xspace}
\usepackage{scalerel}
\usepackage[most]{tcolorbox}
\usepackage{multirow}
\usepackage{scalefnt}
\usepackage{bold-extra}
\usepackage[shortlabels]{enumitem}
\usepackage[tikz]{bclogo}
\usepackage{subcaption}
\usepackage{tikz-feynman}
\usepackage{cancel}
\usepackage{verbatim}
\usetikzlibrary{arrows,shapes}
\usepackage{afterpage}
\usepackage{graphicx,grffile}
\usepackage{xspace}
\usepackage[normalem]{ulem}
\usepackage{makecell}

\newcommand{\incl}{{\tt inclusive}}
\newcommand{\fidYR}{{\tt fiducial-YR}}
\newcommand{\fidATLAS}{{\tt fiducial-ATLAS}}

\newcommand{\stepone}{{Step\,I}}
\newcommand{\steptwo}{{Step\,II}}
\newcommand{\stepthree}{{Step\,III}}

\newcommand\F{${\rm F}$}
\newcommand\FJ{${\rm FJ}$}
\newcommand\FJJ{${\rm FJJ}$}

\newcommand\vh{$VH$}
\newcommand\zh{$ZH$}
\newcommand\wh{$W^\pm H$}
\newcommand\wph{$W^+ H$}
\newcommand\wmh{$W^- H$}

\newcommand{\as}{\alpha_s}

\newcommand{\pt}{{p_{\text{\scalefont{0.77}T}}}}

\newcommand{\ptarg}[1]{{p_{\text{\scalefont{0.77}T,$#1$}}}}
\newcommand{\ptargsqr}[1]{{p^{2}_{\text{\scalefont{0.77}T,$#1$}}}}

\newcommand{\ptrad}{{p_{\text{\scalefont{0.77}T,rad}}}}
\newcommand{\pth}{{p_{\text{\scalefont{0.77}T,$H$}}}}
\newcommand{\ptz}{{p_{\text{\scalefont{0.77}T,$Z$}}}}
\newcommand{\ptw}{{p_{\text{\scalefont{0.77}T,$W$}}}}
\newcommand{\ptwp}{{p_{\text{\scalefont{0.77}T,$W^+$}}}}

\newcommand{\ptbb}{{p_{\text{\scalefont{0.77}T,$bb$}}}}
\newcommand{\ptb}{{p_{\text{\scalefont{0.77}T,$b$-jet}}}}
\newcommand{\ptwph}{{p_{\text{\scalefont{0.77}T,$W^+H$}}}}

\newcommand{\mzh}{{m_{\text{\scalefont{0.77}$ZH$}}}}
\newcommand{\mwh}{{m_{\text{\scalefont{0.77}$WH$}}}}
\newcommand{\mbb}{{m_{\text{\scalefont{0.77}$bb$}}}}
\newcommand{\mee}{{m_{\text{\scalefont{0.77}$e^+e^-$}}}}

\newcommand{\ptbone}{{p_{\text{\scalefont{0.77}T,$b_{\rm hard}$}}}}

\newcommand{\pte}{{p_{\text{\scalefont{0.77}T,$e$}}}}
\newcommand{\ptem}{{p_{\text{\scalefont{0.77}T,$e^-$}}}}
\newcommand{\ptep}{{p_{\text{\scalefont{0.77}T,$e^+$}}}}
\newcommand{\pteone}{{p_{\text{\scalefont{0.77}T,$e_1$}}}}

\newcommand{\ptmiss}{{p_{\text{\scalefont{0.77}T,miss}}}}

\newcommand{\yh}{{y_{\text{\scalefont{0.77}$H$}}}}

\newcommand{\ywp}{{y_{\text{\scalefont{0.77}$W^+$}}}}

\newcommand{\absdybb}{{|\Delta y_{\text{\scalefont{0.77}$b$,$b$}}|}}
\newcommand{\dybb}{{\Delta y_{\text{\scalefont{0.77}$b$,$b$}}}}

\newcommand{\etae}{{\eta_{\text{\scalefont{0.77}$e$}}}}

\newcommand{\etab}{{\eta_{\text{\scalefont{0.77}$b$-jet}}}}

\newcommand{\muF}{{\mu_{\text{\scalefont{0.77}F}}}}
\newcommand{\muR}{{\mu_{\text{\scalefont{0.77}R}}}}

\newcommand{\Q}{{Q_{\text{\scalefont{0.77}$0$}}}}

\newcommand{\noun}[1]{{\scshape #1}}
\newcommand{\MADGRAPH}{\noun{MadGraph v4}}

\newcommand{\POWHEG}{\noun{Powheg}}

\newcommand{\POWHEGBOX}{\noun{Powheg-Box}}
\newcommand{\POWHEGBOXRES}{\noun{Powheg-Box-Res}}
\newcommand{\POWHEGBOXVTWO}{\noun{Powheg-Box-V2}}

\newcommand{\minlo}{{\noun{MiNLO$^{\prime}$}}\xspace}
\newcommand{\minnlo}{{\noun{MiNNLO$_{\rm PS}$}}\xspace}
\newcommand{\unnlops}{{\noun{UNNLOPS}}\xspace}
\newcommand{\geneva}{{\noun{Geneva}}\xspace}

\newcommand{\OpenLoops}{{\noun{OpenLoops}}\xspace}

\newcommand{\PYTHIA}[1]{\noun{Pythia{#1}}\xspace}

\newcommand{\vhatnnlo}{\textsc{vh@nnlo}\xspace}

\newcommand{\citere}[1]{ref.\,\cite{#1}}

\newcommand{\citeres}[1]{refs.\,\cite{#1}}

\newcommand{\eqn}[1]{eq.\,(\ref{#1})}

\newcommand{\fig}[1]{figure\,\ref{#1}}

\newcommand{\Fig}[1]{Figure\,\ref{#1}}

\newcommand{\tab}[1]{table\,\ref{#1}}
\newcommand{\sct}[1]{section~\ref{#1}}

\newcommand{\app}[1]{appendix~\ref{#1}}
\newcommand{\LambdaPWG}{\Lambda_{\rm pwg}}

\usepackage{xcolor}

\newcommand{\tmop}[1]{\ensuremath{\operatorname{#1}}}

\newtcolorbox{empheqboxed}{colback=white!35, 
 colframe=black,
 width=\textwidth,
 sharpish corners,
 top=-2mm, 
 bottom=0pt
}

\title{
Next-to-next-to-leading order event generation for \boldmath{$VH$} production with \boldmath{$H\to b\bar{b}$} decay}

\author[]{Silvia Zanoli,$^{a}$}
\author[]{Mauro Chiesa,$^{b}$}
\author[]{Emanuele Re,$^{c,d}$}
\author[]{Marius Wiesemann$^{a}$}
\author[]{and \hspace{2cm}Giulia Zanderighi\,$^{a,e}$}

\emailAdd{zanoli@mpp.mpg.de}
\emailAdd{mauro.chiesa@unipv.it}
\emailAdd{emanuele.re@mib.infn.it}
\emailAdd{marius.wiesemann@mpp.mpg.de}
\emailAdd{zanderi@mpp.mpg.de}

\affiliation[]{$^{a}$Max-Planck-Institut f\"ur Physik, F\"ohringer Ring 6,
  80805 M\"unchen, Germany}
\affiliation[]{$^{b}$Dipartimento di Fisica, Universit\`a di Pavia, and INFN,
Sezione di Pavia, Via A. Bassi 6, 27100 Pavia, Italy}
\affiliation[]{$^{c}$INFN, Sezione di Milano\,-\,Bicocca, and
  Universit\`a di Milano\,-\,Bicocca, Piazza della Scienza 3, 20126 Milano, Italy}
\affiliation[]{$^{d}$LAPTh, Universit\'e Grenoble Alpes, Universit\'e Savoie Mont Blanc, CNRS, 74940 Annecy, France}
\affiliation[]{$^{e}$Physik-Department, Technische Universit\"at M\"unchen, James-Franck-Strasse 1, 85748 Garching, Germany}

\abstract{We consider the Higgsstrahlung process in hadronic
  collisions and present the computation of next-to-next-to-leading
  order predictions matched to parton showers for both production and
  $H\to b\bar{b}$ decay employing the \minnlo{} method.  We present
  predictions for \zh{} and \wh{} production including spin
  correlations and off-shell effects by calculating the
  full processes $pp \to \ell^+\ell^-H \to \ell^+\ell^-b\bar{b}$, $pp
  \to \nu_\ell\bar\nu_\ell H \to \nu_\ell\bar\nu_\ell b\bar{b}$ and
  $pp \to \ell^\pm \nu_\ell H \to \ell^\pm \nu_\ell b\bar{b}$ in the
  narrow-width approximation for the Higgs boson. For the \wh{}
  process, NNLO+PS accuracy in production and decay is achieved for
  the first time.  Our calculations are validated against earlier
  simulations in the NNLOPS approach that includes NNLO corrections
  via multi-differential reweighting.  The new \minnlo{} generators
  for these processes, which evaluate NNLO corrections on-the-fly in the
  event generation, will supersede those earlier calculations.  Our
  predictions are in good agreement with recent measurements of the
  Higgsstrahlung cross sections.}


\preprint{MPP-2021-204, LAPTH-044/21}

\begin{document}

\maketitle

\section{Introduction}
\label{sec:intro}

The Higgs boson is a unique particle in the Standard Model of Particle
Physics (SM).
Although it interacts with all massive particles, which obtain their
masses precisely through their interaction with the Higgs field, only
with the high-energy proton--proton collisions of the Large Hadron
Collider (LHC) the Higgs boson could be discovered due it its
relatively large mass of $125$\,GeV.  The discovery in 2012 was based
on 5 fb$^{-1}$ of data collected at 7 and 8
TeV~\cite{Aad:2012tfa,Chatrchyan:2012xdj}. A discovery in this early
stage of the LHC was possible since the Higgs boson has a sufficiently
large branching fraction to photons and $Z$ bosons, which lead to
experimentally very clean signatures with an excellent mass
resolution.  Since the discovery studying the properties of the
observed scalar resonance has been one of the cornerstones of the rich
physics programme at the LHC. In particular, the Higgs sector provides
a window to new-physics phenomena that could for instance affect the
Higgs couplings. With all Higgs properties appearing to be SM-like
thus far and without any other clear signs of physics beyond the SM
(BSM), precision tests of the Higgs sector have become a valuable path
towards the observation of new physics.

Higgs-boson production at the LHC proceeds predominantly through gluon
fusion and is induced by a top-quark loop, while the Higgs decay to
bottom quarks ($H\to b\bar{b}$) has the largest branching ratio.
Other Higgs-production modes are suppressed by roughly one order of
magnitude or more with respect to gluon fusion, but they are
indispensable in the study of the Higgs properties to determine as
many of the Higgs couplings as possible.  For instance, the associated
production of a Higgs boson and a vector boson (\vh{} production or
``Higgsstrahlung") or vector-boson fusion (VBF) provide access to
Higgs couplings to vector bosons, increasing the sensitivity to those
couplings in addition to the Higgs decay modes to vector bosons.  Even
more importantly, the $H\to b\bar{b}$ decay, which provides direct
access to the bottom Yukawa coupling, could only be observed
\cite{ATLAS:2018kot,CMS:2018nsn} by exploiting the \vh{} production
mode with a boosted Higgs boson.  In fact, it had been thought for a
long time that Higgsstrahlung would not be accessible at the LHC until
it was proposed to consider $VH$ events where the Higgs boson has
large transverse momentum (i.e.\ it is boosted) and to apply
substructure techniques to identify the bottom quarks and reconstruct
the Higgs boson \cite{Butterworth:2008iy}.  This renders the accurate
modeling of both \vh{} production and the $H\to b\bar{b}$ decay at the
level of fully exclusive events indispensable. While this was already
crucial for the experiments to observe this decay mode, it will become
even more critical for an accurate determination of the bottom Yukawa
coupling in the future.

A remarkable progress in next-to-next-to-leading order (NNLO) QCD
calculations for SM processes has been made in the past years, with
all $2\to 2$ colour-singlet scattering processes computed at this
accuracy
\cite{Ferrera:2011bk,Ferrera:2014lca,Ferrera:2017zex,Campbell:2016jau,Harlander:2003ai,Harlander:2010cz,Harlander:2011fx,Buehler:2012cu,Marzani:2008az,Harlander:2009mq,Harlander:2009my,Pak:2009dg,Neumann:2014nha,Czakon:2021yub,Catani:2011qz,Campbell:2016yrh,Grazzini:2013bna,Grazzini:2015nwa,Campbell:2017aul,Gehrmann:2020oec,Cascioli:2014yka,Grazzini:2015hta,Heinrich:2017bvg,Kallweit:2018nyv,Gehrmann:2014fva,Grazzini:2016ctr,Grazzini:2016swo,Grazzini:2017ckn,deFlorian:2013jea,deFlorian:2016uhr,Grazzini:2018bsd,Baglio:2012np,Li:2016nrr,deFlorian:2019app},
and even first NNLO calculations for genuine $2\to 3$ processes
emerging in the recent past
\cite{Chawdhry:2019bji,Kallweit:2020gcp,Czakon:2021mjy,Chawdhry:2021hkp}.
The Higgsstrahlung process is actually much simpler from a
computational point of view: As far as QCD corrections are concerned
the calculation corresponds to an extension of the Drell-Yan (DY)
process in which the Higgs boson is emitted from the vector
boson. Indeed, the inclusive NNLO cross section was computed almost
twenty years ago \cite{Brein:2003wg,Brein:2011vx,Brein:2012ne}, and
about ten years later fully differential NNLO calculations appeared
\cite{Ferrera:2011bk,Ferrera:2013yga,Ferrera:2014lca,Campbell:2016jau,Ferrera:2017zex,Caola:2017xuq,Gauld:2019yng}.
Part of the NNLO corrections to $ZH$ production is the loop-induced
gluon-fusion ($gg$) process, which is mediated by a top-quark loop,
while other contributions mediated by a top-quark loop have been shown
to be relatively small \cite{Brein:2011vx} and are thus often
neglected in differential calculations. In fact, the effectively only
leading-order (LO) accurate loop-induced $gg$ process dominates the
theoretical uncertainties in $ZH$ analyses, which calls for the
inclusion of NLO corrections to $gg\to ZH$ production, especially
since first approximations suggested that these are quite
large~\cite{Altenkamp:2012sx,Harlander:2014wda,Hasselhuhn:2016rqt,Wang:2021rxu}.
Recently, substantial progress was made in the calculation of the
relevant two-loop amplitude to (on-shell) $gg\to ZH$ production
\cite{Davies:2020drs,Chen:2020gae,Alasfar:2021ppe}, which eventually
led to the first calculation at NLO QCD in a small mass expansion
\cite{Wang:2021rxu}.  Strategies to isolate the $gg\to ZH$ component
in experimental analyses have been suggested in
\citere{Harlander:2018yns}.  The analytic resummation of logarithms at
threshold and due to a jet veto in \vh{} production have been
considered in
\citeres{Dawson:2012gs,Shao:2013uba,Li:2014ria,Harlander:2014wda}.
Recently, also \wh{} production in association with a jet was
calculated at NNLO accuracy in QCD \cite{Gauld:2020ced}.  As far as
NLO electroweak (EW) corrections are concerned, they have been
available for a while
\cite{Ciccolini:2003jy,Denner:2011id,Denner:2014cla}.


Since \vh{} measurements exploit a particularly differential strategy,
requiring a boosted Higgs and substructure techniques, as discussed
before, it is mandatory to combine state-of-the-art higher-order
calculations with full-fledged event simulations through a parton
shower (PS) Monte Carlo generator.  NLO QCD corrections matched to
parton showers are available as a merged calculation for
\vh{}\,+\,0,\,1-jet production within \POWHEG{} \cite{Luisoni:2013cuh}
using the \minlo{} method~\cite{Hamilton:2012np,Hamilton:2012rf}.
Similarly, both NLO QCD and EW corrections were simultaneously matched
to a parton shower in a merged \vh{}+0,1-jet calculation in
\citere{Granata:2017iod}. Predictions at LO+PS for $gg\to ZH$ can be
easily obtained with several tools, including with \POWHEGBOX{}
\cite{Luisoni:2013cuh}.  Merged predictions at LO, up to $gg\to
ZH$+\,1-jet, have been considered in
\citeres{Hespel:2015zea,Goncalves:2015mfa}, and a study on how
different approximations compare has also been performed in
\citere{Amoroso:2020lgh}.
The effects from anomalous couplings in \vh{} production have been
considered within \POWHEG{} in \citere{Mimasu:2015nqa}.  In recent
years, a great effort has been devoted to the matching of NNLO QCD
calculations and parton showers (NNLO+PS), and essentially four
different methods have been introduced: The NNLOPS
method~\cite{Hamilton:2013fea} based on reweighting \minlo{} events
\cite{Hamilton:2012rf}, the \unnlops{} method
\cite{Hoeche:2014aia,Hoche:2014dla}, the \geneva{} method
\cite{Alioli:2013hqa}, and, most recently, the \minnlo{} approach
\cite{Monni:2019whf,Monni:2020nks}.  At first these methods have been
applied to the simplest colour singlet processes, namely Higgs-boson
production~\cite{Hamilton:2013fea,Hamilton:2015nsa,Hoche:2014dla} and
Drell Yan~\cite{Karlberg:2014qua,Hoeche:2014aia,Alioli:2015toa}.
Especially in the NNLOPS, \minnlo{} and \geneva{} approaches a variety
of colour-singlet production processes are now available at NNLO+PS,
including \zh{} \cite{Astill:2018ivh,Alioli:2019qzz}, \wh{}
\cite{Astill:2016hpa,Alioli:2019qzz}, $\gamma\gamma$
\cite{Alioli:2020qrd}, $Z\gamma$
\cite{Lombardi:2020wju,Lombardi:2021wug}, $W^\pm\gamma$
\cite{Cridge:2021hfr}, $ZZ$ \cite{Alioli:2021egp,Buonocore:2021fnj}
and $W^+W^-$ \cite{Re:2018vac,Lombardi:2021rvg} production. Very
recently, also the NNLO+PS matching for the very first production
process with coloured particles in the initial and final state has
been achieved in \citere{Mazzitelli:2020jio}, which extents the
\minnlo{} method to top-quark pair production.

Similarly important in this context is an accurate and fully exclusive
description of the $H\to b\bar{b}$ decay.  The total decay rate for
massless bottom quarks is known to N$^4$LO QCD by
now~\cite{Gorishnii:1990zu, Gorishnii:1991zr, Kataev:1993be,
  Surguladze:1994gc, Larin:1995sq, Chetyrkin:1995pd, Chetyrkin:1996sr,
  Baikov:2005rw,Davies:2017xsp,Davies:2017xsp}. Electroweak
corrections were computed in \citeres{Fleischer:1980ub, Bardin:1990zj,
  Dabelstein:1991ky, Kniehl:1991ze}. We refer to
\citeres{Denner:2011mq,Spira:2016ztx} for a review of calculations for
the inclusive $H\to b \bar b$ branching ratio.  Fully differential
calculations are available at fixed-order through NLO QCD for massive
bottom quarks since a long
time~\cite{Braaten:1980yq,Drees:1990dq,Sakai:1980fa,Janot:1989jf}.
NNLO QCD predictions have been first computed for massless bottom
quarks
\cite{Anastasiou:2011qx,DelDuca:2015zqa,Ferrera:2017zex,Caola:2019pfz},
and the corresponding results retaining massive bottom quarks were
presented in
\citeres{Bernreuther:2018ynm,Behring:2019oci,Behring:2020uzq,
  Somogyi:2020mmk}.  Recently, even the N$^3$LO corrections were
calculated in the massless approach
\cite{Mondini:2019vub,Mondini:2019gid}.  The effect of top-quark loops
have been studied in \citeres{Primo:2018zby,Mondini:2020uyy}.  A
complete calculation at NLO QCD in the SM Effective Field Theory
(SMEFT) was presented in \citere{Cullen:2019nnr}, while anomalous
couplings between the vector bosons in the production of the \vh{}
final state, including the $H\to b\bar{b}$ decay, were considered at
NNLO QCD in \citere{Bizon:2021rww}.  As far as Monte Carlo predictions
are concerned, the simulation of $H\to b\bar{b}$ decay events has been
achieved at NNLO+PS in
\citeres{Bizon:2019tfo,Alioli:2020fzf,Hu:2021rkt}.

In this paper, we present a fully exclusive simulation of \zh{} and
\wh{} production with $H\to b\bar{b}$ decay in the \minnlo{}
framework.  Both production and decay events are simulated at NNLO+PS
accuracy and consistently combined to obtain accurate predictions for
the $\ell^+\ell^-b\bar b$, $\nu_\ell \bar\nu_\ell b\bar b$ and
$\ell^\pm \nu_\ell b\bar b$ final states, including off-shell effects
and spin correlations of the vector bosons and treating the Higgs
boson in the narrow-width approximation.  NNLO+PS accuracy for the
$pp\to \nu_\ell \bar\nu_\ell H\to \nu_\ell \bar\nu_\ell b\bar b$ and
$pp\to \ell^\pm \nu_\ell H \to \ell^\pm \nu_\ell b\bar b$ processes is
achieved for the first time.  Our implementation has been done both
within \POWHEGBOXVTWO{}~\cite{Alioli:2010xd} and within
\POWHEGBOXRES{}~\cite{Jezo:2015aia}, and the ensuing results are
completely compatible within numerical uncertainties.  For the
$\ell^+\ell^-b\bar b$ final state our \minnlo{} predictions are
compared against the NNLOPS results of \citere{Bizon:2019tfo}, while
for $\ell^\pm \nu_\ell H$ production we validate our calculation
against the NNLOPS results presented in \citere{Astill:2016hpa} for
on-shell Higgs bosons.  The new \minnlo{} implementation shall
supersede those NNLOPS implementations in the future within the
\POWHEGBOX{} framework~\cite{Alioli:2010xd}.  Finally, we present a
complete study of NNLO+PS events for both \zh{} and \wh{} production
with $H\to b\bar{b}$ decay, finding agreement with a recent LHC
measurement from the ATLAS collaboration~\cite{ATLAS:2020jwz}.

This manuscript is organized as follows: in \sct{sec:calculation} we
introduce our calculation by discussing the processes
(\sct{sec:process}), sketching the \minnlo{} method
(\sct{sec:minnlo}), describing our practical implementation
(\sct{sec:practical}) and recalling the combination of production and
decay events (\sct{sec:proddec}). An extensive validation of our
\minnlo{} generator is presented in \sct{sec:validation} against
NNLOPS prediction for \wh{} production without decay (\sct{sec:wh})
and for \zh{} production with $H\to b\bar b$ decay (\sct{sec:zh}). In
\sct{sec:results} we present new results for both \zh{} and \wh{}
production including the $H\to b\bar{b}$ decay: After introducing the
setup (\sct{sec:setup}), we discuss inclusive and fiducial cross
sections (\sct{sec:cs}), distributions in the fiducial phase space
(\sct{sec:dist}), the effect of different clustering algorithms in the
definition of bottom-quark jets (\sct{sec:cluster}), and compare
against data in (\sct{sec:data}).  We conclude in
\sct{sec:summary}. While in the main text we focus on results for
\wph{} and \zh{} production, in \app{app:Wmresults} we present results
for \wmh{} production.


\section{Outline of the calculation}
\label{sec:calculation}

\subsection{Description of the processes}
\label{sec:process}

We consider the processes
\begin{align}
pp \to ZH\to  \ell^+\ell^-H,\quad pp \to ZH\to  \nu_\ell\bar\nu_\ell H\quad \textrm{and}\quad pp\to W^\pm H\to \ell^\pm\nu_\ell H\,,
\label{eq:process}
\end{align}
for massless leptons $\ell \in\{e,\mu, \tau\}$, where by $\nu_\ell$ we indicate generally a neutrino or anti-neutrino, including the decay mode
\begin{align}
\label{eq:decy}
H\to b\bar{b}\,.
\end{align}
Both production and decay events are generated with NNLO accuracy
within the \minnlo{} framework, introduced in the next section. Their
combination is then consistently matched with the \PYTHIA{8}
\cite{Sjostrand:2014zea} parton shower, achieving full-fledged
hadron-level NNLO+PS events for the $\ell^+\ell^-b\bar{b}$ and
$\ell^\pm\nu_\ell b\bar{b}$ final states, with the only approximation
that the interference between the final-state bottom quarks and the
initial-state partons vanishes. Indeed, such interference is strongly
suppressed because of the small Higgs width and enters only at second
order in the strong coupling constant, since the resonant Higgs boson
is a colour singlet.  Off-shell effects of the intermediate bosons as
well as spin correlations of the vector boson are fully taken into
account, while the Higgs decay is included in the narrow-width
approximation.  For \zh{} production we have implemented both the
$Z\to \ell^+\ell^-$ and the $Z\to \nu_\ell\bar\nu_\ell$ decay modes,
but, without loss of generality, we focus on the decay to charged
leptons throughout this paper.  For simplicity, we will refer to the
$\ell^+\ell^-b\bar{b}$ and $\ell^\pm\nu_\ell b\bar{b}$ final states as
\zh{} and \wh{} production in the following.

\begin{figure}[t]
  \begin{center}
\begin{subfigure}[b]{0.33\linewidth}
      \centering
      \resizebox{0.95\linewidth}{!}{
      \begin{tikzpicture}
        \begin{feynman}
    \vertex (a1) {\(q\)};
    \vertex[below=1.6cm of a1] (a2){\(\bar q\)};
    \vertex[below=0.8cm of a1] (a3);
    \vertex[right=1.2cm of a3] (a4);

    \vertex[right=1cm of a4] (a5);

    \vertex[right=0.6cm of a5](a6);
    \vertex[above=0.3cm of a6](a7);

    \vertex[right=0.8cm of a7](a8);
    \vertex[above=0.3cm of a8](a9);
    \vertex[left=0.3cm of a9](b9);
    \vertex[below=1.2cm of b9](b10);

    \vertex[right=0.8cm of a9](a10);
    \vertex[above=0.2cm of a10](a11){\(\ell^+\)};
    \vertex[below=0.2cm of a10](a12){\(\ell^-\)};
    \vertex[below=0.6cm of a10](a13){\(b\)};
    \vertex[below=1.4cm of a10](a14){\(\bar b\)};

     \diagram* {
       {[edges=fermion]
         (a1)--(a4)--(a2),
         (a11)--(b9)--(a12),
         (a14)--(b10)--(a13),
       },
       (a4) -- [boson, edge label=\(Z\)] (a5),
       (a5) -- [boson, edge label=\(Z\),inner sep=1.5pt,near end] (b9),
       (a5) -- [scalar, edge label=\(H\)] (b10),
       };

  \end{feynman}
      \end{tikzpicture}}
\vspace{0.1cm}
      \caption{}
        \label{subfig:zzres1}
    \end{subfigure}%
\begin{subfigure}[b]{0.33\linewidth}
      \centering
      \resizebox{0.95\linewidth}{!}{
\begin{tikzpicture}      
  \begin{feynman}
    \vertex (a1) {\(q\)};
    \vertex[below=1.6cm of a1] (a2){\(\bar q\)};
    \vertex[below=0.8cm of a1] (a3);
    \vertex[right=1.2cm of a3] (a4);

    \vertex[right=1cm of a4] (a5);

    \vertex[right=0.6cm of a5](a6);
    \vertex[above=0.3cm of a6](a7);

    \vertex[right=0.8cm of a7](a8);
    \vertex[above=0.3cm of a8](a9);
    \vertex[left=0.3cm of a9](b9);
    \vertex[below=1.2cm of b9](b10);

    \vertex[right=0.8cm of a9](a10);
    \vertex[above=0.2cm of a10](a11){\(\ell^\pm\)};
    \vertex[below=0.2cm of a10](a12){\(\nu_\ell\)};
    \vertex[below=0.6cm of a10](a13){\(b\)};
    \vertex[below=1.4cm of a10](a14){\(\bar b\)};

     \diagram* {
       {[edges=fermion]
         (a1)--(a4)--(a2),
         (a11)--(b9)--(a12),
         (a14)--(b10)--(a13),
       },
       (a4) -- [boson, edge label=\(W^\pm\)] (a5),
       (a5) -- [boson, edge label=\(W^\pm\),inner sep=1.5pt,near end] (b9),
       (a5) -- [scalar, edge label=\(H\)] (b10),
       };

  \end{feynman}
\end{tikzpicture}}
\vspace{0.1cm}
\caption{}
        \label{fig:zzres2}
\end{subfigure}
\begin{subfigure}[b]{0.33\linewidth}
      \centering
      \resizebox{0.95\linewidth}{!}{
      \begin{tikzpicture}
        \begin{feynman}
          \vertex (a1) {\(g\)};
          \vertex[below=1.6cm of a1] (a2){\(g\)};
          \vertex[right=1.2cm of a1] (a3);
          \vertex[right=1.2cm of a2] (a4);
          \vertex[right=1.5cm of a3] (a5);
          \vertex[right=1.2cm of a5] (a7);
          \vertex[right=1cm of a7] (a20);
          \vertex[above=0.3cm of a20] (a9){\(\ell^+\)} ;
          \vertex[below=0.3cm of a20] (a11){\(\ell^-\)};
          \vertex[right=1.5cm of a4] (a6);
          \vertex[right=1.2cm of a6] (a8);
          \vertex[right=1cm of a8] (a21);
          \vertex[above=0.3cm of a21] (a10){\(b\)} ;
          \vertex[below=0.3cm of a21] (a12){\(\bar{b}\)};

          \diagram* {
             {[edges=gluon]
              (a1)--(a3),
              (a2)--(a4),
            },
            {[line width=5mm,edges=fermion]
              (a3)--(a5)--(a6)--(a4)--(a3),
              (a9)--(a7)--(a11),
              (a12)--(a8)--(a10),
            },
            (a5) -- [boson, edge label=\(Z\)] (a7),
            (a6) -- [scalar, edge label=\(H\)] (a8),
          };

        \end{feynman}
        \draw [line width=0.4mm] (a3) -- node[below=5.5mm,font=\large] {$\boldsymbol t$} (a5) -- (a6) -- (a4) -- (a3);
      \end{tikzpicture}}
\caption{}
        \label{fig:zzres3}
\end{subfigure}

\end{center}
  \caption{\label{DiagramsVH} Sample Feynman diagrams for $VH$
    production in the $H\to b\bar{b}$ decay channel with (a,c) $V=Z$
    and $Z\to \ell^+\ell^-$ and (b) $V=W^\pm$ and $W^\pm \to
    \ell^\pm\nu_\ell$, where by $\nu_\ell$ we indicate generally a neutrino
    or anti-neutrino. Panels (a) and (b) are tree-level diagrams in
    the quark-annihilation ($q\bar{q}$) channel at LO, while panel (c)
    shows a loop-induced diagram in the gluon-fusion ($gg$) channel
    entering at NNLO QCD.  }
\label{fig:zzres}  
\end{figure}
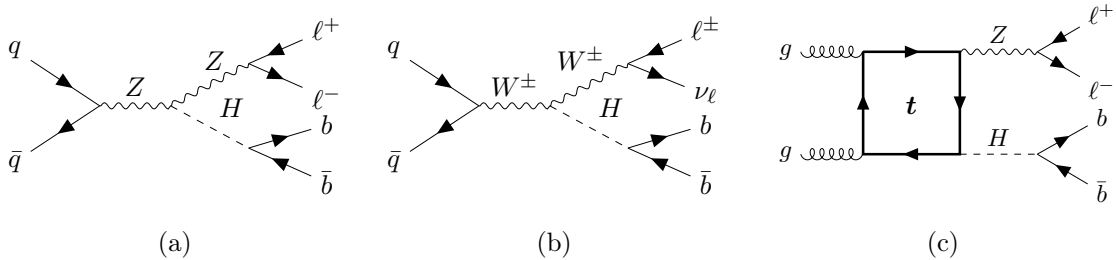

Sample LO diagrams of the two processes are shown in
\fig{DiagramsVH}\,(a) and (b). They are $q\bar{q}$-initiated with an
$s$-channel vector boson emitting a Higgs boson and subsequent decays
of the two bosons. \Fig{DiagramsVH}\,(c) shows a diagram for the
loop-induced process in the $gg$ channel that due to charge
conservation exists only for \zh{}, but not for \wh{} production.
This process enters the $ZH$ cross section at $\mathcal{O}(\as^2)$ and
is therefore a NNLO contribution relative to the $q\bar{q}$-initiated
process. Its dominant contribution proceeds via a top-quark loop,
since all other quark loops are suppressed by their respective Yukawa
couplings to the Higgs boson.  In this work, we focus on the accurate
simulation of the $q\bar{q}$-initiated processes. The loop-induced
$gg$ process can be computed completely independently at LO+PS and
added to the $q\bar{q}$-initiated cross section. We only include the
loop-induced $gg$ contribution when presenting integrated cross
sections in \sct{sec:cs} and when comparing to data in \sct{sec:data}.
In fact, we would like to stress that very recently the $gg\to ZH$
two-loop amplitude for an on-shell $Z$ boson has been calculated
\cite{Davies:2020drs,Chen:2020gae,Alasfar:2021ppe}, which will enable
NLO+PS predictions for the loop-induced $gg$ process in the near
future (with the approximation of the $Z$ boson being on-shell in the
two-loop contribution). Apart from the best predictions for the
$q\bar{q}$-initiated $ZH$ production presented here, this calculation
will be crucial to reduce the theoretical uncertainties in the
experimental analyses of boosted $ZH$ production at the LHC.

We note that there are other contributions to both \wh{} and \zh{}
production that are mediated by a closed top-quark loop radiating the
Higgs boson, which were calculated in \citere{Brein:2011vx}. We adopt
the approximation that the virtual diagrams denoted as $V_{\rm I}$ and
$V_{\rm II}$ in \citere{Brein:2011vx} are neglected in our calculation,
while the real
contributions $R_{\rm I}$ and $R_{\rm II}$, where the latter only
contributes to \zh{} production, are included.  In order to facilitate
the comparison to previous results we have excluded those
contributions to \wh{} production, but included them for \zh{}
production, as discussed below.
We recall that the contribution of the $R_{\rm I/II}$
and $V_{\rm I/II}$ terms to the total cross section is positive. Both $R_{\rm I/II}$ and $V_{\rm I/II}$
contribute to about 0.6\%, giving a total positive correction of about
1.2\% with respect to the total $q\bar{q}$-initiated NNLO result
obtained neglecting these terms.

\subsection[\minnlo{} method]{\boldmath{\minnlo{}} method}
\label{sec:minnlo}

Our calculation of $VH$ production at NNLO+PS relies on the \minnlo{}
method, which was formulated and applied to $2\to 1$ processes in
\citeres{Monni:2019whf,Monni:2020nks} and later extended to generic
colour-singlet processes in \citere{Lombardi:2020wju} and to
heavy-quark pair production in \citere{Mazzitelli:2020jio}.  The
interested reader is referred to those publications for details, while
here we only sketch the main ideas behind the \minnlo{} procedure in a
simplified notation.

The \minnlo method that formulates a matching of NNLO corrections with
a parton shower for a system \F{} of colour-singlet particles proceeds
essentially in three steps: In \stepone{}, a \POWHEG{} calculation
\cite{Nason:2004rx,Nason:2006hfa,Frixione:2007vw,Alioli:2010xd}
generates \F{} in association with one light parton at NLO inclusively
over the second radiation.  In \steptwo{}, an appropriate Sudakov form
factor and higher-order corrections are applied such that the
simulation remains finite in the unresolved limit of the light partons
and becomes NNLO accurate for inclusive \F{} production. \stepthree{}
involves both the exclusive generation of the second radiated parton
(accounted for inclusively in \stepone{}) through the \POWHEG{}
Sudakov and the generation of subsequent emissions through the parton
shower.  Since all emissions are correctly ordered in $p_T$ (when
using $p_T$-ordered showers) and the Sudakov in \steptwo{} matches the
leading logarithms generated by the parton shower, the \minnlo{}
method preserves the (leading logarithmic) accuracy of the parton
shower.

We write the fully differential \minnlo{} cross section symbolically as
\begin{align}
\begin{split}\label{eq:minnlo}
{\rm d}\sigma_{\rm\scriptscriptstyle F}^{\rm MiNNLO_{PS}}={\rm d}\Phi_{\scriptscriptstyle\rm FJ}\,\bar{B}^{\,\rm MiNNLO_{\rm PS}}\,\times\,\left\{\Delta_{\rm pwg}(\Lambda_{\rm pwg})+ {\rm d}\Phi_{\rm rad}\Delta_{\rm pwg}(p_{T,{\rm rad}})\,\frac{R_{\scriptscriptstyle\rm FJ}}{B_{\scriptscriptstyle\rm FJ}}\right\}\,,\\
\bar{B}^{\,\rm MiNNLO_{\rm PS}}\sim e^{-S}\,\Big\{{\rm d}\sigma^{(1)}_{\scriptscriptstyle\rm FJ}\big(1+S^{(1)}\big)+{\rm d}\sigma^{(2)}_{\scriptscriptstyle\rm FJ}+\left(D-D^{(1)}-D^{(2)}\right)\times F^{\rm corr}\Big\}\,,
\end{split}
\end{align}
which corresponds to a \POWHEG{} calculation for \F{} plus one light
parton (\FJ{}), but including NNLO accuracy for \F{} production
through a modification of the $\bar B$ function.  Indeed, $\Delta_{\rm
  pwg}$ is the standard \POWHEG{} Sudakov form factor with a default
cutoff of $\LambdaPWG=0.89$\,GeV, $\Phi_{\tmop{rad}}$ and $\ptrad$ are
phase space and transverse momentum of the second radiation,
respectively, and $B_{\scriptscriptstyle\rm FJ}$ and
$R_{\scriptscriptstyle\rm FJ}$ denote the squared tree-level matrix
elements for \FJ{} and \FJJ{} production, respectively, while $\Phi_{\scriptscriptstyle\rm FJ}$
is the \FJ{} phase space.  In the second
line ${\rm d}\sigma^{(1,2)}_{\scriptscriptstyle\rm FJ}$ are the first-
and second-order term in the $\as$ expansion of the differential \FJ{}
cross section, while $e^{-S}$ denotes the Sudakov form factor in the
transverse momentum of \F{} ($\pt$), with $S^{(1)}$ being the first
term in the expansion of the exponent.  Renormalization and
factorization scales are evaluated as $\muR\sim\muF\sim \pt$ in
\minnlo{} and NNLO accuracy is achieved by the last term in the
$\bar{B}$ function, which is of order
$\as^3(p_{\text{\scalefont{0.77}T}})$ and adds the relevant (singular)
contributions \cite{Monni:2019whf}, while regular contributions at
this order are subleading.  The function $D$ is derived from the
following formula that determines the relevant singular terms in
$\pt{}$
\begin{align}
\label{eq:resum}
{\rm d}\sigma_{\scriptscriptstyle\rm F}^{\rm res}=\frac{{\rm d}}{{\rm d}\pt}\left\{e^{-S}\mathcal{L}\right\}=e^{-S}\underbrace{\left\{-S^\prime\mathcal{L}+\mathcal{L}^\prime\right\}}_{\equiv D}\,,
\end{align}
which can be obtained by suitably manipulating a $\pt{}$ resummation
formula, as explained in section 4 of \citere{Monni:2019whf}.
$\mathcal{L}$ is the luminosity factor up to NNLO, including the
convolution of the collinear coefficient functions with the parton
distribution functions (PDFs) and the squared hard-virtual matrix
elements for \F{} production.

Instead of truncating \eqn{eq:minnlo} at the third order,
i.e.\ evaluating
$\left(D-D^{(1)}-D^{(2)}\right)=D^{(3)}+\mathcal{O}(\as^4)$, as
originally done in \citere{Monni:2019whf}, we keep all terms of
$\mathcal{O}(\as^4)$ and higher, as suggested in
\citere{Monni:2020nks}, in order to preserve the total derivative in
\eqn{eq:resum}.  Finally, the factor $F^{\rm corr}$ in \eqn{eq:minnlo}
ensures that the Born-like NNLO corrections are consistently included
in the \FJ{} \POWHEG{} events by spreading them in \FJ{} phase space.

\subsection{Practical implementation}
\label{sec:practical}

We have implemented \minnlo{} generators for \vh{} production both in
the \POWHEGBOXVTWO{} \cite{Alioli:2010xd} and the \POWHEGBOXRES{}
\cite{Jezo:2015aia} framework.  Both calculations yield fully
compatible results within numerical uncertainties.  Our
\POWHEGBOXVTWO{} implementation uses the \vh{}+jet generators
developed in~\citere{Luisoni:2013cuh}, which were already used in the
former NNLOPS calculations of
\citeres{Astill:2016hpa,Astill:2018ivh}. It then includes NNLO QCD
corrections to \vh{} production by exploiting the original
implementation of the \minnlo{} method for Drell-Yan production of
\citeres{Monni:2019whf,Monni:2020nks}, which is suitable since the
respective two-loop corrections for \vh{} production correspond to
those of the Drell-Yan (DY) process with the Higgs boson being emitted
from the vector boson.  Our \minnlo{} \vh{} generators in
\POWHEGBOXRES{}, on the other hand, are based on the \vh{}+jet
implementation of~\citere{Granata:2017iod}.  The \vh{}+jet generators
have then been upgraded to include NNLO QCD accuracy for \vh{}
production through the general \minnlo{} implementation for
colour-singlet production developed for \POWHEGBOXRES{} in
\citere{Lombardi:2020wju}.

As far as the physical amplitudes are concerned, both calculations
exploit the same results, documented in
\citeres{Luisoni:2013cuh,Granata:2017iod}.  The tree-level real and
double-real matrix elements (i.e.\ for \vh{}+1,2-jet production) are
obtained using the automated interface~\cite{Campbell:2012am} between
\POWHEGBOX{} and \MADGRAPH{}~\cite{Alwall:2007st}. The \vh{}+jet
one-loop amplitude is taken from the analytic calculation
of~\citere{Granata:2017kpe}.
The one-loop and two-loop amplitudes have been obtained by applying
the form-factor correction to the quark vertex function (equivalent to
Drell-Yan production) to the Born amplitude.  We note that the
\vh{}+jet generator of \citere{Granata:2017iod} in \POWHEGBOXRES{} in
principle also allows us to use
\OpenLoops{}~\cite{Cascioli:2011va,Buccioni:2017yxi,Buccioni:2019sur}
for the evaluation of the relevant amplitudes, but we have not made
use of that feature.  Note that, as discussed before, our \vh{}
\minnlo{} generators include the $R_{\rm I}$ and $R_{\rm II}$ real
contributions, first calculated in \citere{Brein:2011vx}, while we
neglect the $V_{\rm I}$ and $V_{\rm II}$ virtual amplitudes, which
contribute below one percent.

The calculation of $D$ in \eqn{eq:resum}, which is the central
ingredient to include the NNLO QCD corrections through \minnlo{},
involves the evaluation of several convolutions with the parton
distribution functions (PDFs), which are performed through
\noun{hoppet}~\cite{Salam:2008qg}.  Moreover, the collinear
coefficient functions require the computation of polylogarithms, for
which we employ the \noun{hplog} package~\cite{Gehrmann:2001pz}.

For the calculation of the decay of the Higgs boson to bottom quarks
at NNLO+PS we follow closely the procedure of \citere{Bizon:2019tfo}
using the NNLOPS method.  Thus, pure $H\to b\bar b$ events are
generated, such that the inclusive $H\to b\bar b$ decay width is NNLO
QCD accurate and observables involving an extra jet are NLO QCD
accurate. While for the generation of decay events we could also apply
the \minnlo{} method, we stick to the NNLOPS approach based on
including NNLO corrections through an a-posteriori reweighting.  This
is because in the present case the reweighting is completely
straightforward as one needs to reweight to one single number,
i.e.\ to the NNLO decay width of the Higgs boson to bottom quarks.
All details about the reweighting are given in section\,2.2 of
\citere{Bizon:2019tfo}. In particular, we follow the procedure and
settings recommended there, i.e.\ we use the three-jet resolution
variable in the Cambridge algorithm (with $y_{3,\rm pow} = 2$ and
$y_{3\rm ref} = e^{-4}$) to smoothly turn off the reweighting of
events accompanied by hard radiation, which are already NLO accurate.
All relevant amplitudes have been computed in \citere{DelDuca:2015zqa}. 

Finally, we report the most relevant non-standard settings that we
have used in this paper.
To turn off spurios higher-order logarithmic terms at large transverse
momenta we use a modified logarithm for $\pt > m_{VH}/2$, where
$m_{VH}$ is the invariant mass of the Higgs and vector-boson system,
which smoothly vanishes at $\pt = m_{VH}$, as introduced
in \citere{Mazzitelli:2020jio}. The hard scale in the logarithm is
consistently changed to $m_{VH}/2$ as explained in
\citere{Mazzitelli:2021mmm}.
At small transverse momenta, we use the standard \minnlo{} scale
setting given in eq.\,(14) of \citere{Monni:2020nks} with $\Q
=0$\,GeV. We regularize the Landau singularity by freezing the strong
coupling and the PDFs for scales below $0.8$\,GeV.
Furthermore, we activate the option {\tt largeptscales\,1} to set the
scales entering the cross section at large transverse momenta~\cite{Monni:2020nks}.  
We turn on the \POWHEGBOX{} option
\texttt{doublefsr 1}, which was introduced and discussed in detail in
\citere{Nason:2013uba}.
In our interface to the parton-shower we have kept all settings to the
standard ones, in particular for the recoil scheme.

\subsection[Combining $VH$ production and $H\to b\bar{b}$ decay at NNLO+PS]{Combining \boldmath{$VH$} production and \boldmath{$H\to b\bar{b}$} decay at NNLO+PS}
\label{sec:proddec}

In order to combine \vh{} production at NNLO+PS with the decay of the
Higgs boson to bottom quarks at NNLO+PS, we follow the procedure
outlined in \citere{Bizon:2019tfo}, which we summarize here.  First,
$\ell^+\ell^-H$ and $ \ell^\pm\nu_\ell H$ production events as well as
$H\to b\bar b$ decay events are generated, using the calculations
outlined in the previous section.

Once available production and decay events are merged
following a two-step procedure. First, one replaces, in each
production event, the Higgs boson with the decay products of a Higgs
boson in a decay event. This procedure is explained in all detail in
section\,3.1 of \citere{Bizon:2019tfo}.
Note that decay events are generated on-shell, while we include the
Higgs width in the production events. The procedure performs a rescaling
of the decay event in such a way that momentum is conserved exactly
but small off-shell effects suppressed by $\Gamma_H/m_H$ are
neglected.
Second, the weights of the combined events are set using the 
production and decay weights. In particular, when multiple weights
are present (e.g. to probe scale variations) all possible combinations
are included in the final event file.
More precisely, the combined weight ($w_{\rm full}$) of a production
event with weight $w_{\rm prod}$ and a decay event with weight $w_{\rm
  dec}$ is given by

\begin{equation}
w_{\rm full} = w_{\rm proc} \cdot {\rm Br}_{H \to b
\bar{b}}\cdot \frac{w_{\rm dec}}{\Gamma_{H\to b\bar b}}\,, 
\label{eq:full}
\end{equation}
where ${\rm Br}_{H \to b\bar{b}}$ is the branching ratio of the Higgs
boson to bottom quarks, which is treated as an input, while
$\Gamma_{H\to b\bar b}$ is the $H\to b\bar b$ decay width, which is
computed from the decay events. Accordingly, the last factor in
Eq.~\eqref{eq:full} integrates to one exactly if one is fully
inclusive over the Higgs decay products. This implies that regardless
of the accuracy of the last factor (\minlo{} vs. \minnlo{}) the
integrated inclusive cross section will be exactly equal to the cross
section obtained from the production events times the branching ratio
of the Higgs boson to bottom quarks. Similarly, the precise value of
the bottom Yukawa coupling used in the Higgs decay cancels out exactly
in the ratio of the last factor in \eqn{eq:full}, since we evaluate
the Yukawa coupling at a fixed scale $m_H$.

In order to match the final events to a parton shower one needs to
take into account that the starting scale of the shower for production
and decay is different. In our combined event file we store the
starting scale ({\tt scalup}) of the production stage, while, when
interfacing to the shower we recompute on-the-fly the veto scale for
the decay ({\tt scalup}$_{\tt dec}$) using the decay kinematics, as
explained in section\,3.2 of \citere{Bizon:2019tfo}.

In order to respect the radiation bound coming from the production
stage we simply use the {\tt scalup} variable stored in the combined
event files, as usually done when interfacing \POWHEG{} with \PYTHIA{}. In
order to constrain shower radiation off the decay products, we
implement a vetoed shower, i.e. we first allow \PYTHIA{} to generate
emissions off the Higgs decay products in all available phase space
and, once the shower is completed, we check the hardness of the
splittings that were generated in the decays.
To do so, we go through all splittings that originated from the Higgs
decay products generated by \POWHEG{} and, for each, we compute the
corresponding hardness using
\begin{equation}
    t = 2 \, p_{\rm rad} \cdot p_{\rm em} \, \frac{E_{\rm rad}}{E_{\rm em}}\,,
\label{eq:hardenss}
\end{equation}
where $E_{\rm rad}$ ($E_{\rm em}$) and $p_{\rm rad}$ ($p_{\rm em}$)
are the energy and momentum of the radiating (emitted) particle,
respectively.  If for each of these splittings the hardness is smaller
than {\tt scalup}$_{\tt dec}$, we accept the showered event, otherwise
we reject it and we attempt to shower the event again, until the above
requirement is fulfilled. After 1000 shower attempts, we reject the
event.

Finally, we note that when interfacing to the shower, we make sure
that the automatic matrix element corrections to the decay of the
parton shower program are switched off, since they are already
included with better accuracy in our combined events.

\section{Validation against NNLOPS results}
\label{sec:validation}

$VH$ production at NNLO+PS has already been calculated with the NNLOPS
method, which combines the original \minlo{} approach
\cite{Hamilton:2012rf} with a reweighting procedure fully differential
in the Born phase space to include the NNLO corrections. While such
reweighting constitutes a substantial technical obstacle and requires
an additional handling of the generated \minlo{} events to include
NNLO accuracy a posteriori, the physical results are sound.  Precisely
the complications related to the reweighting are solved by the newer
\minnlo{} method, which includes the NNLO corrections directly during
the generation of the events.  NNLOPS results are available so far for
$WH$ production without \cite{Astill:2016hpa} and for $ZH$ production
with $H\to b\bar{b}$ decay \cite{Bizon:2019tfo}. It is therefore very
useful to validate our \minnlo{} calculation by comparing to those
predictions, which will be done in the following. We will show that
\minnlo{} and NNLOPS results are fully compatible, so that our new
\minnlo{} \vh{} generators supersede the previous NNLOPS ones.

\subsection[\wh{} production]{\boldmath{\wh{}} production}
\label{sec:wh}
 
We start by considering \wh{} production keeping the Higgs boson
stable, and we compare our \minnlo{} predictions against the results
of \citere{Astill:2016hpa}.  Since only results for \wph{} are
presented in \citere{Astill:2016hpa} we consider only this case.

We consider proton--proton collisions at the LHC with a center of mass
energy of $\sqrt{s}=13$\,TeV and present predictions for the process
$pp \to e^+ \nu_e H$. The input parameters are identical to the ones
in \citere{Astill:2016hpa}. In particular, we use the NNLO set of the
MMHT2014nnlo68cl parton densities corresponding to a value of
$\alpha_s(m_Z)$ = 0.118, and we set $m_W$ = 80.399\,GeV, $\Gamma_W$ =
2.085\,GeV and $m_H$ = 125.0\,GeV. Furthermore we set $\alpha_{EM}$ =
1/132.3489 and $\sin^2{\theta_W}$ = 0.2226. The Higgs boson is
considered to be on-shell and, in order to compare with
\citere{Astill:2016hpa}, contributions where the Higgs boson is
radiated off a heavy quark loop are neglected. As already mentioned, these contributions
have been shown to be of the order of
1\%~(see e.g. section\,I.5.2.c of \citere{LHCHiggsCrossSectionWorkingGroup:2016ypw}).  The central
renormalization and factorization scales are set following the
\minnlo{} procedure. The perturbative uncertainties are obtained from
a 7-point scale variation, i.e. by varying $\muR$ and $\muF$ around
the central scale by a factor of two with the constraint $1/2 \le \muR
/ \muF \le 2$. The phase-space definition is fully inclusive, without
any fiducial cuts applied.

The total cross sections are $96.72(4)^{+1.9\%}_{-0.6\%}$\,fb for
\minnlo{} and $96.69(3)^{+1.3\%}_{-1.3\%}$\,fb for NNLOPS.  Thus they
agree at the sub-permille level, which is even within the numerical
errors and far below the scale uncertainties.  There is a small
difference in scale uncertainties, which are of the same size, but
slightly asymmetric in the case of \minnlo{}.  This is a consequence
of the different settings of perturbative scales and the treatment of
terms beyond accuracy in the two calculations, and it is therefore not
unexpected. 

\begin{figure}[p]
\begin{center}\vspace{-0.2cm}
\begin{tabular}{cc}
\includegraphics[width=.31\textheight]{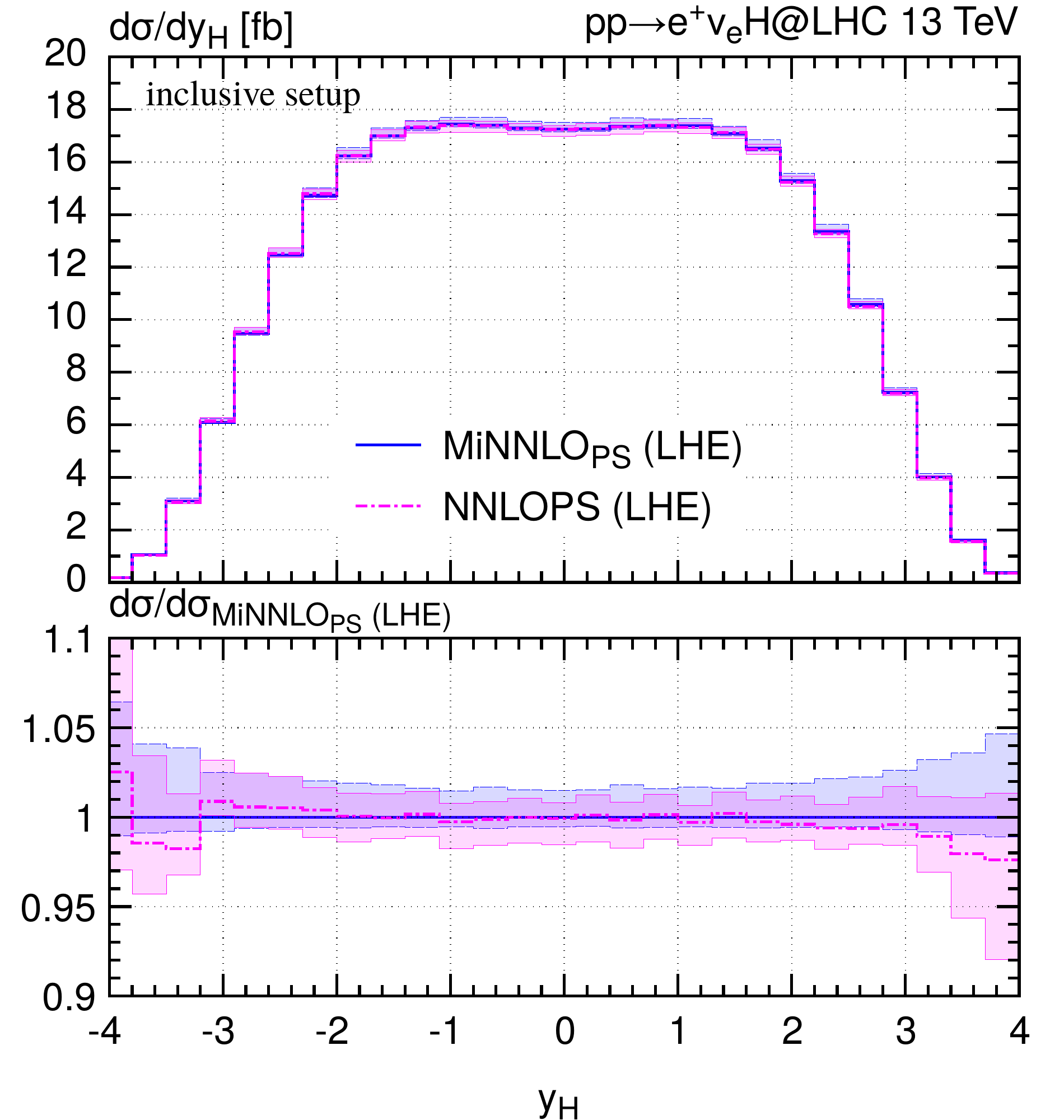} 
&
\includegraphics[width=.31\textheight]{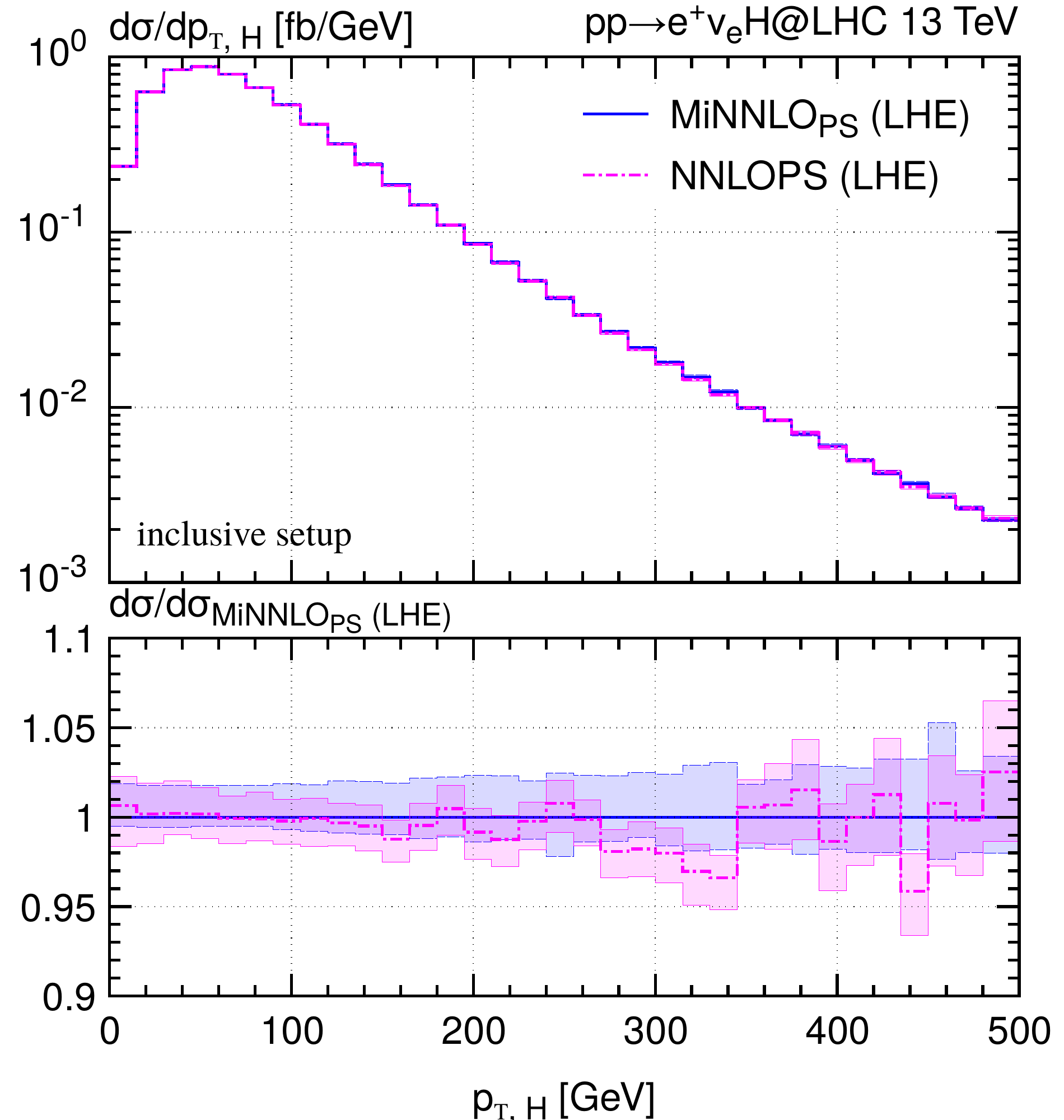}
\end{tabular}\vspace{-0.15cm}
\begin{tabular}{cc}
\includegraphics[width=.31\textheight]{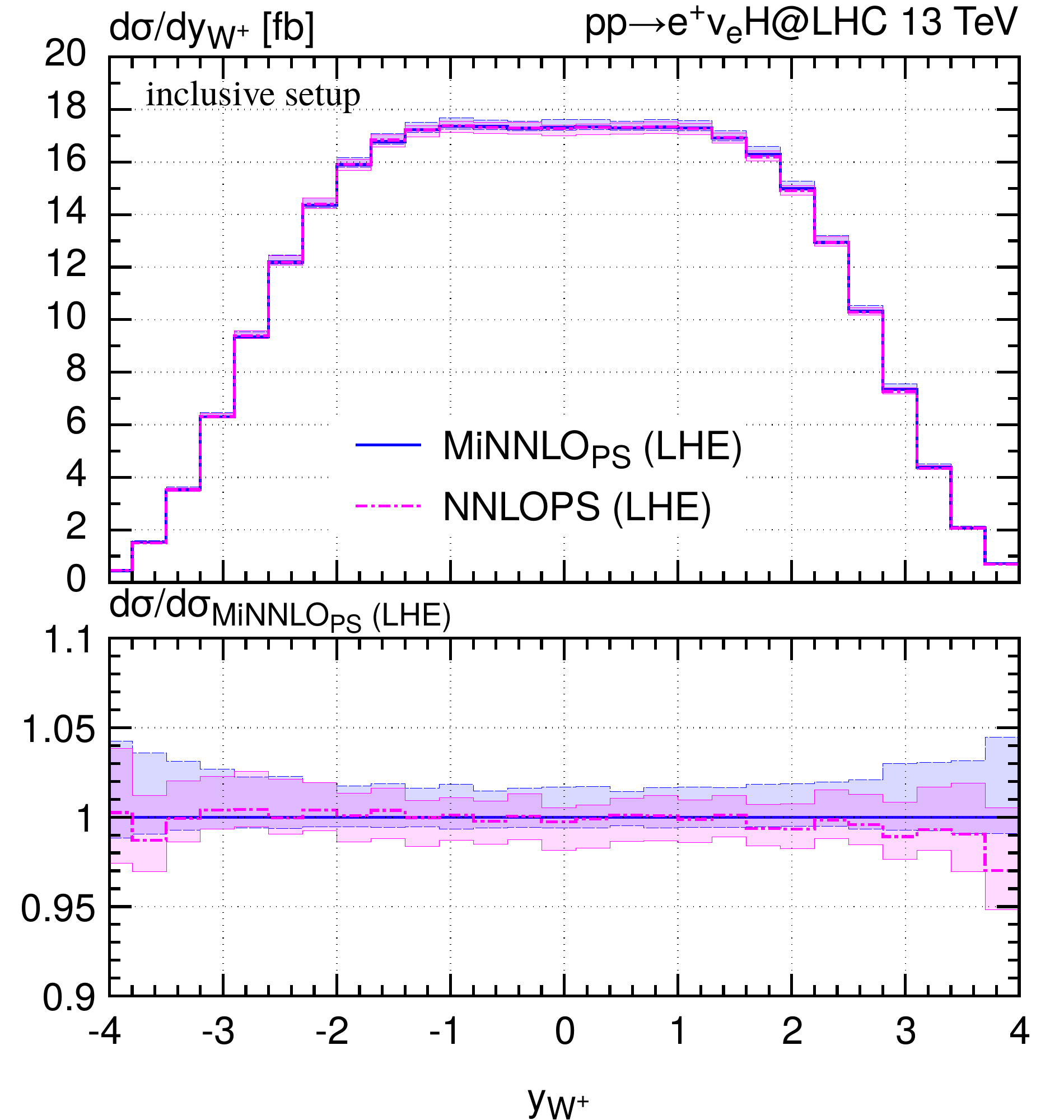}
&
\includegraphics[width=.31\textheight]{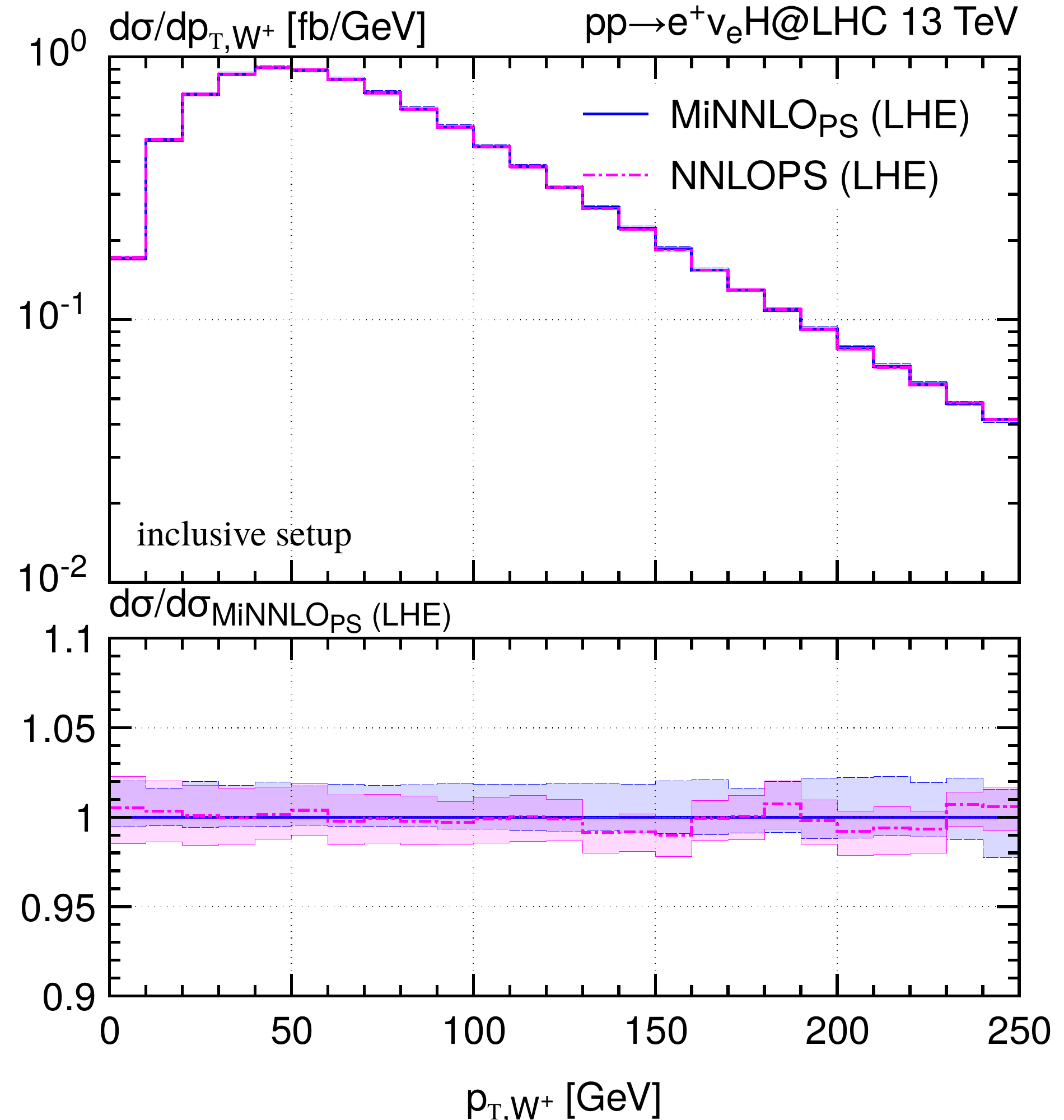}
\end{tabular}\vspace{-0.15cm}
\begin{tabular}{cc}
\includegraphics[width=.31\textheight]{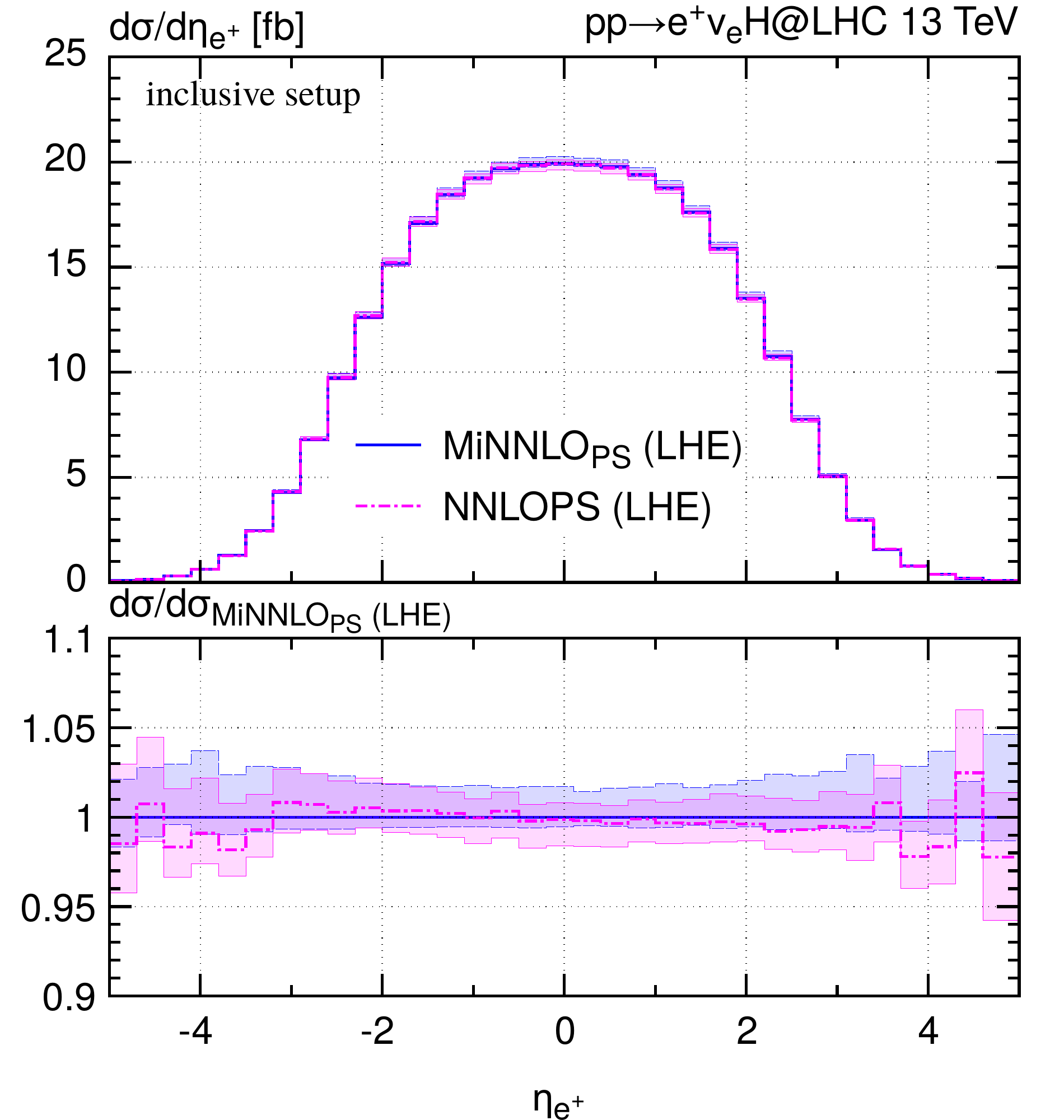}
&
\includegraphics[width=.31\textheight]{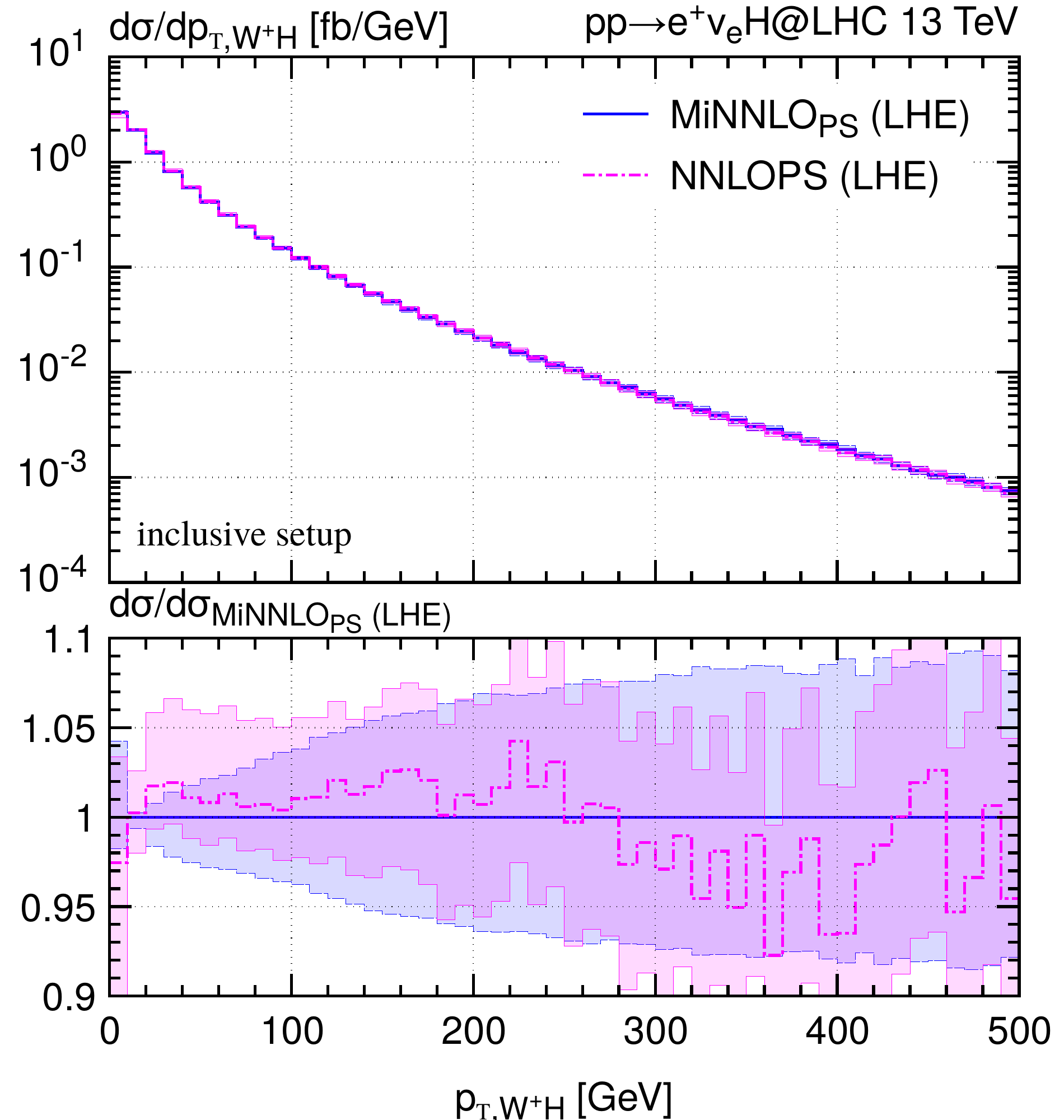} 
\end{tabular}
\caption{\label{fig:validationwh} Comparison of \minnlo{} and NNLOPS predictions for $e^+\nu_e H$ production.} 
\end{center}
\end{figure}
\afterpage{\clearpage}

\Fig{fig:validationwh} compares \minnlo{} and NNLOPS results at the
level of Les Houches events (LHEs) for various differential
distributions.  In particular, the Higgs rapidity ($\yh$) and
transverse momentum ($\pth$), the rapidity ($\ywp{}$) and the
transverse momentum ($\ptwp$) of the $W^+$ boson, as well as the
rapidity of the positron ($\etae$) and the
transverse momentum of the $W^+H$ system ($\ptwph$) are shown. For all
rapidity distributions the central predictions are in excellent
agreement between \minnlo{} and NNLOPS at the level of $\sim 1\%$, up
to statistical fluctuations in the tails. As already observed for the
total cross section, some slight differences in the scale-uncertainty
bands are visible.  Also for the transverse-momentum spectra of the
Higgs and $W^+$ boson we find decent agreement between \minnlo{} and
NNLOPS.  The central predictions differ by less than $1$-$2$\%, except
for some statistical fluctuations.  Finally, the differences in the
transverse-momentum spectrum of the $W^+H$ system are slightly larger,
reaching up to $3$-$4$\%. However, this distribution is effectively
only NLO accurate at $\ptwph\gg0$ and subject to matching ambiguities
at intermediate $\ptwph$ values, which is indicated also by the
enlarged uncertainty bands. Indeed, any difference is fully covered by
the scale variations.
While the two NNLO+PS methods have the same nominal accuracy, the
difference in the size of the uncertainty band of the $\ptwph$
distributions in the region $30\lesssim \ptwph \lesssim 90$ GeV can be
attributed to the fact that the two NNLO+PS methods differ in how the
two-loop NNLO corrections are distributed on the $\ptwph$ spectrum
(i.e. through the use of modified logarithms in \minnlo{} and through
a $\pt$-dependent multi-differential rescaling in NNLOPS). In
particular, as already noticed in~\citere{Monni:2019whf}, the shape of
the \minnlo{} uncertainty band more closely resembles that of a
resummed result matched to a fixed-order prediction.

We have considered many other observables, and we found the same level
of agreement for all of them.  In conclusion, with agreement typically
at the level of $1-2$\% and at most at the level of a few percent, all
within scale variations, the \minnlo{} and NNLOPS calculations provide
remarkably compatible results, and both calculations can be considered
to be equivalent, up to small subleading corrections.

\subsection[$Z H$ production with $H\to b\bar{b}$ decay]{\boldmath{$Z H$} production with $H\to b\bar{b}$ decay}
\label{sec:zh}
The situation for \zh{} production is very similar.  First of all, the
inclusive cross sections obtained with the new implementation for the
production stage (i.e.\ before including the $H\to b\bar{b}$ decay and
without fiducial cuts) have been thoroughly validated against the
completely independent fixed-order calculation implemented in MCFM.
We find an agreement at the sub-percent level for the inclusive cross
section: The central values are $26.567(1)$\,fb for MCFM,
$26.51(1)$\,fb for \minnlo{} using the \POWHEGBOXRES{} implementation,
and $26.63(9)$\,fb using the \POWHEGBOXVTWO{} one, where the numbers
in brackets refer to statistical uncertainties only.  Given that scale
uncertainties are at the one-percent level, and that the calculations
differ by terms beyond accuracy, the agreement is remarkably good.

\renewcommand\arraystretch{1.4}
\begin{table}[b]
  \centering
  \begin{tabular} {c}
    \Xhline{1pt}
    fiducial cuts for $pp\to ZH \to e^{+}e^{-}b\bar{b}$ \\
    \hline
    $\pte > 7$\,GeV, $\pteone > 27$\,GeV, $|\etae| < 2.5$ \\
    $81\,{\rm GeV} < \mee < 101$\,GeV  \\
    $\ge$2 $b$-jets (flavour-$\it{k}_T$\,\cite{Banfi:2006hf}, R=0.4)\\
    $\ptb > 27$\,GeV, $|\etab| < 2.5$ \\
    \Xhline{1pt}
  \end{tabular}
  \caption{\label{tab:cutsvalidation}Fiducial cuts for $e^{+}e^{-}b\bar{b}$ production from \citere{Bizon:2019tfo}.}
  \label{fidcutsvalidation}
\end{table}
\renewcommand\arraystretch{1}

We continue by comparing \minnlo{} predictions to the results of
\citere{Bizon:2019tfo}, which correspond to the combination of $pp\to
e^+e^-H$ production and $H\to b\bar{b}$ decay at NNLO+PS.  Our
comparison is thus performed for $e^+e^-b\bar{b}$ events at the LHE
level.  As before, we consider $\sqrt{s}=13$\,TeV proton--proton
collisions at the LHC. The input parameters and fiducial cuts are
identical to those in \citere{Bizon:2019tfo}.  In particular we use
the PDF4LHC15\textunderscore{nnlo}\textunderscore{mc} parton
distribution functions, corresponding to a value of $\alpha_s(m_Z)$ =
0.118. We set $m_Z$ = 91.1876\,GeV, $\Gamma_Z$ = 2.4952\,GeV, $m_W$ =
80.398\,GeV, $\Gamma_W$ = 2.141\,GeV, $m_H$ = 125.0\,GeV, $\Gamma_H$ =
4.14\,MeV, and $G_F$ = 1.166387$\times \rm{10^{-5} \,GeV^{-2}}$. The
EW coupling is computed as $\alpha_{EM}=\sqrt{2}G_F
m_W^2(1-{m_W^2}/{m_Z^2})/\pi$ and the mixing angle as
$\cos^2{\theta_W}$ = ${m_W^2}/{m_Z^2}$.  The branching ratio for $H
\to b \bar{b}$ is set to Br$_{H \to b \bar{b}}$ =
0.5824. 
For the evaluation of heavy-quark loops we use the bottom-quark pole
mass $m_b$ = 4.92\,GeV and the top-quark pole mass $m_t$ = 173.2\,GeV.
In the $H\to b\bar{b}$ decay, the bottom Yukawa coupling is evaluated
in the $\rm{\overline{MS}}$ scheme at a scale equal to the Higgs boson
mass $m_H$ with a value of $y_b(m_H)$ = 1.280 $\times \rm{10^{-2}}$.
We would like to stress, however, that given the way production and
decay events are combined in \eqn{eq:full} the precise value of the
Yukawa coupling has no impact, as already pointed out in
\sct{sec:proddec}.  Scale uncertainties are obtained through customary
7-point scale variations with the constraint $1/2 \le \muR/\muF \le
2$, while correlating the scale variation factors in production and
decay.  The fiducial cuts are are summarized in
\tab{fidcutsvalidation}. If more than two bottom-flavoured jets
($b$-jets) are present in the final state, the pair with the invariant
mass closer to the Higgs-boson mass is chosen.

\begin{figure}[t!]
\begin{center}\vspace{-0.2cm}
\begin{tabular}{cc}
\includegraphics[width=.31\textheight]{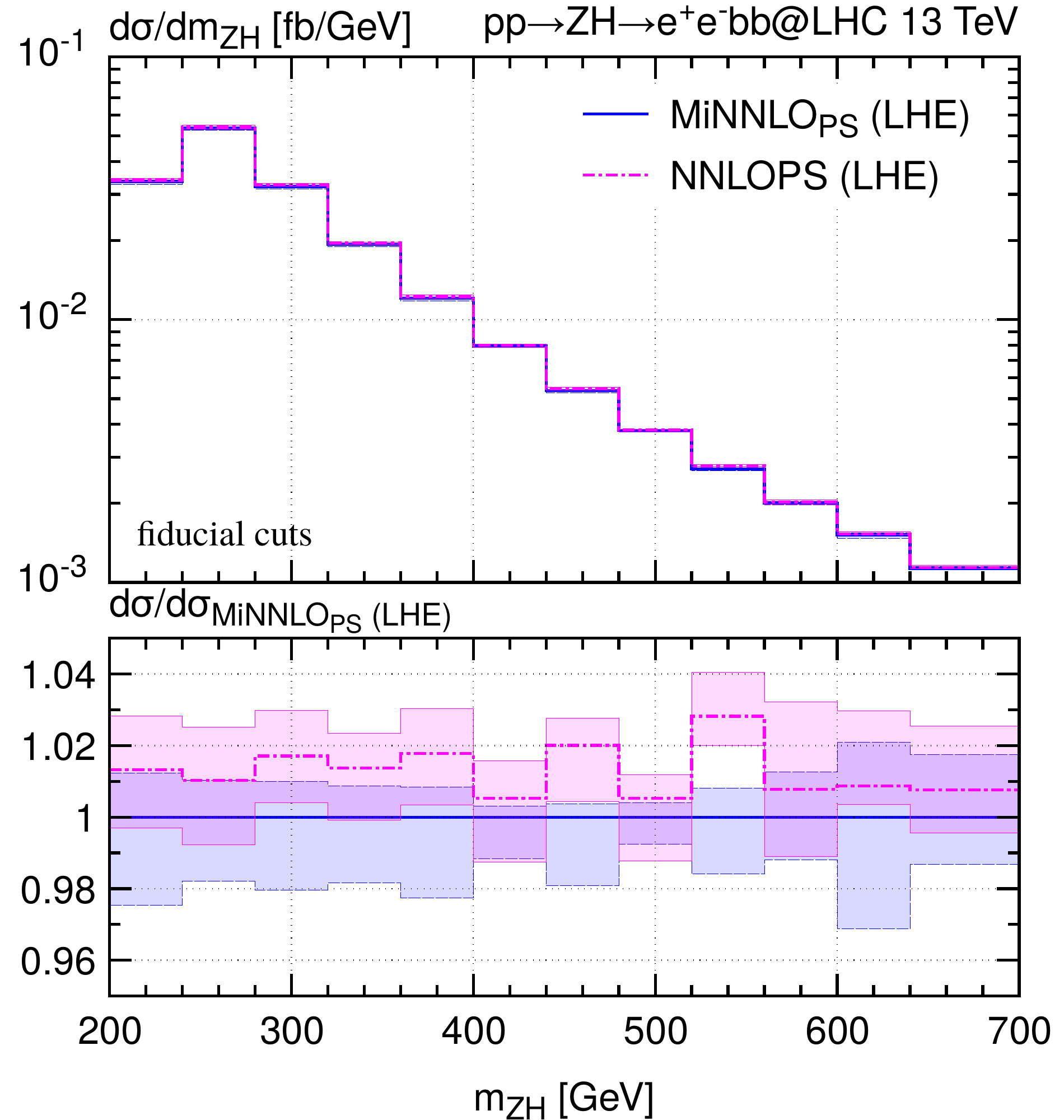}
&
\includegraphics[width=.31\textheight]{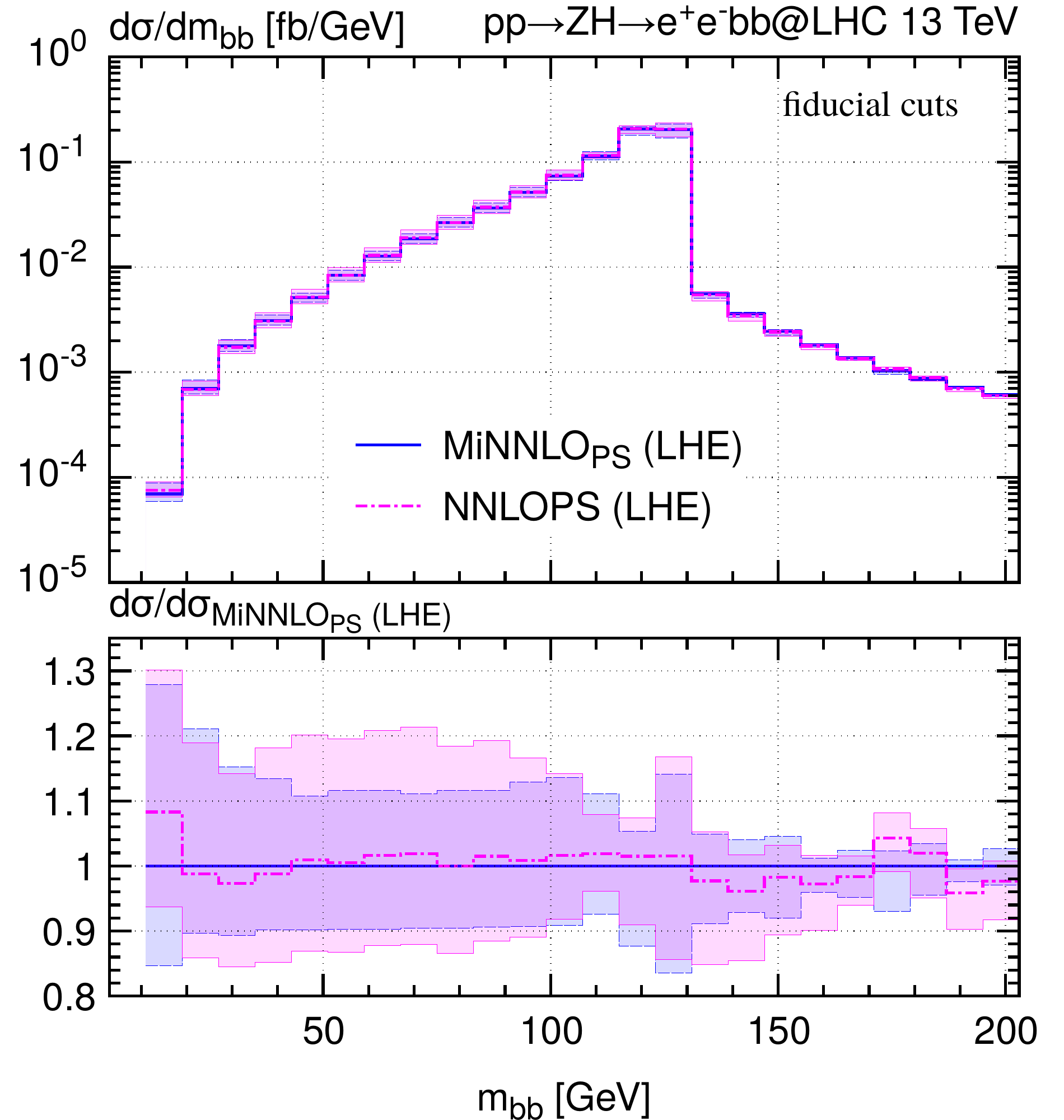}
\end{tabular}\vspace{-0.15cm}
\begin{tabular}{cc}
\includegraphics[width=.31\textheight]{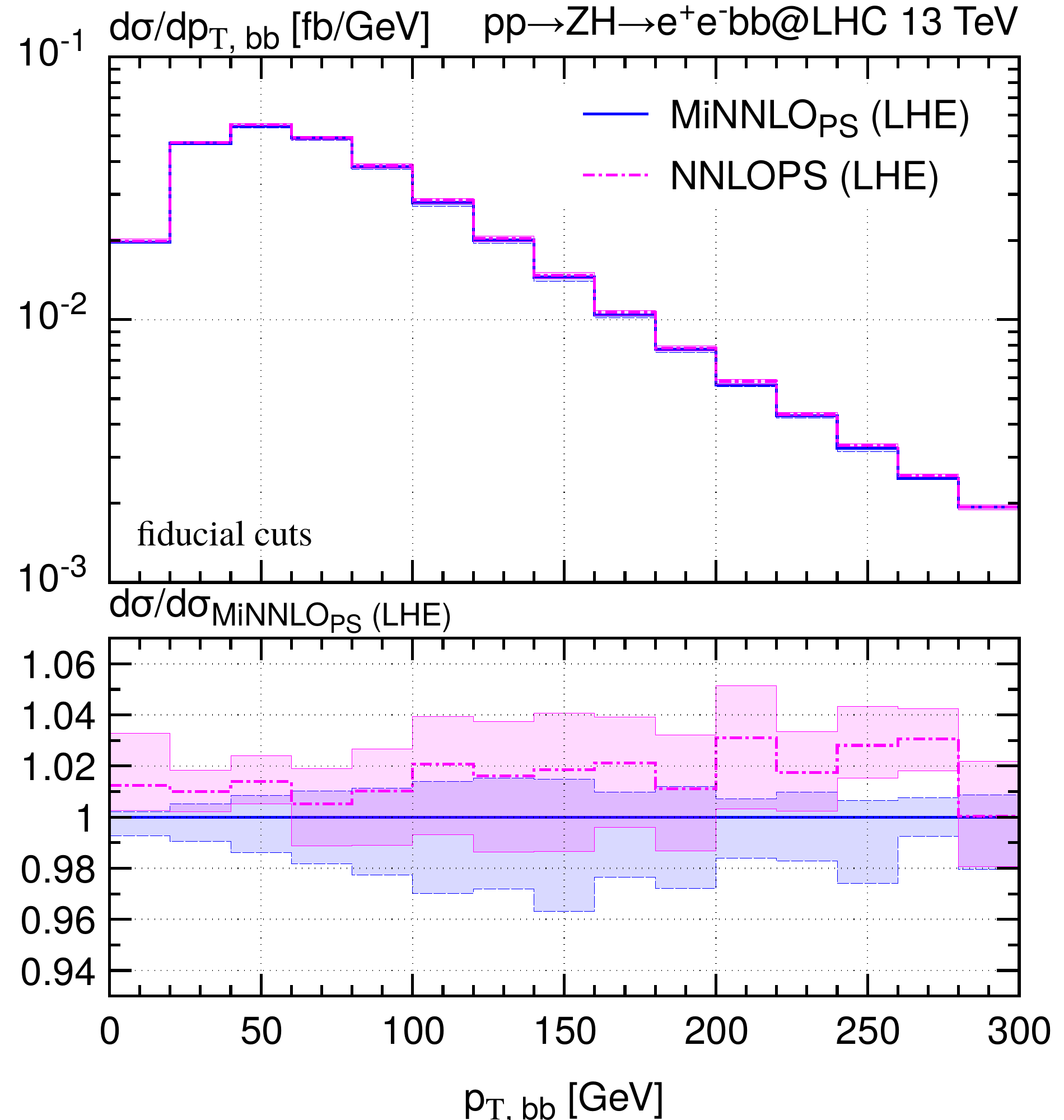} 
&
\includegraphics[width=.31\textheight]{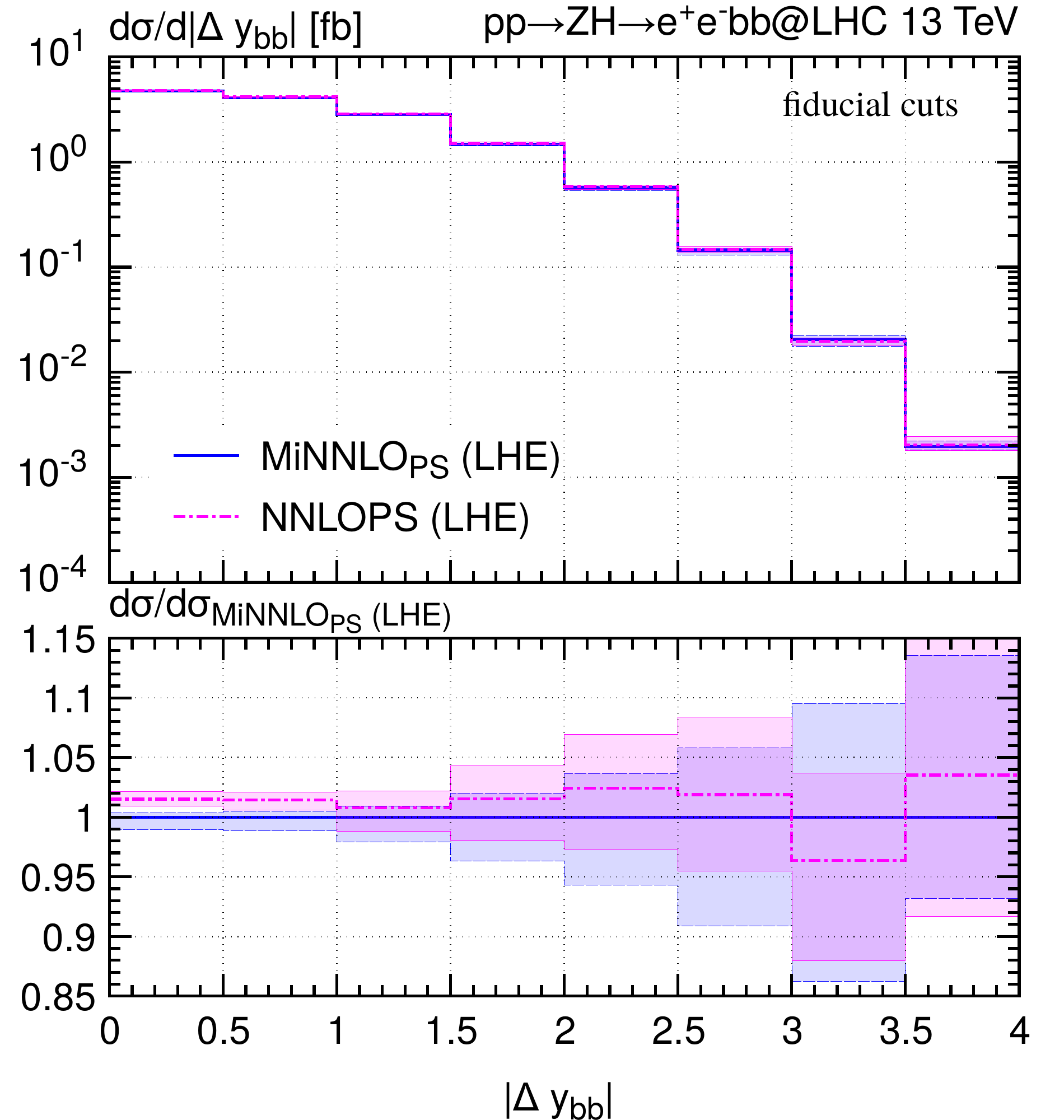}
\end{tabular}\vspace{-0.15cm}
\caption{\label{fig:validationzh} Comparison of \minnlo{} and NNLOPS predictions for $e^+e^- b\bar{b}$ production with the fiducial cuts quoted in \tab{tab:cutsvalidation}.}
\end{center}
\end{figure}

The fiducial cross sections are $6.261(7)^{+0.9\%}_{-1.8\%}$\,fb for
\minnlo{} and $6.348(6)^{+1.2\%}_{-1.4\%}$\,fb for NNLOPS. Because of
the more exclusive setup, the difference between the \minnlo{} and
NNLOPS predictions is slightly larger than for the inclusive cross
section and for $W^+H$ production in the previous section. Still, the
fiducial cross sections agree at the level of $1.4$\%, which is fully
covered by the scale uncertainties. We would like to stress that this
level of agreement in the fiducial phase-space volume of the
$e^+e^-b\bar{b}$ final state is quite remarkable, given the very small
scale dependence of this process.


As far as differential distributions are concerned, in
\fig{fig:validationzh} we consider the invariant mass of the \zh{}
system ($\mzh{}$) and observables related to the bottom quarks from
the Higgs decay. In particular, the invariant mass of the $b$-jet pair
($\mbb$), its transverse momentum ($\ptbb$), and the rapidity
difference between the two $b$-jets ($\absdybb$) are shown.  We note
that the two $b$ here refer to the two $b$-jets with invariant mass
closest to the Higgs mass, as discussed above.  Also here we find a
good agreement between \minnlo{} and NNLOPS predictions for all
observables within the scale-variation bands, with differences, by and
large, lower than $1$-$2$\%. Besides statistical fluctuations, these
differences are induced by the $1.4$\% shift in the fiducial cross
section noticed before, while the agreement in terms of shapes is even
better.

\section{Phenomenological results}
\label{sec:results}

In this section we present phenomenological results for \vh{}
production with $H\to b\bar b$ decay in proton--proton collisions at
the LHC with a center of mass energy of $\sqrt{s}=13$\,TeV. More
precisely, we consider both \minlo{} and \minnlo{} results for the
processes $pp \to e^\pm \nu_e b\bar{b}$ and $pp \to e^{+} e^{-}
b\bar{b}$, and compare our state-of-the-art \minnlo{} predictions to
recent ATLAS data~\cite{ATLAS:2020jwz}. Besides integrated inclusive
and fiducial cross sections we also discuss differential distributions
in the fiducial phase space, and we analyze the effects of using
different jet-clustering algorithms.

\subsection{Input parameters and settings}
\label{sec:setup}
We use the NNLO set of the NNPDF31 \cite{Ball:2017nwa} parton
densities with $\alpha_s (m_Z)$=0.118. Our input parameters are set to
the following PDG values \cite{ParticleDataGroup:2020ssz}: $G_F$ =
1.166379$\times \rm{10^{-5}\,GeV^{-2}}$, $m_W$ = 80.379\,GeV,
$\Gamma_{W}$ = 2.085\,GeV, $m_Z$ = 91.1876\,GeV, $\Gamma_Z$ =
2.4952\,GeV, $m_H$ = 125.09\,GeV and $\Gamma_H$ = 4.1 MeV. We compute
the EW coupling as $\alpha_{EM}=\sqrt{2}G_F
m_W^2(1-{m_W^2}/{m_Z^2})/\pi$ and the mixing angle as
$\cos^2{\theta_W}$ = ${m_W^2}/{m_Z^2}$. The $H\to b\bar{b}$ branching
ratio is set to ${\rm Br}_{H\to b\bar{b}}$ = 0.5824. Similarly to what
is done in the validation section, in $pp \to HZ \to e^{+} e^{-}
b\bar{b}$ we include contributions in which the Higgs boson is
radiated from a heavy quark loop, setting the pole mass of the bottom
quark to $m_b$ = 4.78\,GeV and the pole mass of the top quark to $m_t$
= 172.5\,GeV.  The bottom Yukawa coupling is set to $y_b(m_H)$ = 1.280
$\times \rm{10^{-2}}$, which, however, as pointed out in
\sct{sec:proddec}, cancels out in the ration of \eqn{eq:full} when
combining production and decay events.  We employ \PYTHIA{8}
\cite{Sjostrand:2014zea} with the Monash tune \cite{Skands:2014pea}
for all matched predictions.  We do not include any effect from
hadronization, underlying event modelling or a QED
shower, except for the comparison to data in \sct{sec:data}.

Besides a fully inclusive setup, which we refer to as \incl{}, we
consider two sets of fiducial cuts. The first one is inspired by the
CERN Yellow Report \cite{LHCHiggsCrossSectionWorkingGroup:2016ypw} and
is refereed to as \fidYR{}.
The second set of cuts is taken from \citere{ATLAS:2020jwz} and is
referred to as \fidATLAS{}. Both sets of fiducial cuts are summarized
in \tab{fidcuts}.

\renewcommand\arraystretch{1.4}
\begin{table}[h]
  \centering
  \begin{tabular} {c | c }
    \Xhline{1pt}
   \multicolumn{2}{c}{\fidYR{} \cite{LHCHiggsCrossSectionWorkingGroup:2016ypw}} \\
    \hline
    $pp\to W^{\pm}H \to e^{\pm} \nu_{e}b\bar{b}$ & $pp\to ZH \to e^{+}e^{-}b\bar{b}$ \\
    \hline
    $\pte > 15$\,GeV, $\ptmiss > 15$\,GeV, $|\etae| < 2.5$  & $\pte > 15$\,GeV, $|\etae| < 2.5$ \\
                                                 & $75\,{\rm GeV} < \mee < 105$\,GeV \\
    $\ge$2 $b$-jets (anti-$\it{k}_T$\,\cite{Cacciari:2008gp}, R=0.4) & $\ge$2 $b$-jets (anti-$\it{k}_T$\,\cite{Cacciari:2008gp}, R=0.4) \\
    $\ptb > 25$\,GeV, $|\etab| < 2.5$ & $\ptb > 25$\,GeV, $|\etab| < 2.5$ \\
    \Xhline{1pt}
    \multicolumn{2}{c}{\fidATLAS{} \cite{ATLAS:2020jwz}} \\
    \hline
    $pp\to W^{\pm}H \to \ell^{\pm} \nu_{\ell}b\bar{b}$ & $pp\to ZH \to \ell^{+}\ell^{-}b\bar{b}/\nu_\ell\bar\nu_\ell b\bar{b}$ \\
    \hline
    $|\yh| < 2.5$ & $|\yh| < 2.5$ \\
    \text{categories:} & \text{categories:} \\
    $\ptw \in$ [250,400]\,GeV,
    & $\ptz \in$ [250,400]\,GeV,\\
    $\ptw \in$ [400, $\infty$]\,GeV\,
    & $\ptz \in$ [400, $\infty$]\,GeV\,\\
    \Xhline{1pt}
  \end{tabular}
  \caption{\label{tab:cuts}Fiducial cuts used in the setup \fidYR{} and setup \fidATLAS{}.}
  \label{fidcuts}
\end{table}
\renewcommand\arraystretch{1}

\subsection{Inclusive and fiducial cross sections}
\label{sec:cs}

\renewcommand\arraystretch{1.4}
\begin{table}[h]
  \centering
  \begin{tabular} {l | c | c}
    \Xhline{1pt}
   \multicolumn{3}{c}{\boldmath{ $pp\to W^+H \to e^{+}\nu_{e}b\bar{b}$}} \\
    \hline
    $\sigma$ [fb] & \incl & \fidYR \\
    \hline
    $\rm{MiNLO'}$ & ${54.04}^{+6.6\%}_{-3.6\%}$ & ${20.13}^{+2.3\%}_{-3.1\%}$ \\
    $\rm{MiNNLO_{PS}}$ & ${57.44}^{+1.7\%}_{-0.8\%}$ & ${21.27}^{+1.3\%}_{-1.3\%}$ \\
    \Xhline{1pt}
    \multicolumn{3}{c}{\boldmath{$pp\to W^-H \to e^{-}\bar{\nu}_{e}b\bar{b}$}} \\
    \hline
    $\sigma$ [fb] & \incl & \fidYR \\
    \hline
    $\rm{MiNLO'}$ & ${33.82}^{+6.6\%}_{-3.6\%}$ & ${13.07}^{+2.4\%}_{-3.3\%}$ \\
    $\rm{MiNNLO_{PS}}$ & ${35.87}^{+1.5\%}_{-0.7\%}$ & ${13.77}^{+1.5\%}_{-1.6\%}$ \\
    \Xhline{1pt}
    \multicolumn{3}{c}{\boldmath{$pp\to ZH \to e^{+}e^{-}b\bar{b}$}} \\
    \hline
    $\sigma$ [fb] & \incl & \fidYR \\
    \hline
    $\rm{MiNLO'}$ & ${14.88}^{+6.7\%}_{-3.7\%}$ & ${5.21}^{+2.2\%}_{-3.0\%}$ \\
    $\rm{MiNNLO_{PS}}$ (no $gg \to ZH$) & ${15.79}^{+1.8\%}_{-0.9\%}$ & ${5.48}^{+1.2\%}_{-1.2\%}$ \\
    $\rm{MiNNLO_{PS}}$ (with $gg \to ZH$) & ${16.99}^{+3.6\%}_{-2.3\%}$ & ${6.07}^{+3.4\%}_{-2.9\%}$ \\    
    \Xhline{1pt}    
  \end{tabular}
  \caption{\label{tab:cs} Integrated cross sections for the $e^{+}\nu_{e}b\bar{b}$, $e^{-}\bar{\nu}_{e}b\bar{b}$ and $e^{+}e^{-}b\bar{b}$ final states in the \incl{} and the \fidYR{} setup.} 
  \label{xsections}
\end{table}
\renewcommand\arraystretch{1}

We start by discussing integrated cross section both in the fully
inclusive case and including fiducial cuts.  In \tab{tab:cs} we report
\minlo{} and \minnlo{} prediction for both the neutral and the
charged-current \vh{} production processes with $H\to b\bar{b}$
decay. For \zh{} production we report separately the numbers with and
without the gluon-fusion contribution that is induced by a closed
heavy-quark loop (top and bottom).
The inclusive cross sections are in good agreement with the NNLO predictions from \vhatnnlo{}~\citeres{Brein:2003wg,Brein:2011vx,Brein:2012ne,Harlander:2018yio} that are obtained in the 
on-shell approximation with $\muR=\muF=m_{VH}$. Assuming branching ratios consistent with our calculation 
(${\rm Br}_{H\to b\bar{b}}$ = 0.5824, ${\rm Br}_{W^\pm\to e^\pm\nu_e}$ = 0.10898 and ${\rm Br}_{Z\to e^+e^-}$ = 0.033628), 
they read
$\sigma_{W^+H\to e^+\nu_e b\bar{b}} = 58.3^{+0.2\%}_{-0.3\%}$ fb, $\sigma_{W^-H\to e^-\nu_e b\bar{b}} = 36.3^{+0.2\%}_{-0.3\%}$ fb, $\sigma_{ZH\to e^+e^- b\bar{b}}^{{\rm no}\ gg \to ZH} = 16.0^{+0.2\%}_{-0.3\%}$ fb, and $\sigma_{ZH\to e^+e^- b\bar{b}}^{{\rm with}\ gg \to ZH} = 17.0^{+1.5\%}_{-1.2\%}$ fb.\footnote{As far as
    the terms mediated by a closed top-quark loop
    are concerned,
    the cross sections computed with \vhatnnlo{} have been obtained switching off both the
    $R_{\rm I}$ and $V_{\rm I}$ terms for the \wh{} case, whereas for the \zh{} case we kept
    the $R_{\rm I/II}$ contributions, just like in our calculation.} 
We note that the inclusive on-shell calculation differs from the \minnlo{} also with respect to
the scale settings and the different treatment of terms beyond accuracy. 

Furthermore, we find in  \tab{tab:cs} that \minnlo{} induces a correction of about
5-6\% with respect to \minlo{}, both in the fully inclusive and in the
fiducial phase space volume, disregarding the loop-induced $gg$
contribution in the case of \zh{} production.  While for the inclusive
cross sections \minlo{} and \minnlo{} predictions are compatible
within scale variations, the \minnlo{} corrections in the fiducial
phase space are generally not covered by plain scale variations when
correlating the scales in production and decay.  If we vary the scales
in production and decay in an uncorrelated manner, the unnaturally
small \minlo{} uncertainty band in the fiducial setup increases and
the predictions become compatible within scale variations. The scale
uncertainties of \minnlo{} are significantly smaller than the ones of
\minlo{} in either case, with a reduction by more than a factor of
two.  Considering the loop-induced $gg\to ZH$ contribution, its effect
is quite significant. It contributes about almost 8\% in the inclusive
and almost 11\% in the fiducial phase space to the \minnlo{} cross
sections. As pointed out before, the contribution from loop-induced
$gg$ diagrams dominates the current theoretical uncertainties. Thus,
including NLO QCD corrections to the $gg\to ZH$ process in the
matching to parton showers is a crucial next step.\footnote{Recently,
  substantial progress was made in the calculation of the relevant
  two-loop amplitude to (on-shell) $gg\to ZH$ production
  \cite{Davies:2020drs,Chen:2020gae,Alasfar:2021ppe} and a first
  calculation at NLO QCD in a small mass expansion was achieved
  \cite{Wang:2021rxu}}

\subsection{Distributions in the fiducial phase space}
\label{sec:dist}

\begin{figure}[p!]
\begin{center}\vspace{-0.2cm}
\begin{tabular}{cc}
\includegraphics[width=.31\textheight]{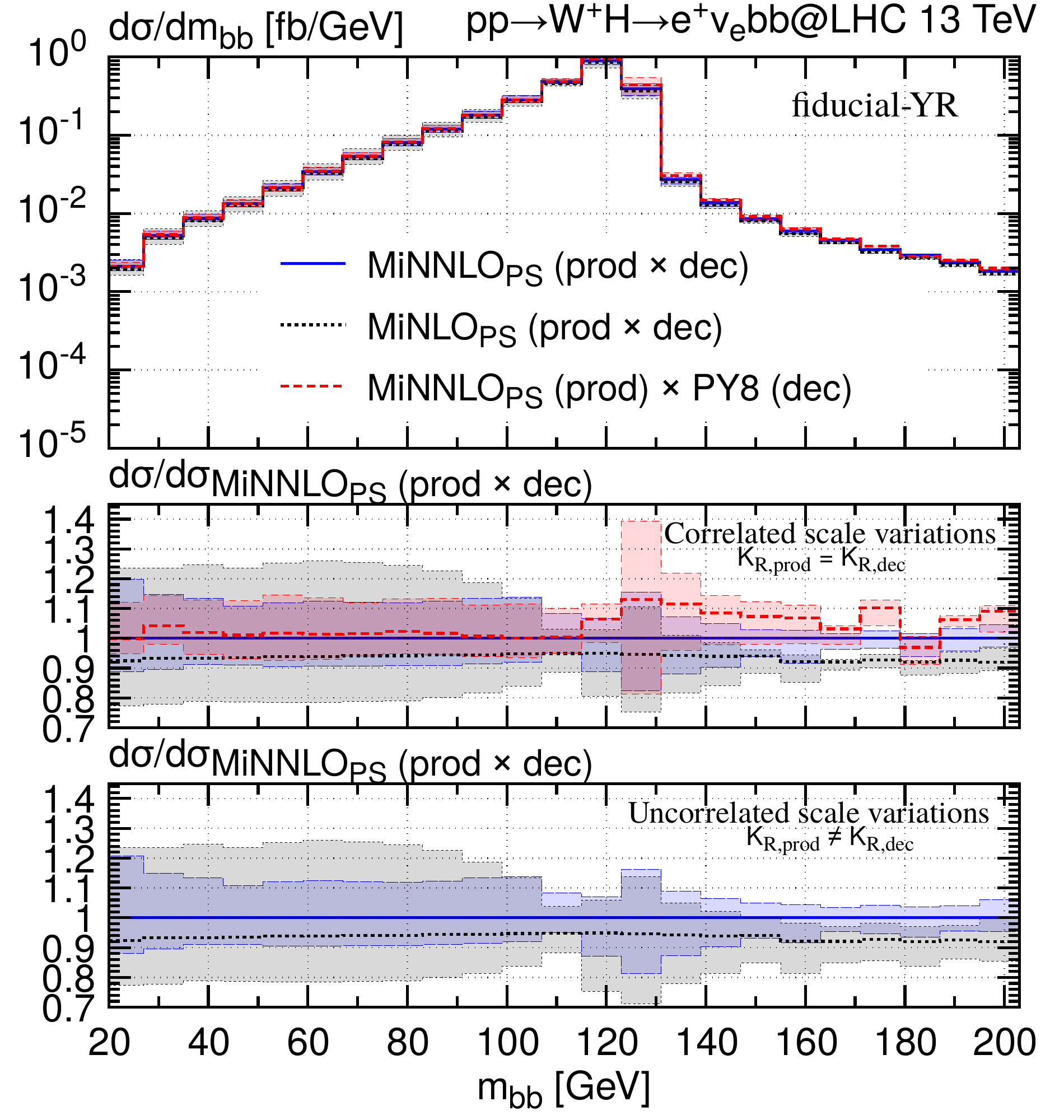} 
&
\includegraphics[width=.31\textheight]{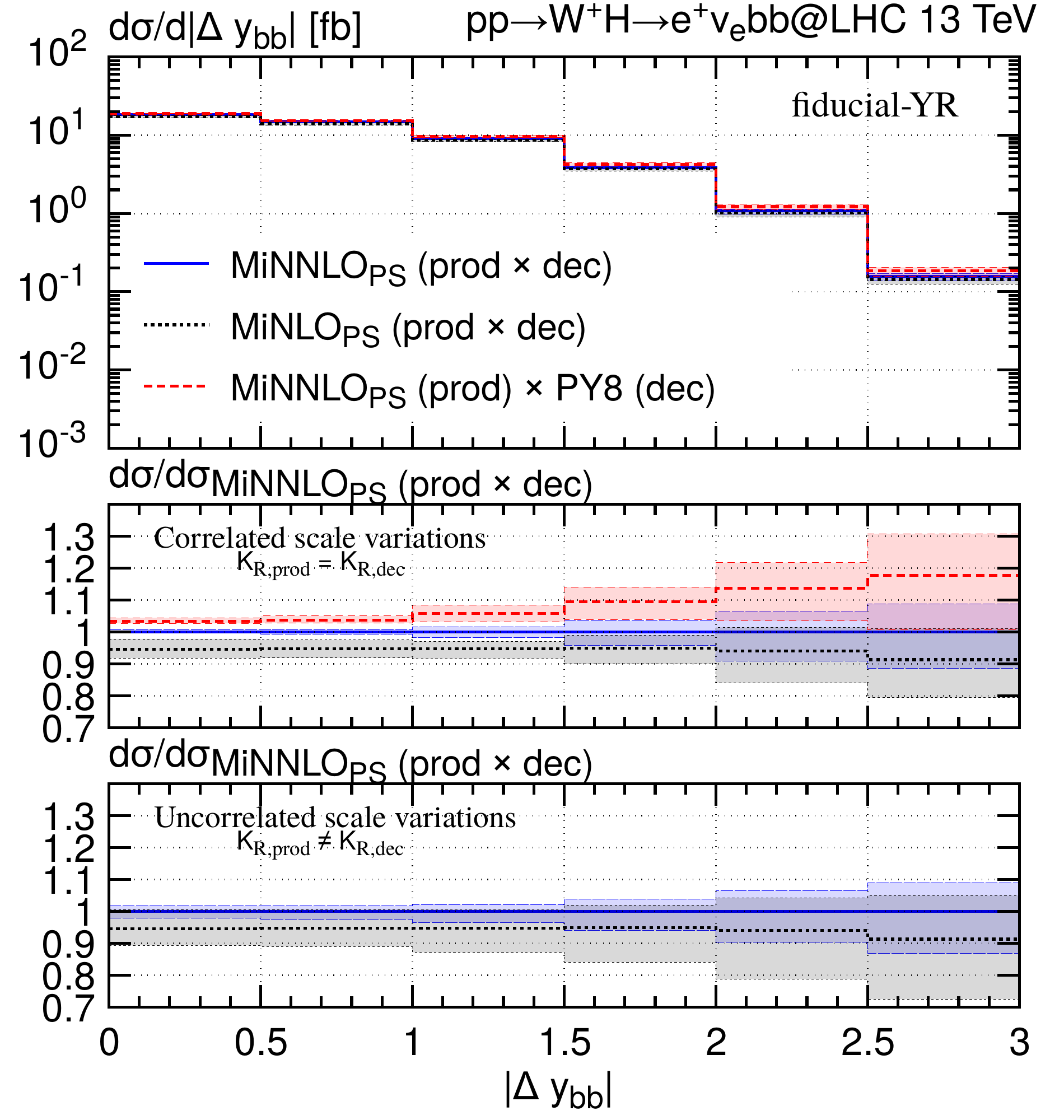}
\end{tabular}\vspace{-0.15cm}
\begin{tabular}{cc}
\includegraphics[width=.31\textheight]{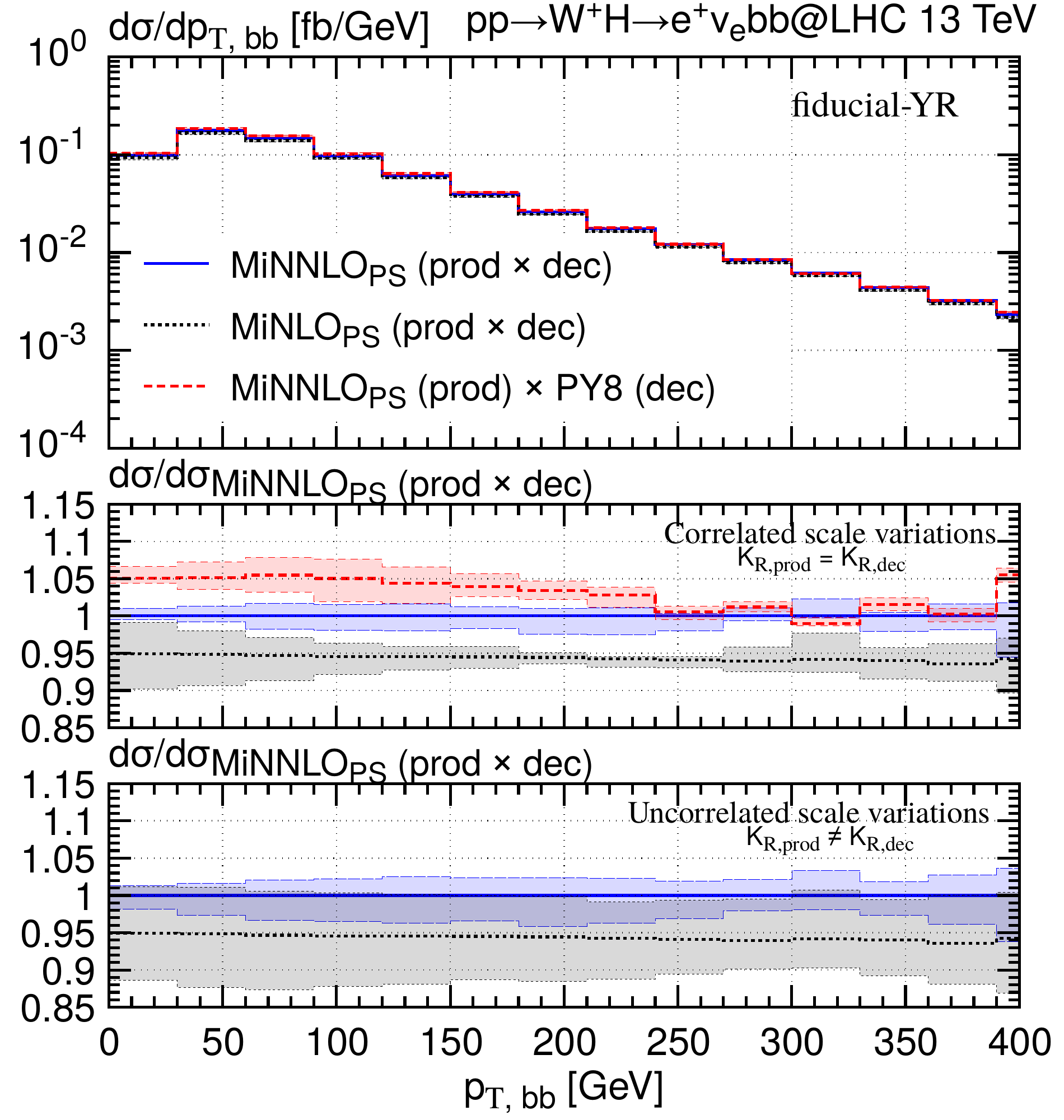}
&
\includegraphics[width=.31\textheight]{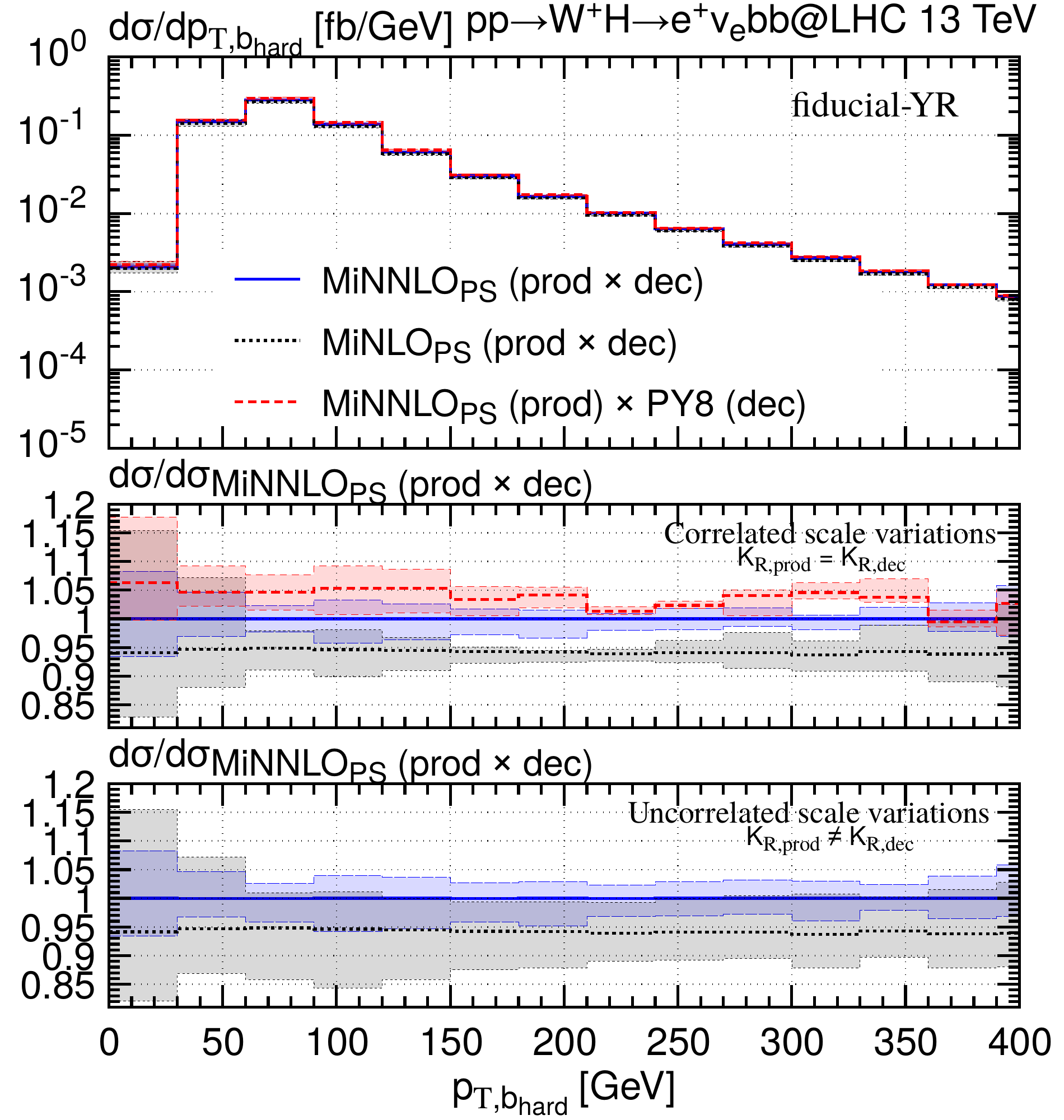}
\end{tabular}\vspace{-0.15cm}
\begin{tabular}{cc}
\includegraphics[width=.31\textheight]{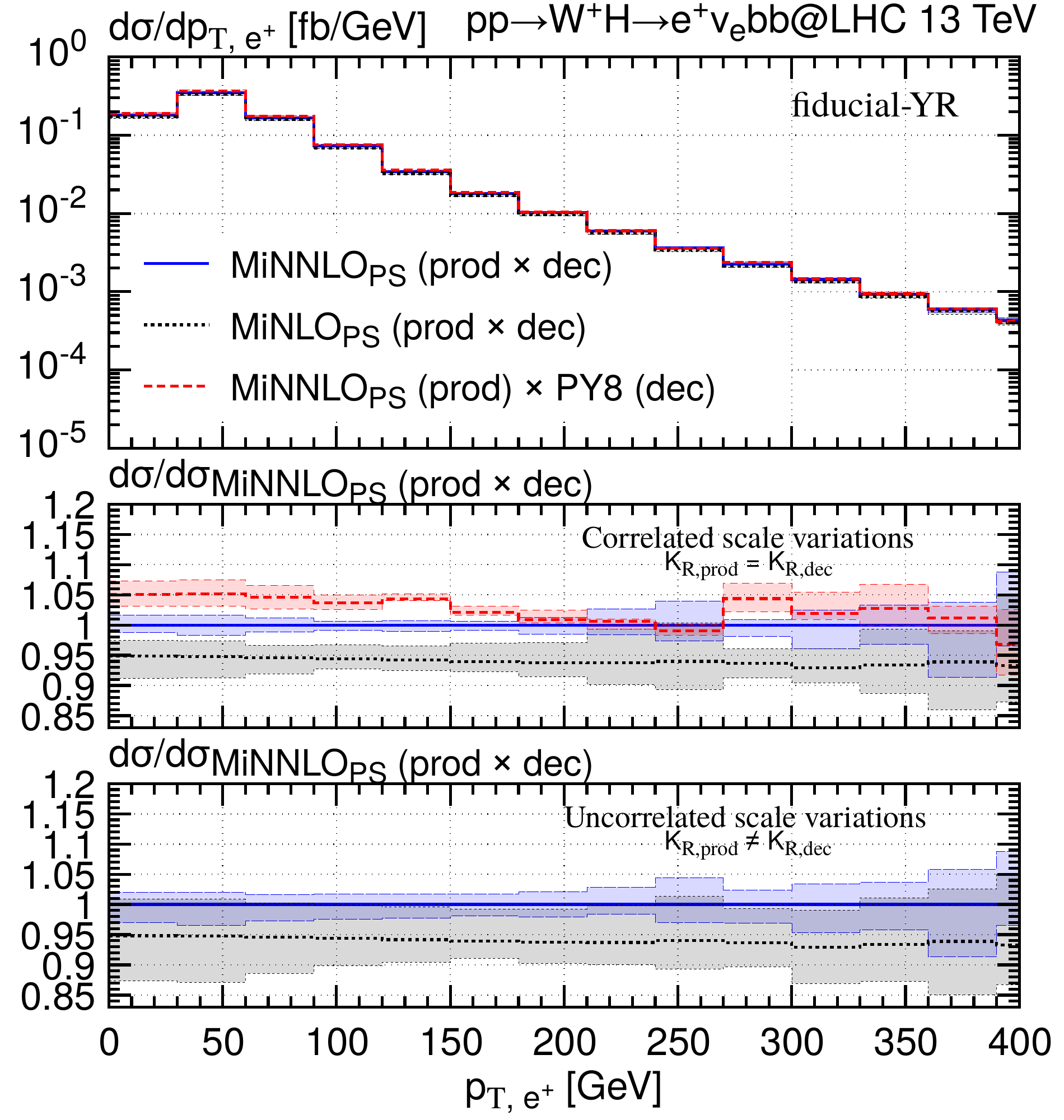}
&
\includegraphics[width=.31\textheight]{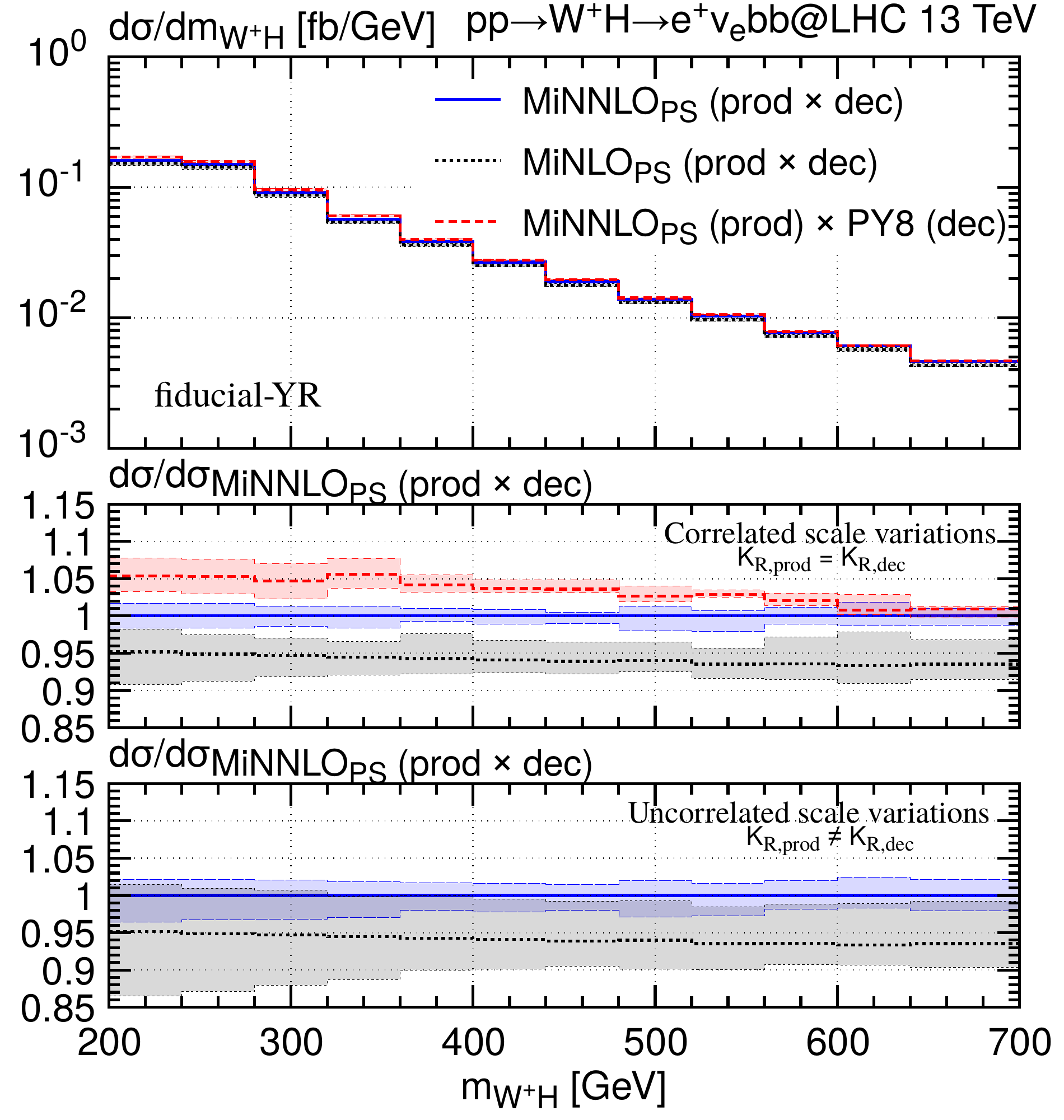} 
\end{tabular}
\caption{\label{fig:resultswh} Differential distributions for $e^{+}\nu_{e}b\bar{b}$ production with \fidYR{} cuts. $K_{\rm R,prod}$ and $K_{\rm R,dec}$ refer to the variation factors of the renormalization scales of production and decay, respectively. }
\end{center}
\end{figure}
\afterpage{\clearpage}

\begin{figure}[p!]
\begin{center}\vspace{-0.2cm}
\begin{tabular}{cc}
\includegraphics[width=.31\textheight]{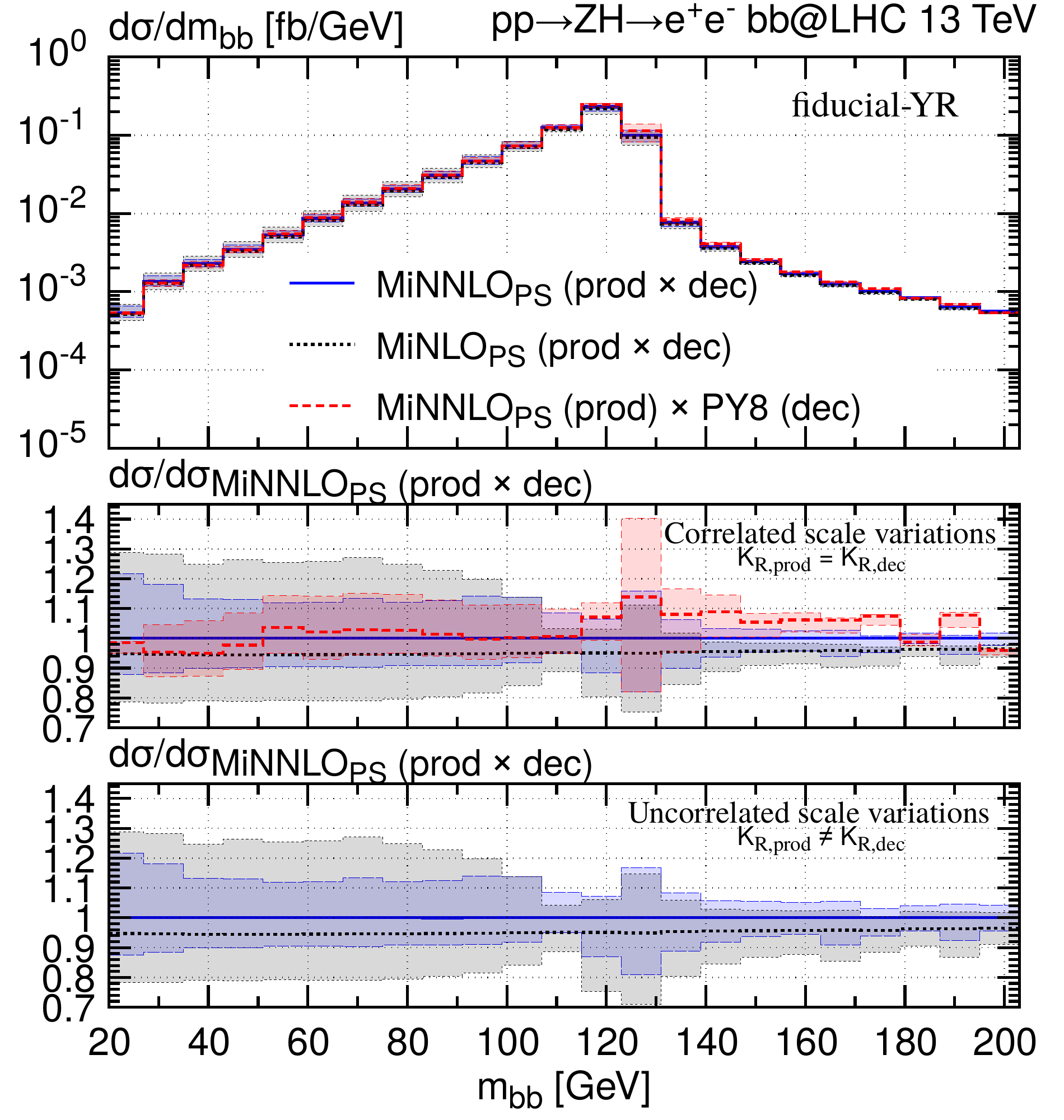} 
&
\includegraphics[width=.31\textheight]{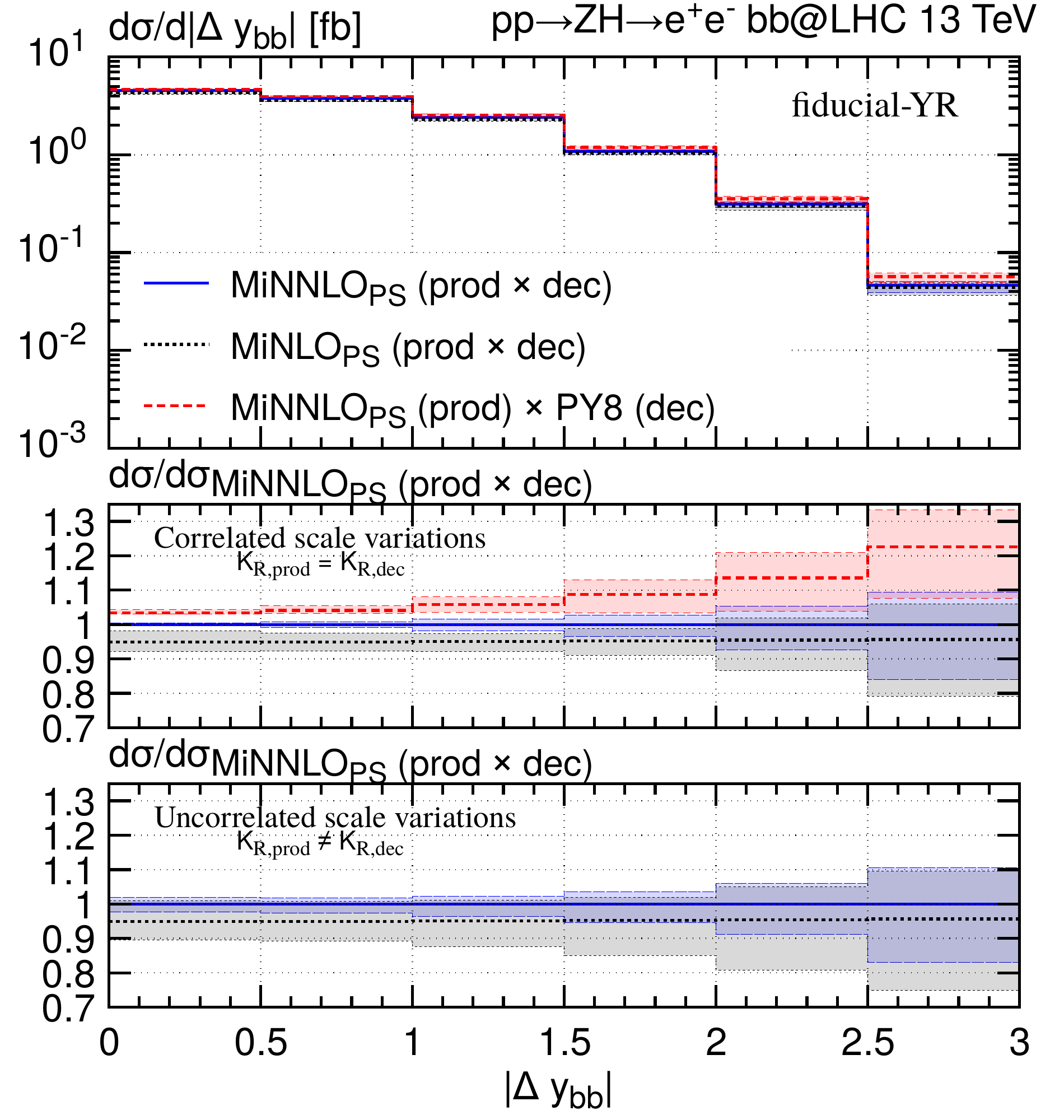}
\end{tabular}\vspace{-0.15cm}
\begin{tabular}{cc}
\includegraphics[width=.31\textheight]{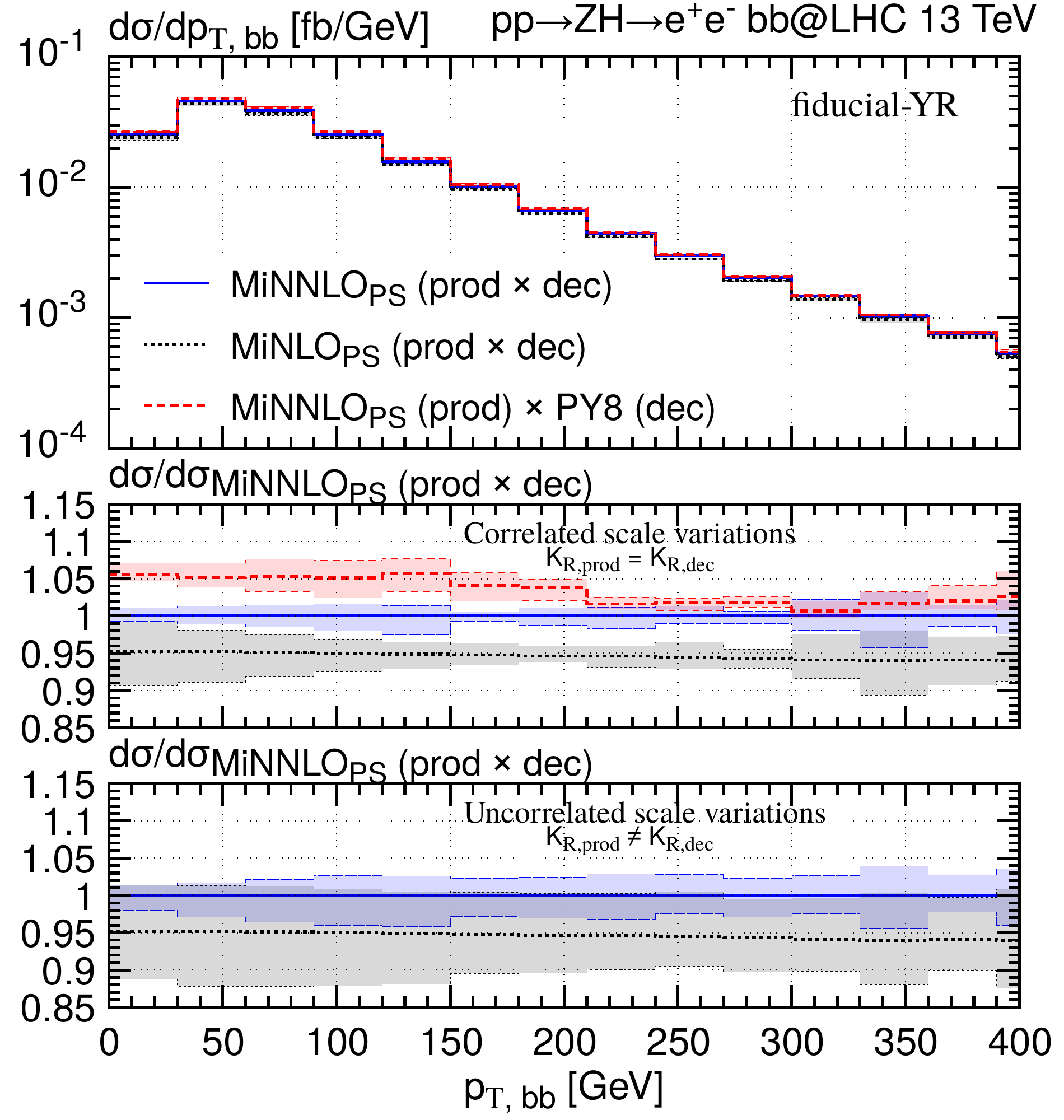}
&
\includegraphics[width=.31\textheight]{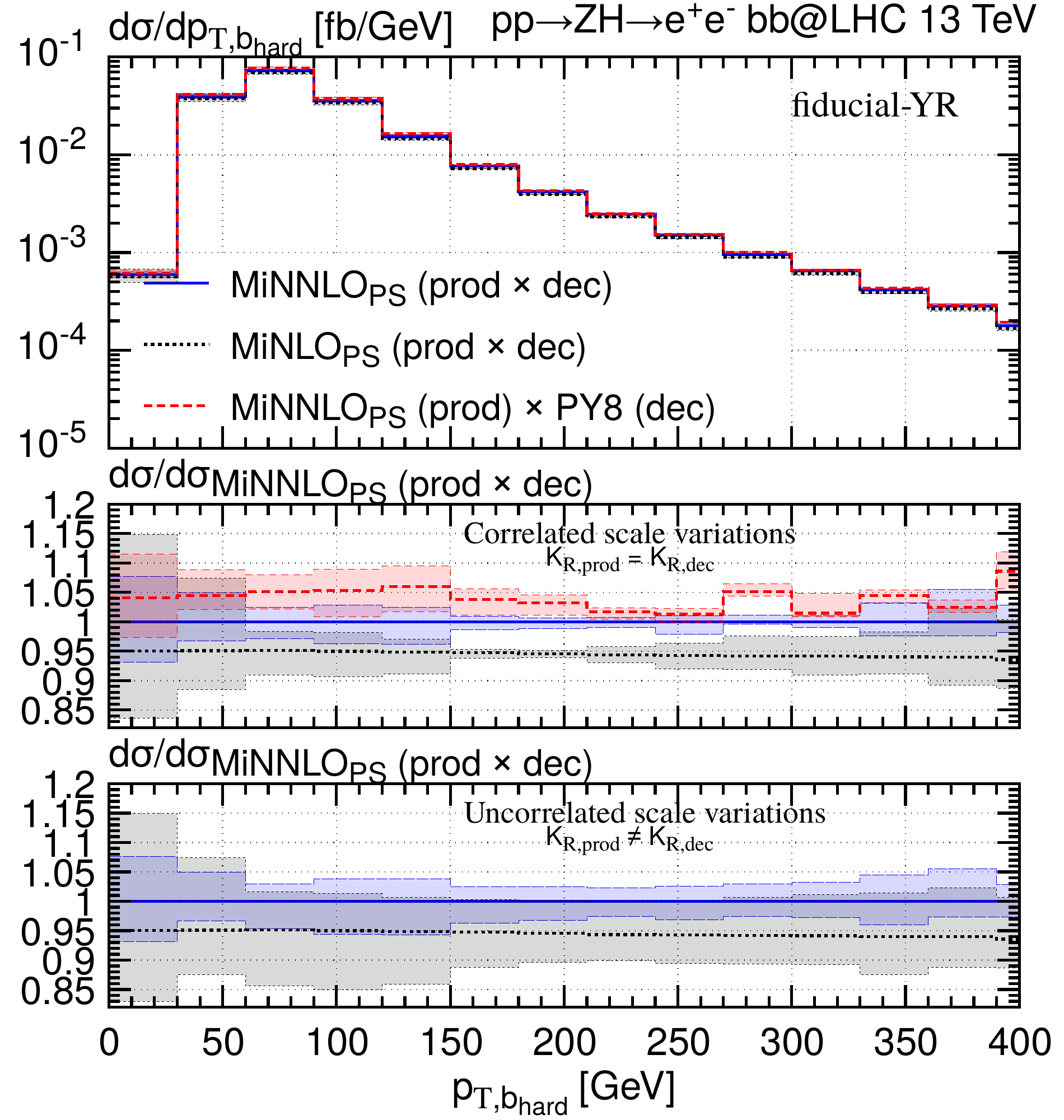}
\end{tabular}\vspace{-0.15cm}
\begin{tabular}{cc}
\includegraphics[width=.31\textheight]{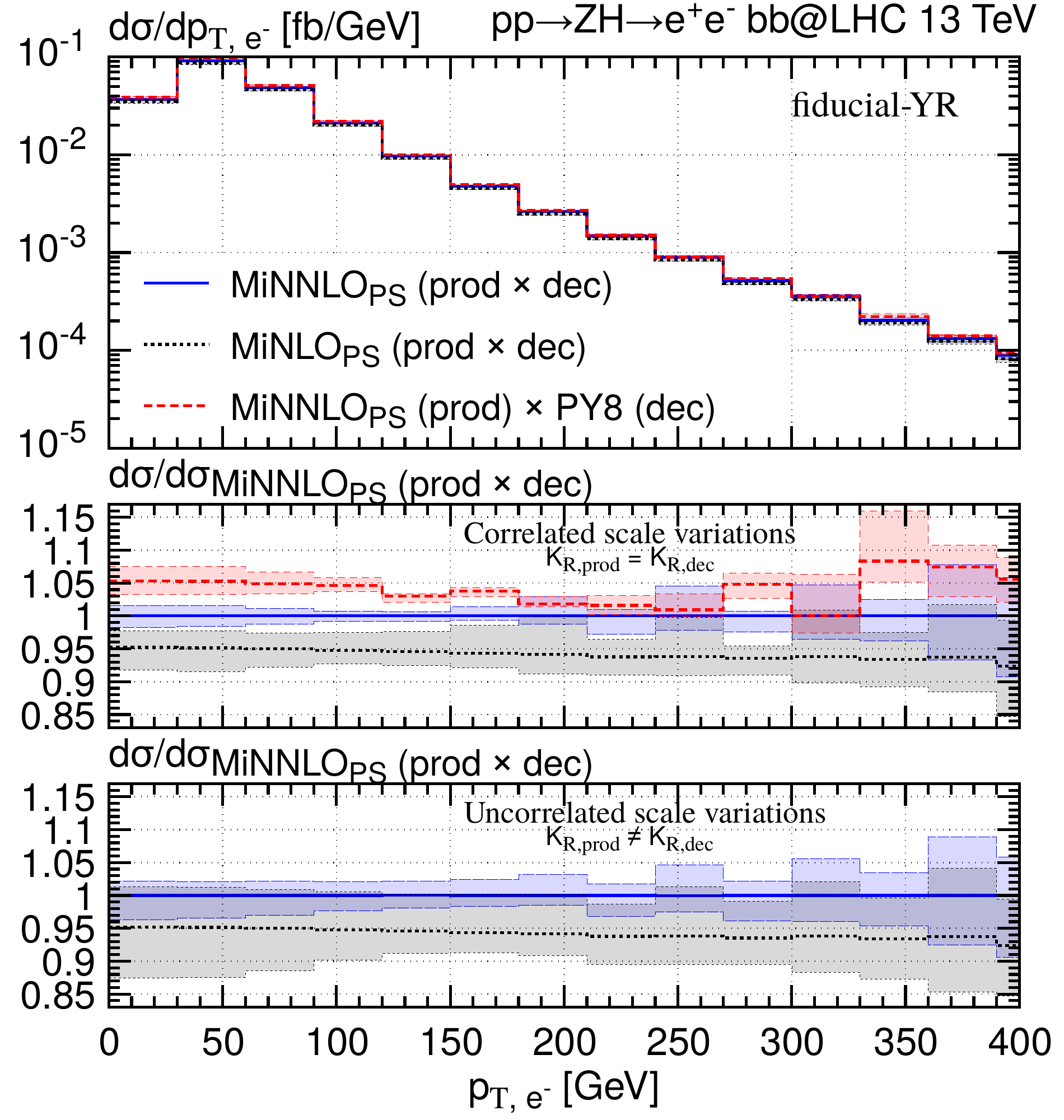}
&
\includegraphics[width=.31\textheight]{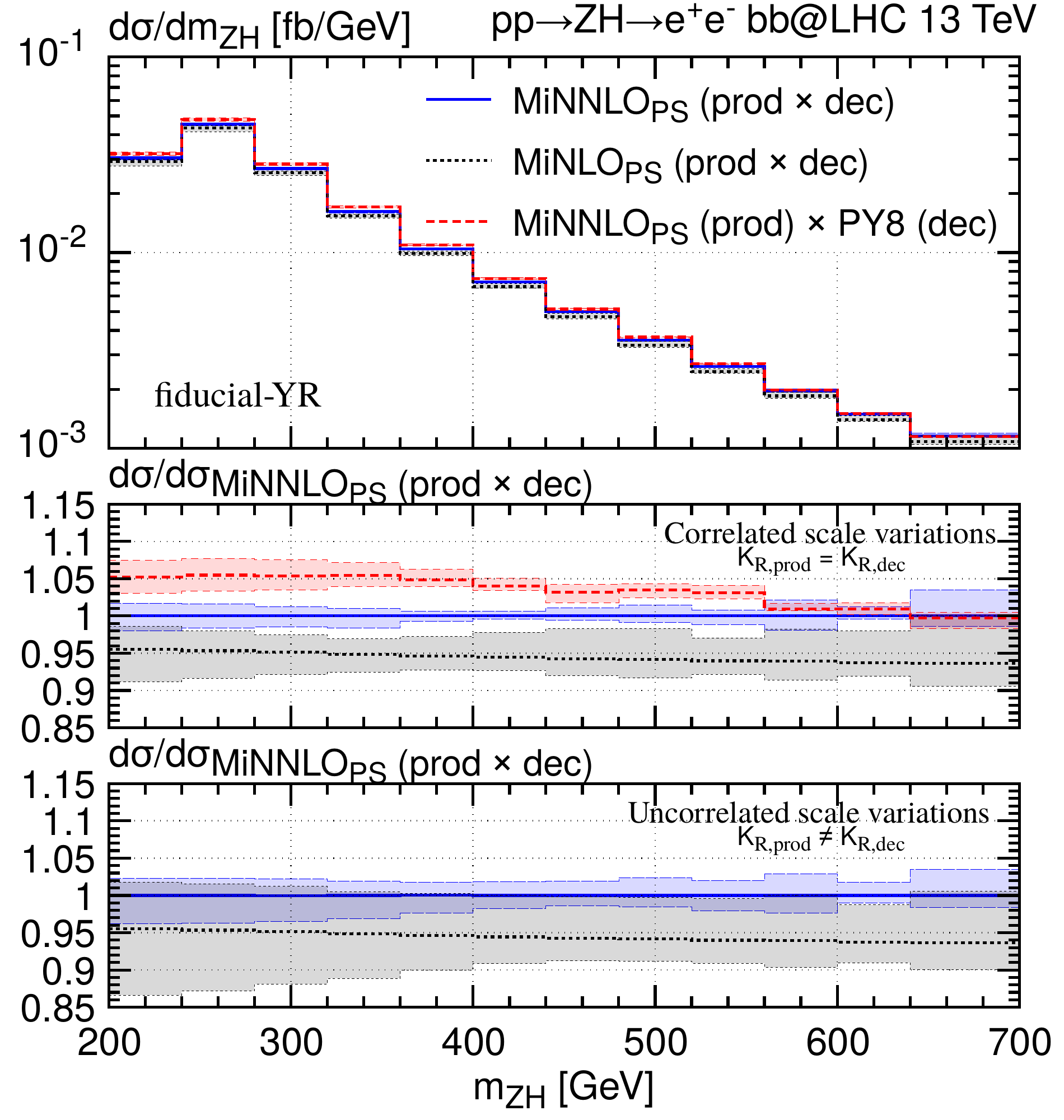} 
\end{tabular}
\caption{\label{fig:resultszh} Differential distributions for $e^{+}e^- b\bar{b}$ production with \fidYR{} cuts.$K_{\rm R,prod}$ and $K_{\rm R,dec}$ refer to the variation factors of the renormalization scales of production and decay, respectively. }
\end{center}
\end{figure}
\afterpage{\clearpage}

We continue our presentation of phenomenological results by
considering differential distributions in the \fidYR{} phase space
defined in \tab{tab:cuts}.  Figures\,\ref{fig:resultswh} and
\ref{fig:resultszh} show various observables for $e^+\nu_e
b\bar{b}\,(W^+H)$ and $e^+e^- b\bar{b}\,(ZH)$ production,
respectively.\footnote{The corresponding results for $e^-\nu_e
  b\bar{b}\,(W^-H)$ production are shown in \app{app:Wmresults}.}

In particular, the invariant mass ($\mbb$) and transverse momentum
($\ptbb$) of the bottom-quark pair, the absolute rapidity difference
between the two bottom quarks ($\absdybb$), the transverse momentum of
the hardest bottom quark ($\ptbone$), the transverse momentum of the
positron/electron ($\ptep$/$\ptem$), and the invariant mass of the
colour-singlet system in the final state ($\mwh$/$\mzh$) are shown. We
recall that with the two bottom quarks (from the Higgs decay) we
actually denote the two jets with at least one bottom quark whose
invariant mass is closest to the Higgs-boson mass.  The plots are
organized as follows: The main frame includes the full \minnlo{}
prediction (blue, solid), the full \minlo{} prediction (black,
dotted), and the \minnlo{} prediction with LO decay through \PYTHIA{8}
(red, dashed) as absolute cross sections of the differential
distributions. The lower two insets show the ratio to the \minnlo{}
central prediction, where in the first inset scale uncertainties are
evaluated by assuming that the $7$-point scale dependence in
production and decay is fully correlated, while in the second inset
uncorrelated scale variations in production and decay are performed,
but excluding the extreme variations (one scale up, the other down).
When the Higgs decay is simulated by \PYTHIA{8}, the scale variation
is also handled by \PYTHIA{8} using its so-called ``automated
parton-shower variation'' facility~\cite{Mrenna:2016sih}. In
particular, we vary the renormalisation scale for final-state
radiation by a factor 2 around the central value, correlating this
variation with the one performed in the production stage.  The
relative behaviour observed in both the \wh{} and the \zh{} results is
very similar. Thus, we can discuss the sets of plots simultaneously.

Looking at the $m_{b\bar{b}}$ distribution, we notice that \minlo{}
and \minnlo{} predictions are very similar as far as the shape is
concerned, while the \minnlo{} corrections increase the cross section
by roughly 5\%, as already observed for the integrated fiducial cross
section in the previous section. \minlo{} and \minnlo{} predictions
are fully compatible within scale variations, regardless of whether
scale variations are assumed to be correlated or
uncorrelated. Interestingly, for this observable the two cases lead to
very similar uncertainties, which can be seen by comparing the first
and the second ratio inset.  Moreover, we observe clearly reduced
scale uncertainties for \minnlo{} compared to \minlo{} below the
Higgs-mass threshold by roughly a factor of two, but similarly large
uncertainties above the threshold.  Using \PYTHIA{8} to include the
$H\to b\bar b$ decay at LO compares rather well to using the full
\minnlo{} prediction for production and decay.

For the rapidity difference of the two bottom quarks \minlo{} and
\minnlo{} predictions are not compatible at central rapidities when
assuming correlated uncertainties. In particular, the \minnlo{} result
features a very small band at the $\pm 1$\% level. By uncorrelating
the scale variation in production and decay more realistic scale bands
are obtained, leading to fully compatible \minlo{} and \minnlo{}
predictions, where the uncertainties of the latter increase only to a
very few percent. This behaviour can be observed also for the
remaining distributions, where it is visible almost over the entire
plotted ranges. This justifies a more conservative uncertainty
estimate by taking uncorrelated scale variations in production and
decay.  We also observe that the LO $H\to b\bar b$ decay through
\PYTHIA{8} is insufficient to describe this observable accurately,
especially in the forward $\dybb$ region. Moreover, in the central
rapidity region, the uncertainty bands of the \PYTHIA{8}-decayed
results are not compatible with the \minnlo{} ones.

Considering the other differential observables it is clear that
similar features appear. For all of them \minnlo{} induces a mostly
flat $\sim 5\%$ correction on top of \minlo{} and leads to a
significant reduction of the scale uncertainties. Note that for
correlated scale uncertainties some accidental cancellations in the
scale variations occur, which are particularly visible from the
\minlo{} scale-uncertainty band around $200$\,GeV in both the
$p_{T,bb}$ and $\ptbone$ distributions.  The red curve that includes
the $H\to b\bar b$ decay through \PYTHIA{8} at LO is in shape rather
similar to the full \minnlo{} result, but it is roughly 5\% above the
full \minnlo{} prediction and thereby outside the quoted scale
uncertainties in various phase-space regions.

\subsection{Impact of jet-clustering algorithms}
\label{sec:cluster}

\begin{figure}[t]
\begin{center}\vspace{-0.2cm}
\begin{tabular}{cc}
\includegraphics[width=.31\textheight]{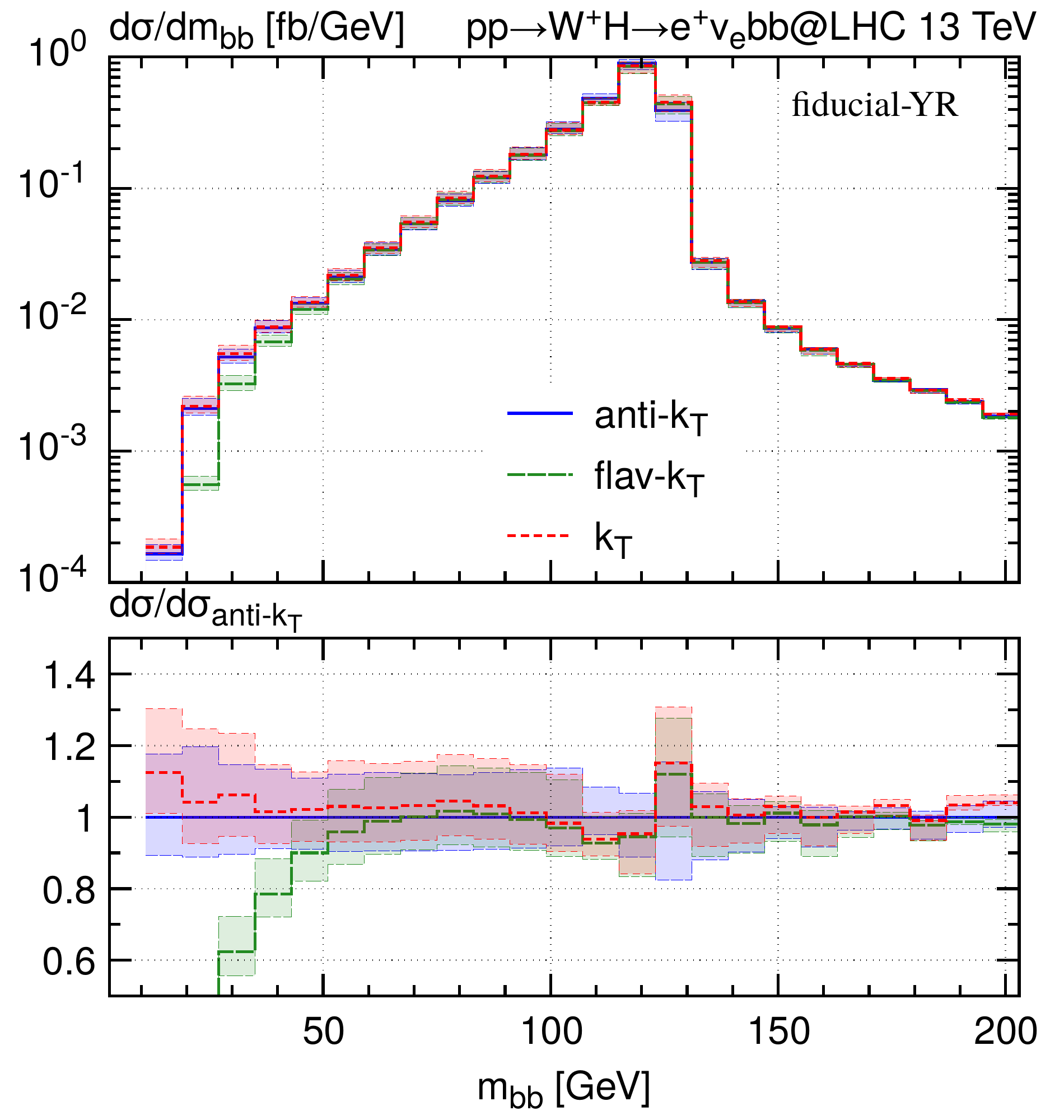} 
&
\includegraphics[width=.31\textheight]{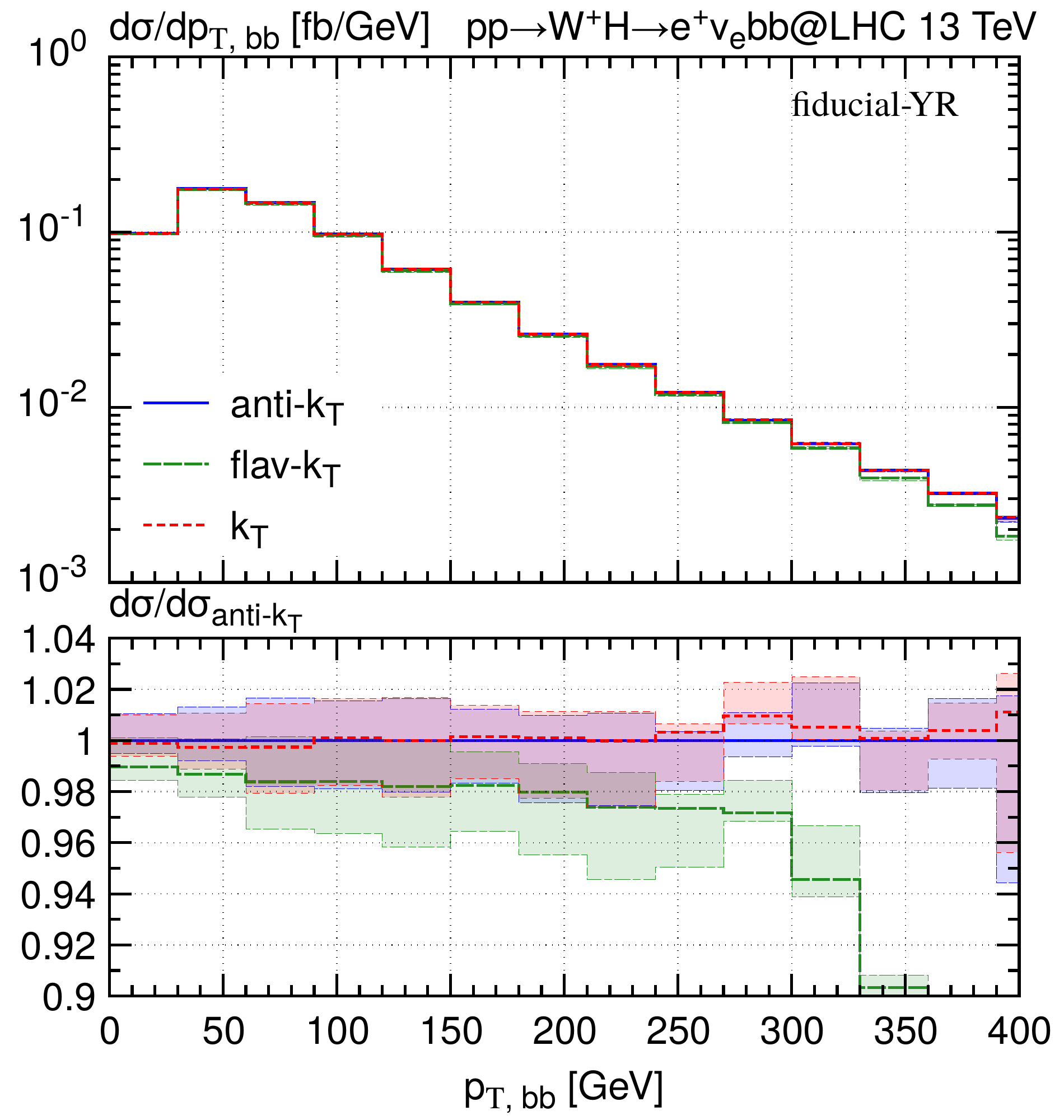} 
\end{tabular}\vspace{-0.15cm}
\begin{tabular}{cc}
\includegraphics[width=.31\textheight]{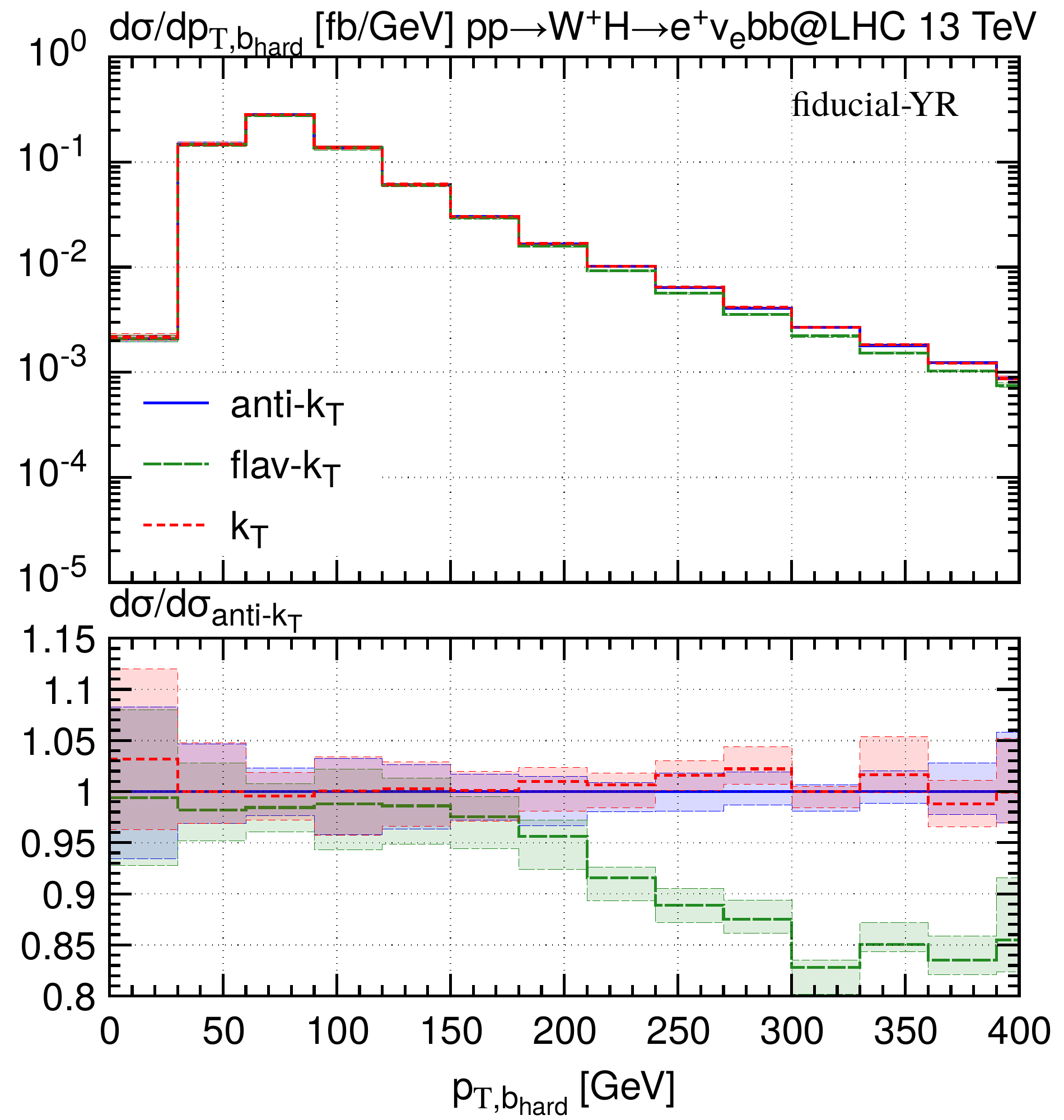} 
&
\includegraphics[width=.31\textheight]{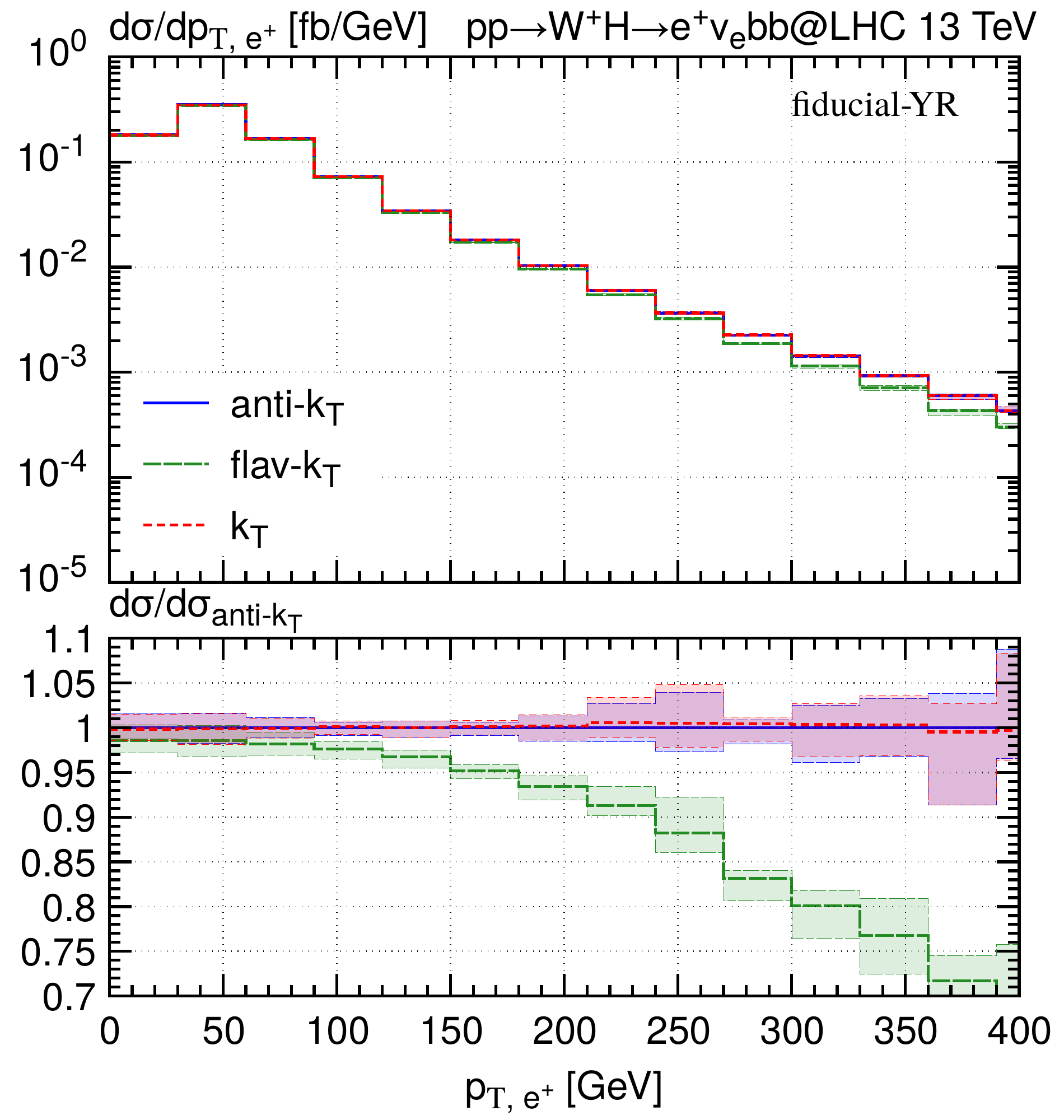} 
\end{tabular}
\caption{\label{fig:clustering} Differential distributions for
  $e^{+}\nu_{e}b\bar{b}$ production with \fidYR{} cuts using different
  jet-clustering algorithms, all with $R=0.4$.}
\end{center}
\end{figure}

In this section we address the impact of different jet-clustering
algorithms in the definition of $b$-jets on differential
distributions.  Our calculations are based on matrix elements with
massless bottom quarks in the final state. Therefore, performing an
event simulation is particularly important as it allows us to include
a finite bottom-quark mass in the generation of Les Houches events or
through \PYTHIA{} when showering the events. This makes it possible to
perform any standard jet clustering to define $b$-jets, which in a
massless calculation would otherwise lead to infrared (IR) unsafe
results.  For massless $b$-jets, for instance in a fixed-order
calculation, it is instead possible to use the flavour-$k_T$ algorithm
of \citere{Banfi:2006hf} for an IR-safe definition of $b$-jets.

Figure\,\ref{fig:clustering} shows the $\mbb$, $\ptbb$, $\ptbone$, and
$\ptep$ distributions for \wph{} production in the \fidYR{}
phase-space defined in \tab{tab:cuts}.
The different curves correspond to using the anti-$k_T$
\cite{Cacciari:2008gp} (blue, solid), the flavour-$k_T$
\cite{Banfi:2006hf} (green, long-dashed), and the $k_T$
\cite{Catani:1993hr} (red, short-dashed) jet-clustering algorithm, all
with $R=0.4$. In the bottom panels we show the ratio to the anti-$k_T$
result.
It is easy to see that the anti-$k_T$ and
$k_T$ algorithms lead to very compatible results with small
differences only in certain phase-space regimes, in particular for the
$\mbb$ distribution. There is a 5-10\% difference between anti-$k_T$
and $k_T$ clustering both at small $\mbb$ and at the $m_H$ threshold.
Such differences are not unexpected and they originate from soft
gluons being more likely to be clustered in the $b$-jets with the
$k_T$ algorithm.

As far as the flavour-$k_T$ algorithm is concerned, the differences to
the other two approaches are much more prominent in some regions of
phase-space. The flavour-$k_T$ algorithm assumes $b$-jets to originate
only from jets that contain an uneven number of bottom quarks, since
an even number of bottom quarks in one jet typically originates from
$g\to b\bar{b}$ splittings, which in the flavour-$k_T$ algorithm shall
not be considered as $b$-jets.  The reason for adopting such treatment
stems mostly from fixed-order calculations in the massless
approximation of bottom quarks, i.e.\ in the five-flavour scheme
(5FS), where the standard clustering algorithms lead to definitions of
bottom-quark jets that are not IR safe, as discussed before.  Clear
differences of the flavour-$k_T$ results to the non-flavour specific
clustering algorithms are observed at small $\mbb$ and at large values
of $\ptbb$, $\ptbone$, and $\ptep$. Additionally, around the $\mbb\sim
m_H$ threshold the flavour-$k_T$ curve follows the one of the standard
$k_T$ algorithm and shows a similar difference to the anti-$k_T$ one,
as both the flavour and the standard $k_T$ algorithm use the same
measure in the ordering of the jet clustering.  In order to explain
these differences we focuse on a comparison between $k_T$ and
flavour-$k_T$ algorithms since the underlying distance measure is more
closely related.

We remind the reader how the flavour-$k_T$ algorithm works. The
clustering procedure is the same as any sequential algorithm but the
distance measure is modified as follows:
\begin{equation}
  \label{eq:dij-flavour}
  d_{ij}^{(F)} = \dfrac{(\Delta \eta_{ij}^2 + \Delta \phi_{ij}^2)}{R^2} \times\left\{
    \begin{array}[c]{ll}
      \max(\ptargsqr{i}, \ptargsqr{j})\,, & \quad\mbox{softer of $i,j$ is flavoured,}\\      
      \min(\ptargsqr{i}, \ptargsqr{j})\,, & \quad\mbox{softer of $i,j$ is flavourless.}
    \end{array}
  \right.
\end{equation}
where \textit{softer} means with lower $\pt$.  The distance with the
beam is defined separating explicitly the beam moving towards positive
rapidities $B$ (right-moving beam) and the opposite beam towards
negative rapidities $\bar B$ (left-moving beam),
\begin{align}
  \label{eq:diB-flavour}
  d_{iB}^{(F)} = \left\{
    \begin{array}[c]{ll}
      \max(\ptargsqr{i}, \ptargsqr{B})\,, & \quad\mbox{$i$ is flavoured,}\\
      \min(\ptargsqr{i}, \ptargsqr{B})\,, & \quad\mbox{$i$ is flavourless.}
    \end{array}
  \right. \\
  d_{i\bar B}^{(F)} = \left\{
    \begin{array}[c]{ll}
      \max(\ptargsqr{i}, \ptargsqr{\bar B})\,, & \quad\mbox{$i$ is flavoured,}\\
      \min(\ptargsqr{i}, \ptargsqr{\bar B})\,, & \quad\mbox{$i$ is flavourless.}
    \end{array}
  \right.
\end{align}
where $\ptarg{B}$ ($\ptarg{\bar B}$) represents the hard transverse
scale associated with the right-moving (left-moving) beam, defined as
follows:
\begin{equation}
  \label{eq:ktB}
  \ptarg{B} (\eta) =
  \sum_i \ptarg{i} \left( \Theta(\eta_i - \eta) +
    \Theta(\eta - \eta_i) e^{\eta_i - \eta}\right)\,,
\end{equation}
\begin{equation}
  \label{eq:ktBbar}
  \ptarg{\bar B} (\eta) = 
  \sum_i \ptarg{i} \left( \Theta(\eta - \eta_i) +
    \Theta(\eta_i - \eta) e^{\eta - \eta_i}\right)\,.
\end{equation}
If the smallest distance is a $d_{ij}$, then the two (pseudo)
particles are clustered together into a (pseudo) particle whose
flavour is given by the sum of the two individual flavours. In
particular, when a bottom-flavoured particle and anti-particle end up
in the same jet, they will give rise to a non-flavoured jet. If the
smallest distance is a $d_{ib}$, the (pseudo) particle is considered
as a final jet and removed from the clustering sequence.

The large discrepancy between flavour-$k_T$ and $k_T$ results at small
$\mbb$, which is dominated by small $\Delta R_{bb}$ values, is due to
a different treatment of particular kinematic regions in the two
algorithms.
Specifically, when two $b$-quarks are separated by $\Delta R_{bb} >
R$, the $k_T$ algorithm typically reconstructs two $b$-jets. On the
other hand, the flavour-$k_T$ algorithm compares the relative distance
$d_{ij}$ with the distance of each bottom quark to the beam
$d_{bB}^{(F)}$ ($d_{b\bar B}^{(F)}$).  The latter quantity is by
constuction larger than the individual transverse momenta of the
bottom and anti-bottom quarks. Since $d_{ij}$ can be smaller than
$d_{bB}^{(F)}$ ($d_{b\bar B}^{(F)}$) even for $\Delta R_{bb} > R$, the
two bottom quarks can be clustered even if $\Delta R_{bb} > R$.
Similar arguments apply also in the region of large transverse
momenta, since when the Higgs boson is highly boosted, its decay
products tend to be collimated.

Apart from that, we note that even for bottom quarks with $\Delta
R_{bb} < R$ the output of the two clustering algorithms can be
different: in this case, both algorithms tend to cluster the two
bottom quarks together, giving rise to a $b$-jet in the $k_T$
algortihm and a gluon-jet in the flavour-$k_T$ one. It is thus more
likely that the event passes the fiducial cuts in the $k_T$ case,
because of the presence of one additional secondary bottom quark. In
this context a soft $b$ is often sufficient as it can be promoted to a
hard $b$-jet by clustering with a hard gluon. This clustering
mechanism is instead strongly suppressed in the flavour-$k_T$ one.  In
this case the anti-$k_T$ and $k_T$ algorithms do not actually
reconstruct the two hard $b$-jets originating from the Higgs-boson
decay.

Overall, it is clear that in various phase-space regions the
flavour-$k_T$ clustering leads to significant changes. It is therefore
not advisable to use this approach for theoretical predictions, unless
the same clustering is used also on the experimental side, which is
usually not the case. In fact, the application of the flavour-$k_T$
algorithm to experimental data is challenging, as discussed in
\citere{Banfi:2007gu}.  Alternatively, one should include a
theoretical estimate of the size of the discrepancy due to the use of
different algorithms in the theory simulations and measurements.
While in fixed-order calculations in the 5FS such approach is
mandatory to obtain IR-safe results as far as bottom-quark tagging is
concerned, in exclusive event simulations matched to parton showers
this problem can be circumvented by including the bottom-quark mass
through reshuffling the final-state momenta of the bottom quarks, so
that flavour-$k_T$ clustering can be avoided.  This is an example why
matching high accuracy calculations at fixed order and parton showers
is important and advantageous. One should bear in mind, however, that
such reshuffling is not completely unambiguous and comes with some
uncertainties on the bottom-quark kinematics.  Thus, ideally the
bottom-quark kinematics should be described using massive bottom
quarks at amplitude level , i.e.\ a four-flavour scheme (4FS)
calculation, but this is not always feasible at high accuracy with
current technology and also comes with other shortcomings
\cite{Ridolfi:2019bch}.  Furthermore, it is important to be aware that
in certain configurations a hard reconstructed $b$-jet can come from a
soft bottom quark and that the two selected $b$-jets may actually not
originate from the Higgs-boson decay. These occurrences are less
likely when using the flavour-$k_T$ algorithm.

\subsection{Comparison to ATLAS data}
\label{sec:data}

\begin{figure}[h]
  \begin{center}
  \includegraphics[width=.45\textheight]{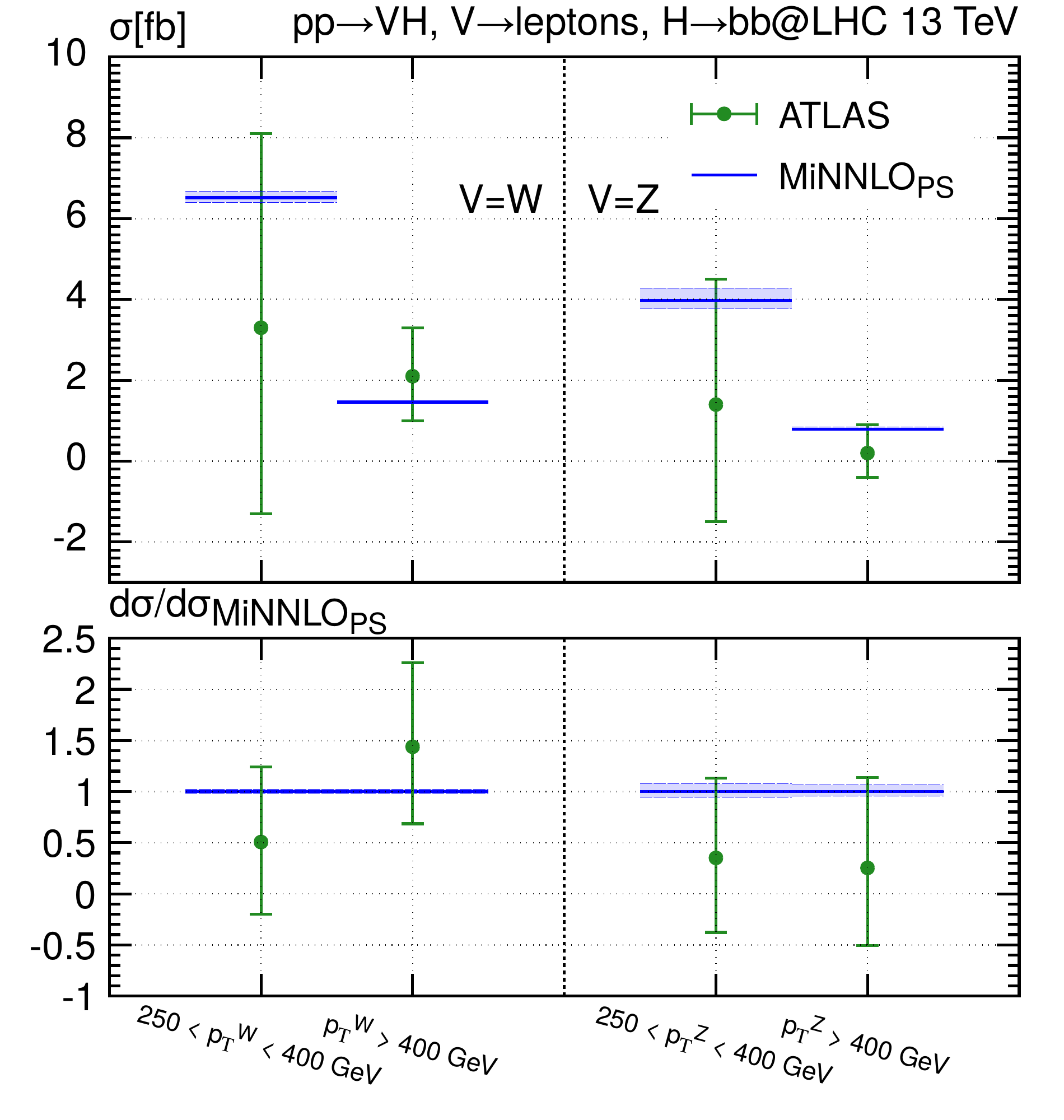}
  \caption{\label{fig:data} Comparison of \minnlo{} predictions to ATLAS data \cite{ATLAS:2020jwz}. }
  \end{center}
\end{figure}

Figure\,\ref{fig:data} compares our \minnlo{} predictions to recent
ATLAS data \cite{ATLAS:2020jwz}. The respective cross sections are
reported in \tab{tab:data}. The results correspond to \wh{} and \zh{}
production with all leptonic final states, i.e.\ $\ell^\pm\nu_\ell
b\bar{b}$, $\ell^+\ell^- b\bar{b}$ and $\nu_\ell\bar\nu_\ell b\bar{b}$
with $\ell \in\{e,\mu,\tau\}$. For this comparison we have included 
hadronization effects through \PYTHIA{8}, which however have a 
negligible impact.
It is clear that the measured \vh{} cross sections are fully compatible with our predictions within
uncertainties. However, one must bear in mind that this measurement
requires relatively large lower cuts on the transverse momentum of the
vector boson. Therefore, the experimental error is quite large, being
dominated mostly by the limited statistics. In fact, within the
current experimental uncertainties the measured cross sections are
compatible with zero within one standard deviation.  It will be
interesting to see how this comparison will develop when the
experiments have collected more data. Moreover, the original
publications \cite{ATLAS:2018kot,CMS:2018nsn}, where the $H\to b\bar
b$ decay was observed, involved a cut of only 150\,GeV on the
transverse momentum of the vector boson, but no unfolding to fiducial
cross sections was performed. Therefore, repeating the exclusive
data--theory comparison at the fiducial level including a cut of only
150\,GeV on vector-boson transverse momentum would be desirable in the
future.

\renewcommand\arraystretch{1.5}
\begin{table}[t]
  \centering
  \begin{tabular} {l | c | c}
    \Xhline{1pt}
    \multicolumn{3}{c}{\boldmath{$pp\to W^\pm H \to  \ell^\pm \nu_\ell b\bar{b}$}} \\    
    \hline
    $\sigma$ [fb]  & $p_{T}^{W} \in$ [250,400]\,GeV  & $p_{T}^{W} \in$ [400, $\infty$]\,GeV  \\    
    \hline
    \minnlo{} & ${6.52}^{+2.4\%}_{-1.8\%}$  & ${1.46}^{+2.5\%}_{-1.9\%}$  \\
    ATLAS \cite{ATLAS:2020jwz}  & ${3.3}^{+3.6 (\rm{Stat.}) + 3.2 (\rm{Syst.})}_{-3.4 (\rm{Stat.}) - 3.0 (\rm{Syst.})}$ & ${2.1}^{+1.0 (\rm{Stat.)} + 0.6 (\rm{Syst.})}_{-0.9 (\rm{Stat.}) -0.5 (\rm{Syst.})}$  \\
    \Xhline{1pt}
    \multicolumn{3}{c}{\boldmath{$pp\to ZH \to (\ell^+\ell^-,\nu_\ell\bar\nu_\ell)b\bar{b}$}} \\        
    \hline
    $\sigma$ [fb]  & $p_{T}^{Z} \in$ [250,400]\,GeV  & $p_{T}^{Z} \in$ [400, $\infty$]\,GeV  \\    
    \hline    
    \minnlo{} & ${3.98}^{+7.6\%}_{-5.4\%}$ & ${0.79}^{+6.5\%}_{-4.2\%}$  \\
    ATLAS \cite{ATLAS:2020jwz} & ${1.4}^{+2.4 (\rm{Stat.}) + 1.9 (\rm{Syst.})}_{-2.3 (\rm{Stat.}) -1.7(\rm{Syst.})}$ & ${0.2}^{+0.6 (\rm{Stat.}) + 0.3 (\rm{Syst.})}_{-0.5(\rm{Stat.}) - 0.3(\rm{Syst.})}$ \\        
    \Xhline{1pt}    
  \end{tabular}
  \caption{\label{tab:data}Fiducial cross sections in the \fidATLAS{} setup compared to data.}
  \label{xsectionsATLAS}
\end{table}
\renewcommand\arraystretch{1}

\section{Conclusions}
\label{sec:summary}

We have presented novel predictions for \vh{} production with $H\to
b\bar b$ decay within the \minnlo{} framework.  NNLO QCD accuracy is
retained for both production and decay, while the matching to the
\PYTHIA{8} parton shower ensures a fully realistic exclusive
description of the process at the level of hadronic events.  For the
\wh{} process NNLO+PS accuracy is achieved for the first time for
production and decay.

Our results have been validated against earlier calculations using the
NNLOPS approach, which exploits a numerically intensive
multidifferential reweighting to upgrade \minlo{} events to NNLO QCD
accuracy a posteriori.  We find fully compatible results within scale
uncertainties when comparing our \minnlo{} predictions to earlier
NNLOPS results for \wh{} production with an on-shell Higgs boson and
for \zh{} production including the $H\to b\bar b$ at NNLO+PS.  The
agreement is particularly remarkable considering the fact that the
scale uncertainties of both predictions are at the level of only 1-2\%
in most phase-space regions. Given that the reweighting procedure of
the NNLOPS approach induces some (technical) limitations and for \vh{}
production it relies on certain approximations in describing the
vector-boson decay, our new \minnlo{} \vh{} generator, which includes
NNLO corrections on-the-fly directly during the usual \POWHEG{} event
generation, lifts those shortcomings and can be considered to
supersede the previous NNLOPS implementation.  We stress, however,
that the NNLOPS predictions can still be useful to assess residual
uncertainties.

Phenomenological results have been discussed in detail for both the
$pp \to e^\pm \nu_e b\bar{b}$ and $pp \to e^{+} e^{-} b\bar{b}$
processes at the level of inclusive and fiducial cross section as well
as for differential distributions in the fiducial phase space. We find
that the NNLO corrections induced by the \minnlo{} procedure with
respect to \minlo{} increase both the inclusive and fiducial cross
sections by roughly $5-6$\%, while the corrections are relatively flat
in phase space for the observables under consideration. Moreover, due
to the relatively small scale dependence of the \vh{} processes,
uncorrelating scale variations in production and decay to estimate the
perturbative uncertainties is crucial to obtain \minlo{} and \minnlo{}
results that are compatible within their scale bands. Regardless of
this, we observe a clear reduction of scale uncertainties, by roughly
a factor of two, when including the \minnlo{} corrections.  As far as
the $H\to b\bar b$ decay is concerned, we have shown that including
the full NNLO QCD corrections induces important effects with respect
to a LO description through \PYTHIA{8} for various observables: For
the cuts under consideration, they shift the fiducial cross section by
about $-5$\% and induce relevant shape effects. In some cases, these
effects are not captured by the scale-variation uncertainty from
\PYTHIA{8}.
Finally, our results are in
agreement with current ATLAS data, keeping in mind that with the
current experimental accuracy the measured \vh{} cross section is
still compatible with zero.

While at the level of the matrix elements our calculations involve
massless bottom quarks, kinematic bottom-mass effects are included
approximately through a reshuffling of the bottom quarks at LHE level.
In this setup, we have compared the anti-$k_T$, $k_T$ and
flavour-$k_T$ algorithms. The anti-$k_T$ and $k_T$ algorithms give
rather similar results, and in the bulk region of the phase space this
is also the case for the flavour-$k_T$ algorthim. However, in some
phase-space regions, such as at small $\mbb$ and at large transverse
momenta, we observe substantial differences in the flavour-$k_T$ case.
Since experimental analyses usually employ the standard (anti-)$k_T$
algorithm, it is important to be aware of these differences and
features, especially when using different jet-clustering algorithms in
the theoretical calculations and in experimental measurements.

We have performed largely independent implementations in both
\POWHEGBOXVTWO{} and \POWHEGBOXRES{}, whose results are fully
consistent. We reckon that our \minnlo{} \vh{} generators will be very
useful for future analyses of these processes in Run III and, in
particular, for the High Luminosity LHC phase. Therefore, we will make
all relevant codes to simulate NNLO+PS events for the $e^\pm \nu_e
b\bar{b}$ and $e^{+} e^{-} b\bar{b}$ final states publicly available
within the \POWHEGBOXRES{} framework.

\section*{Acknowledgements}

We are grateful to Daniele Lombardi for various fruitful discussions
and to Carlo Oleari for some clarifications. We thank Wojcek Bizon for
his support with the original results of \citere{Bizon:2019tfo} and
for providing the relevant event samples as well as
Francesco Tramontano for assistance in the use of the VHNNLO code.
GZ thanks Gavin Salam for discussions on the comparison of different jet-algorithms. 
We have used the Max Planck
Computing and Data Facility (MPCDF) in Garching to carry out part of
the simulations related to this work, as well as the CNRS/IN2P3
Computing Center in Lyon. 
 Silvia Zanoli is supported by the
International Max Planck Research School (IMPRS) on ``Elementary
Particle Physics''.
This research was supported in part by the National Science Foundation under Grant No. NSF PHY-1748958
in the context of the KITP programme ``New Physics from Precision at High Energies''.

\appendix
\section{Results for \wmh{} production}
\label{app:Wmresults}

\begin{figure}[p!]
\begin{center}\vspace{-0.2cm}
\begin{tabular}{cc}
\includegraphics[width=.31\textheight]{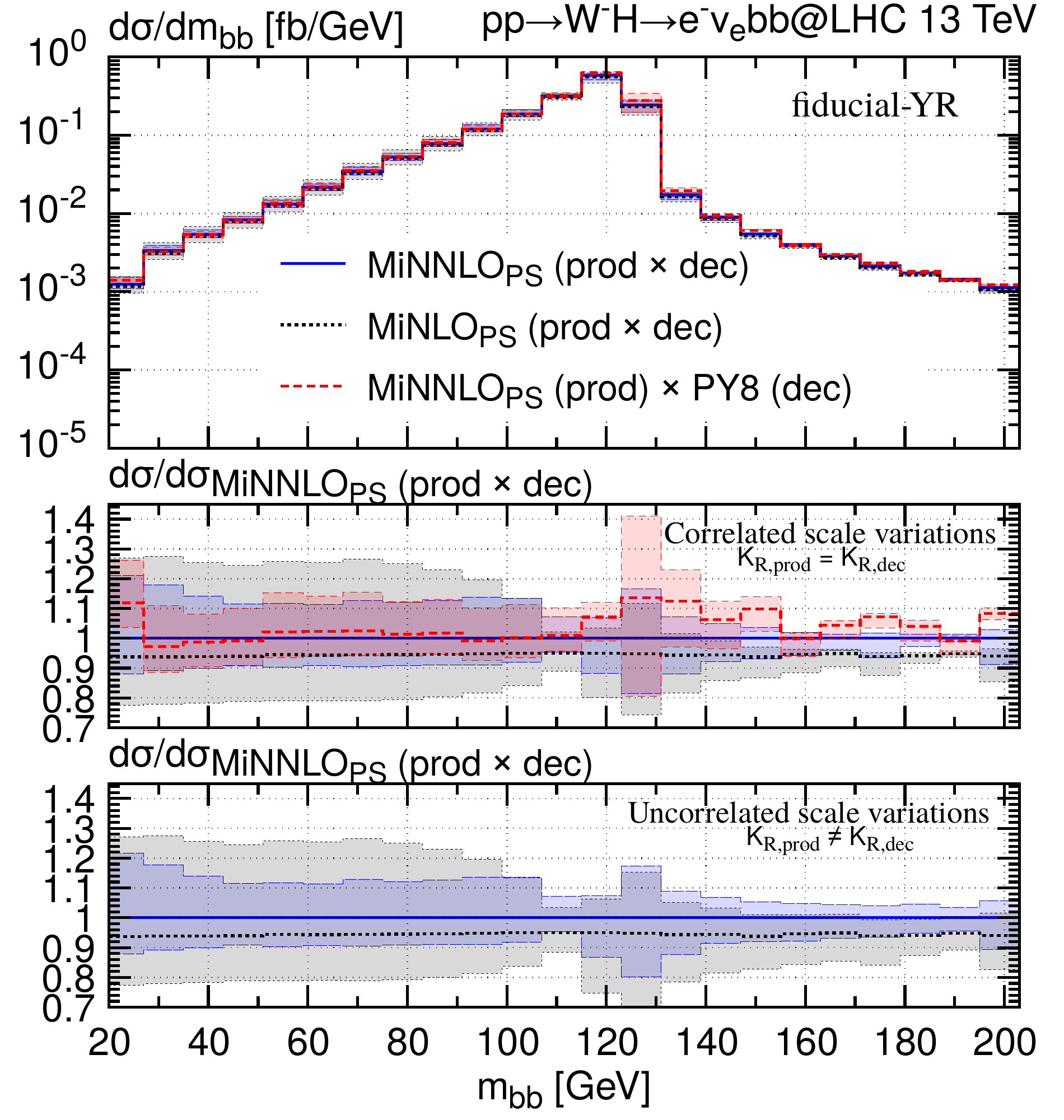} 
&
\includegraphics[width=.31\textheight]{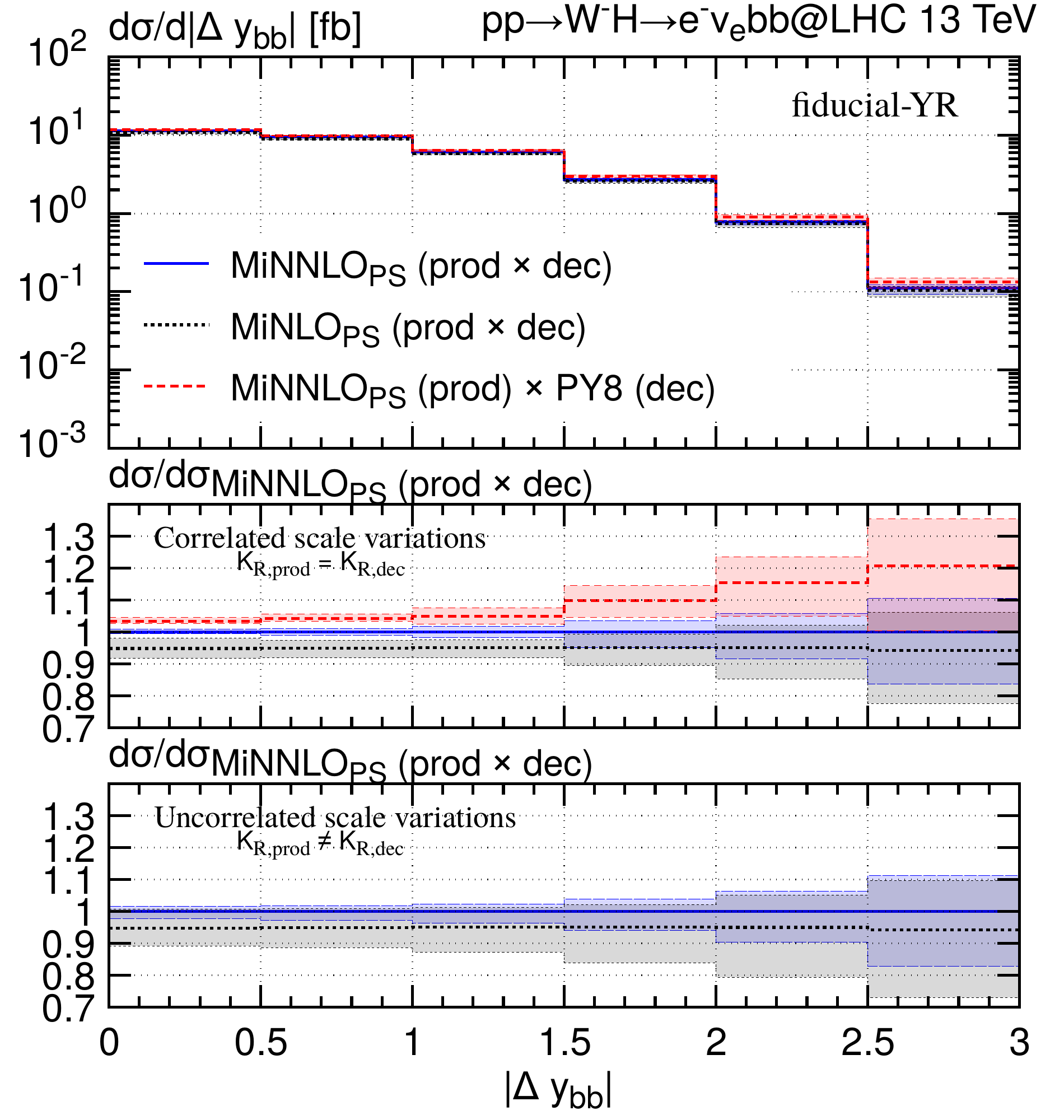}
\end{tabular}\vspace{-0.15cm}
\begin{tabular}{cc}
\includegraphics[width=.31\textheight]{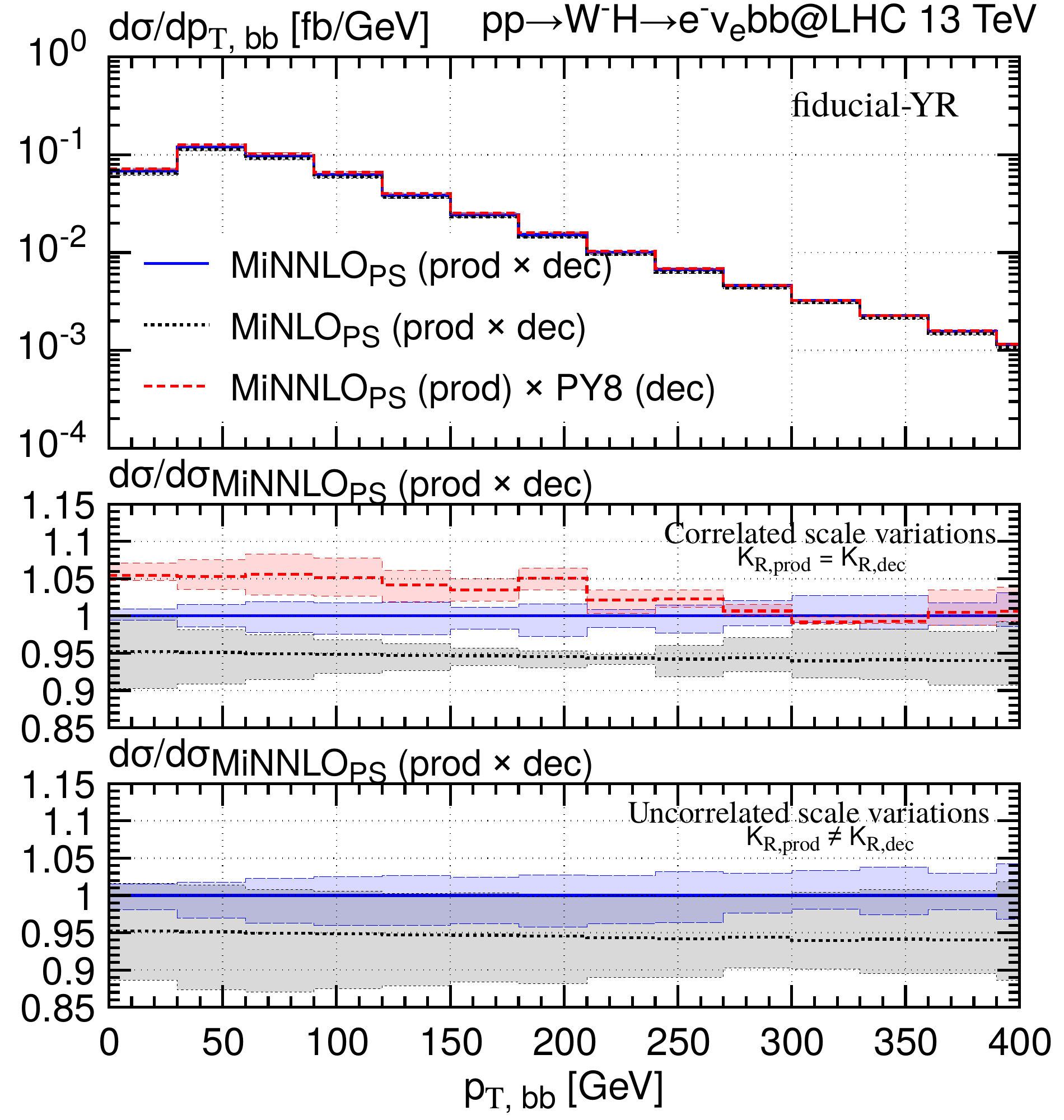}
&
\includegraphics[width=.31\textheight]{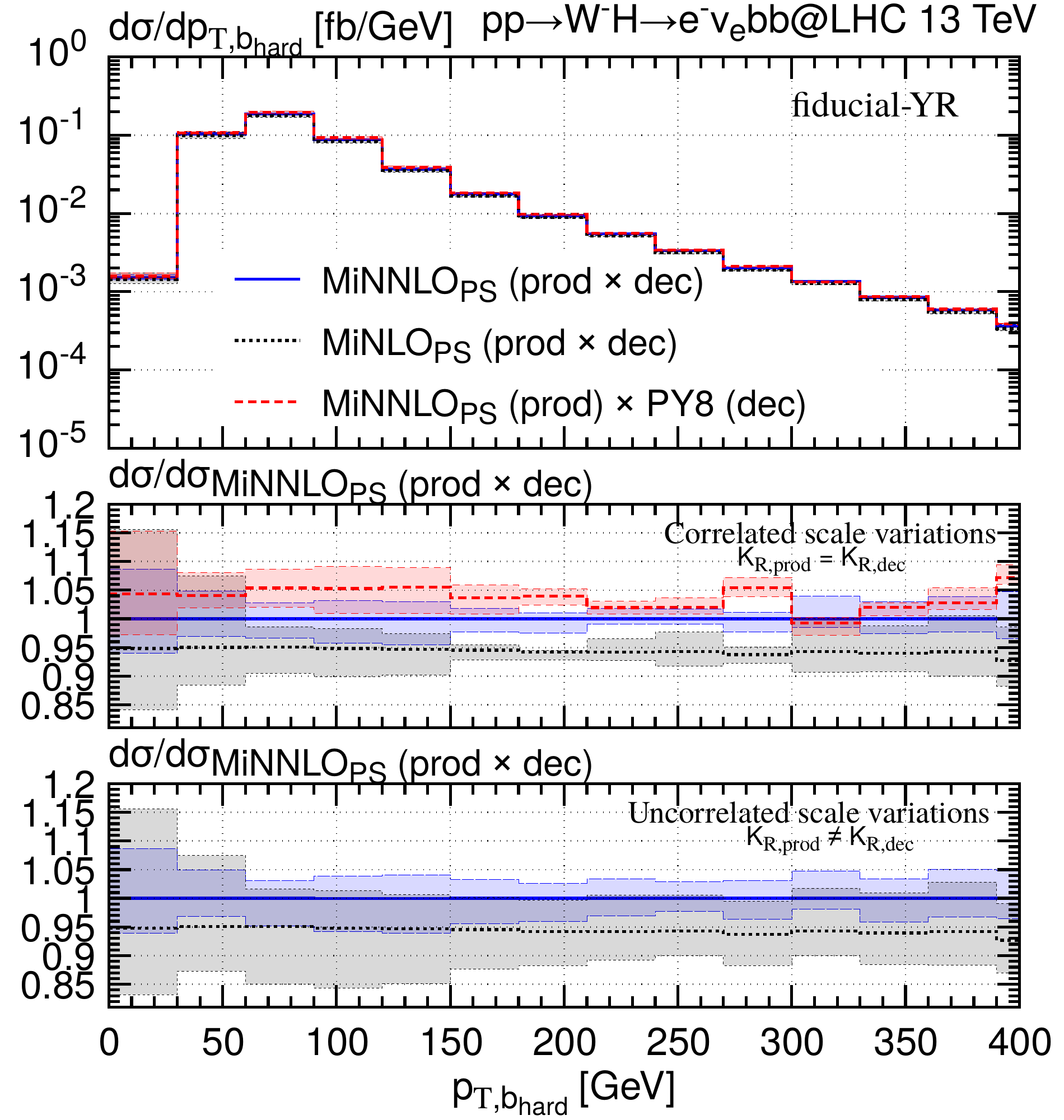}
\end{tabular}\vspace{-0.15cm}
\begin{tabular}{cc}
\includegraphics[width=.31\textheight]{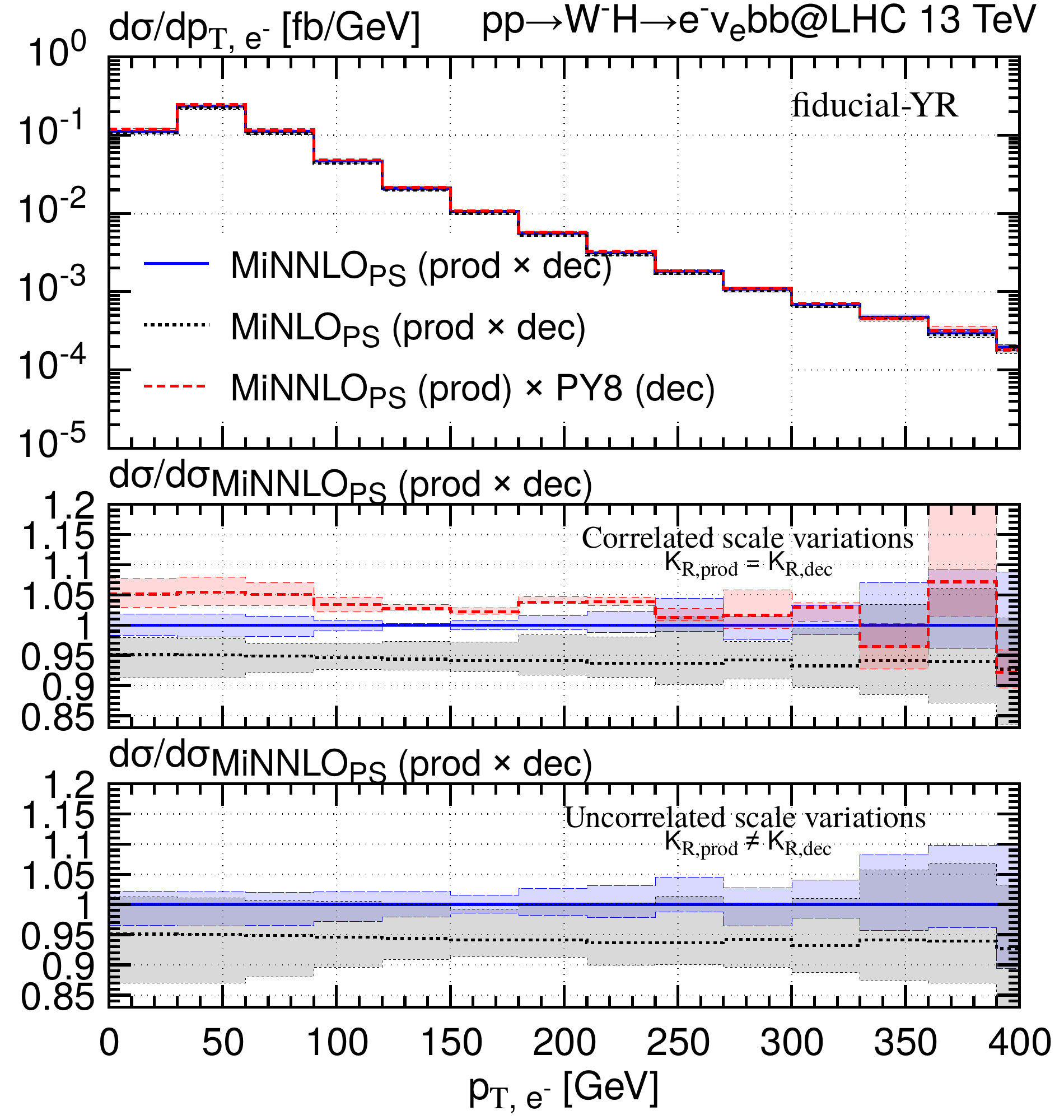}
&
\includegraphics[width=.31\textheight]{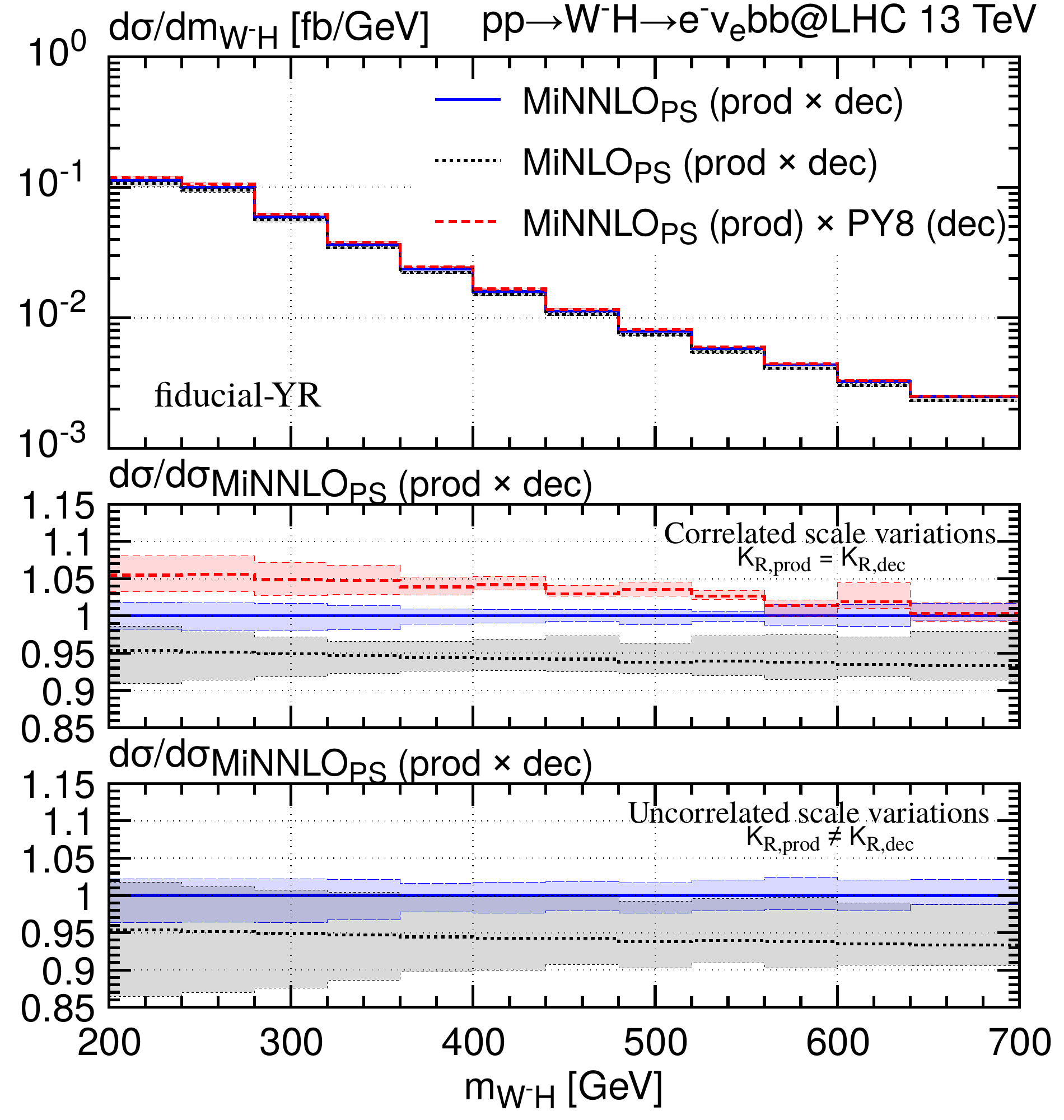} 
\end{tabular}
\caption{\label{fig:resultswhappendix}  Differential distributions for $e^{-}\bar \nu  b\bar{b}$ production with \fidYR{} cuts. $K_{\rm R,prod}$ and $K_{\rm R,dec}$ refer to the variation factors of the renormalization scales of production and decay, respectively. }
\end{center}
\end{figure}

In this section, we report results for \wmh{}
productions. Figure\,\ref{fig:resultswhappendix} shows plots
corresponding to those presented for \wph{} production in
\sct{sec:dist}.  Since the same conclusions that were drawn for \wph{}
production in that section hold also for the \wmh{} process we refrain
from commenting on the results further.

\afterpage{\clearpage}

\bibliography{MiNNLO}
\bibliographystyle{JHEP}

\end{document}